\documentclass[a4paper,12pt]{article}
\pdfoutput=1
\usepackage{graphicx,subfigure,amsmath,amssymb,multirow}
\usepackage{cite}
\usepackage{mathtools}
\usepackage{float}

\renewcommand{\baselinestretch}{1.5}
\newlength{\dinwidth}
\newlength{\dinmargin}
\def\be{\begin{equation}}
\def\ee{\end{equation}}
\def\ba{\begin{eqnarray}}
\def\ea{\end{eqnarray}}
 \def\la{ \langle}
  \def\ra{ \rangle}
     \def\e{ \epsilon}
      \def\r{ \gamma}
       \def\lbd{\lambda}
        \def \d {{\rm d}}
           \def\w{\omega}
            \def\u{\mu}
              \def\a{\alpha}
  \def\b{\beta}
\def\v{\nu}
   \def\sg{\sigma}
      
     \def\ve{ \varepsilon}
          \def\qb{{ \bf q}_\bot}
           \def\kb{{ \bf k}_\bot}

\usepackage{color}

\allowdisplaybreaks

\begin{document}
\title{\bf  Tensor form factors of $P\to P,\,S,\,V$ and $A$ transitions within the standard and the covariant light-front approaches}
\author{Qin Chang$^{a,b}$\footnote{changqin@htu.edu.cn}, Xiao-Lin Wang$^{a}$ and Li-Ting Wang$^{a}$ \\
{ $^a$\small Institute of Particle and Nuclear Physics, Henan Normal University, Henan 453007, China}\\
{ $^b$\small Institute of Particle Physics and Key Laboratory of Quark and Lepton Physics~(MOE) }\\[-0.2cm]
{ \small Central China Normal University, Wuhan, Hubei 430079, China}
}
\date{}

\maketitle
\begin{abstract}
In this paper, we  investigate  the tensor form factors of $P\to P,\,S,\,V$ and $A$  transitions within the standard light-front~(SLF) and the covariant light-front~(CLF) quark models~(QMs). The self-consistency and Lorentz covariance of  CLF QM are analyzed  via these quantities, and the effects of zero-mode are discussed. For the $P\to V$ and $A$ transitions, besides the inconsistence between the results extracted via longitudinal and transverse polarization states, which is caused by the residual $\w$-dependent spurious contributions, we find  and analyze a ``new'' self-consistence problem of the traditional CLF QM, which is caused by the different strategies  for dealing deal with the trace term in CLF matrix element. A possible solution to the problems of traditional CLF QM  is discussed and confirmed numerically. Finally, the theoretical predictions for the tensor form factors of some $c\to q,\,s$   and $b\to q,\,s\,,c$~($q=u,d$) induced  $P\to P,\,S,\,V$ and $A$  transitions are updated within the CLF QM with a self-consistent scheme.
\end{abstract}
\newpage
\section{Introduction }
The heavy-to-light exclusive weak decays provide a fertile ground for testing the Standard Model~(SM) and looking for physics beyond it.   In the calculation of the amplitudes of these decays, some nonperturbative quantities, such as decay constant, distribution amplitudes and form factors, are essential  and important inputs. For instance,  the dominant contribution to the amplitude of $b\to s\r$ radiative decay  is proportional to the form factors associated with tensor current.  
These quantities can be evaluated in many different approaches, such as  Wirbel-Stech-Bauer model~\cite{Wirbel:1985ji}, lattice QCD~\cite{Daniel:1990ah}, QCD sum rules~\cite{Shifman:1978bx,Shifman:1978by} and light-front quark models~(LF QMs)~\cite{Terentev:1976jk,Berestetsky:1977zk,Cheng:1997au,Carbonell:1998rj,Jaus:1999zv}.

The LF QMs can be roughly classified into two types: the standard light-front~(SLF) QM~\cite{Terentev:1976jk,Berestetsky:1977zk} and the covariant light-front~(CLF) QM~\cite{Cheng:1997au,Carbonell:1998rj,Jaus:1999zv}. The SLF QM  is a relativistic quark model based on the LF formalism~\cite{Dirac:1949cp} and LF quantization of QCD~\cite{Brodsky:1997de}, it provides a conceptually simple and phenomenologically feasible framework for evaluating nonperturbative quantities. However, the matrix element evaluated in this approach lacks manifest Lorentz covariance, and therefore, it is replaced later by the CLF QM. A popular framework for the CLF QM  is developed by Jaus~\cite{Jaus:1999zv} with  the help of a manifestly covariant Bethe-Saltpeter~(BS) approach as a guide to the calculation. In this approach, the  zero-mode contributions can be well determined, and the result of the matrix element is expected to be covariant because  the  $\w$-dependent spurious contributions, where $\w^\u=(0,2,{\bf 0}_\bot)$ is the light-like four-vector used to define  light-front by $\w\cdot x=0$  and the $\w$-dependent contributions may violate the covariance, can be eliminated  by inclusion of zero-mode contributions~\cite{Jaus:1999zv}. The LF QMs  have been widely used to evaluate some nonperturbative quantities  of hadrons, and are further applied to phenomenological researches~\cite{Jaus:1989au,Jaus:1989av,Jaus:1991cy,Jaus:1996np,Cheng:1996if,ODonnell:1996sya,Cheung:1996qt,Choi:1996mq,Choi:1997iq,Choi:1998jd,Ji:2000fy,DeWitt:2003rs,Choi:2007yu,Choi:2007se,Barik:1997qq,Hwang:2000ez,Hwang:2010hw,Hwang:2001zd,Hwang:2009cu,Geng:2016pyr,Chang:2016ouf,Chang:2017sdl,Chang:2018aut,Chang:2018mva,Bakker:2000pk,Bakker:2002mt,Bakker:2003up,Choi:2004ww,Choi:2009ym,Choi:2010ha,Choi:2010zb,Choi:2011xm,Choi:2010be,Choi:2014ifm,Choi:2017uos,Choi:2017zxn,Ryu:2018egt,Hwang:2001hj,Hwang:2001wu,Hwang:2010iq,Cheung:2014cka,Wang:2008ci,Wang:2008xt,Shen:2008zzb,Wang:2009mi,Cheng:2017pcq,Kang:2018jzg,Verma:2011yw,Shi:2016gqt,Wang:2018duy,Jaus:2002sv,Cheng:2003sm,Cheng:2004cc,Cheng:2004ew,Wei:2009np,Ke:2017eqo,Wang:2017mqp,Zhao:2018zcb,Zhao:2018mrg,Chua:2018lfa,Hu:2020mxk,Choi:2019wqx,Dhiman:2019ddr,Kumar:2019eck,Ke:2019smy,Ke:2019lcf,Ke:2020ief}.  In this paper,  we shall pay our attention to the form factors related to the tensor current  matrix elements.

The tensor form factors of $B\to \pi\,,K\,,\rho$ and $K^*$ transition have been evaluated in the SLF QM with $\qb=0$ frame~\cite{Geng:2001de}.  Within the CLF QM, the  tensor form factors of $B_{u,d}\to V\,,A$ and $T$ transitions are calculated in Ref.~\cite{Cheng:2004yj} and are corrected in Refs.~\cite{Cheng:2010yj,Cheng:2009ms}; the corrected theoretical results are further applied to  the phenomenological studies of some radiative  $B$ and $B_s$ decays~\cite{Cheng:2009ms}  and radiative  $D$ and $D_s$ decays~\cite{Shen:2013oua}. It is worth checking these previous results of tensor form factors, and evaluating the transitions which are not considered before.  In addition, it should be noted that above-mentioned  works are performed within the traditional CLF QM~\cite{Jaus:1999zv}, which however has covariance and self-consistence problems.

It has been noted for a long time that the traditional CLF approach~\cite{Jaus:1999zv} suffers from a  self-consistence problem in the vector meson system. For instance, the CLF results for the decay constant of vector meson, $f_V$, obtained via  longitudinal ($\lbd=0$) and  transverse ($\lbd=\pm$) polarization states are inconsistent with each other, {\it i.e.} $[f_V]^{\lbd=0}\neq [f_V]^{\lbd=\pm}$~\cite{Cheng:2003sm},  because the former receives an additional contribution characterized by the $B_1^{(2)}$ function. Some analyses has been  made in Ref.~\cite{Choi:2013mda}, and  the authors present a possible solution to the self-consistence problem by introducing  a modified correspondence between the  covariant BS  approach and the LF approach~(named as type-II scheme~\cite{Choi:2013mda}), which requires an additional $M\to M_0$ replacement relative to the traditional correspondence scheme~(named as type-I scheme~\cite{Choi:2013mda}).

In our previous works~\cite{Chang:2018zjq,Chang:2019mmh,Chang:2019obq},  the self-consistence problem has also been studied in detail via  $f_{P,V,A}$ and form factors of $P\to (P,V)$ and $V\to V$ transitions associated with the (axial-)vector current, and the modified type-II  correspondence scheme as a solution to the  self-consistence problem~\cite{Choi:2013mda}  is carefully  tested. Besides, we have also found that: the covariance of the traditional CLF QM in fact can not be maintained strictly due to the residual $\w$-dependent contributions; the self-consistence and covariance problems have the same origin and can be resolved  simultaneously by employing the modified type-II scheme. In this paper, we would like to extend our previous works on above issues to the tensor form factors of $P\to P,\,S,\,V$ and $A$ transitions, and update the theoretical results within a self-consistence scheme.  In addition, we will also show  another ``new'' self-consistence problem of the CLF QM, which has not been noted before.

Our paper is organized as follows. In section 2, we  review briefly the SLF and the CLF QMs for convenience of discussion, and then present our theoretical results for the tensor form factors of  $P\to P,\,S,\,V$ and $A$ transitions. In section 3, the self-consistency and covariance of CLF QM are discussed in detail, and our numerical results for the tensor form factors of some $c\to q,\,s$   and $b\to q,\,s\,,c$~($q=u,d$) induced  $P\to P,\,S,\,V$ and $A$  transitions are presented. Finally, our summary is given in section 5. Some previous theoretical results are collected in appendix A for  convenience of discussion and comparison, and the values of input parameters used in the computation are collected in  appendix B.

\section{ Theoretical Framework and Results}
The hadronic matrix elements associated with tensor operators are commonly factorized in terms of tensor form factors as
\begin{align}
\label{eq:defPP}
\langle P''(p{''})|\bar q''_1 \sigma^{\mu\nu}q'_1|P'(p{'}) \rangle
=& i(P^{\mu}q^{\nu}-P^{\nu}q^{\mu})\frac{F_T(q^2)}{M'+M''}\,,\\
\label{eq:defPS}
\langle S''(p'')|\bar q''_1 \sigma^{\mu\nu}\r_5q'_1|P'(p{'}) \rangle
=& i(P^{\mu}q^{\nu}-P^{\nu}q^{\mu})\frac{U_T(q^2)}{M'+M''}\,,
\end{align}
for $P\to P$ and $P\to S$ transitions, respectively, where $P=p'+p''$, $q=p'-p''$ and $M^{\prime(\prime\prime)}$ is the mass of initial~(final) state. For the $P\to V$ and $P\to A$ transitions, the tensor form factors are defined as
\begin{align}
\label{eq:defPV}
\la  V(p'',\e) | \bar{q}''_1  \sg^{\u\v}   \ q'_1 |P(p') \ra =
&-\ve^{\u\v\a\b}\bigg\{ -\e_{\a }^*P_\b T_1(q^2)+\frac{M'^2-M''^2}{q^2}\e_{\a }^*q_\b\left[ T_1(q^2)- T_2(q^2)\right]\nonumber\\
&-\frac{\e^{ *}\cdot q}{q^2} P_\a q_\b\left[T_1(q^2)- T_2(q^2)-\frac{q^2}{M'^2-M''^2} T_3(q^2) \right]\bigg\}\,,\\
\label{eq:defPA}
\la  \,^{i}\!A(p'',\e) | \bar{q}''_1  \sg^{\u\v}\r_5   \ q'_1 |P(p') \ra =
&\ve^{\u\v\a\b}\bigg\{ -\e_{\a }^*P_\b T_1^{(i)}(q^2)+\frac{M'^2-M''^2}{q^2}\e_{\a }^*q_\b\left[ T_1^{(i)}(q^2)- T_2^{(i)}(q^2)\right]\nonumber\\
&-\frac{\e^{ *}\cdot q}{q^2} P_\a q_\b\left[T_1^{(i)}(q^2)- T_2^{(i)}(q^2)-\frac{q^2}{M'^2-M''^2} T_3^{(i)}(q^2) \right]\bigg\}\,,
\end{align}
where, $\ve^{0123}=1$; $^{i}\!A$ with $i=1$ and $3$ denote $^{2S+1}\!L_J$$=$$^1\!P_1$ and $^3\!P_1$ states, respectively; and for the form factors in Eq.~\eqref{eq:defPA},   the superscript ``$(i)$'' with  $i=1$ and $3$ are added  in order to  distinguish  $P$$\to$$^{1}\!A$ and $P$$\to$$ ^{3}\!A$ transitions. The definitions, Eqs.~\eqref{eq:defPV} and \eqref{eq:defPA}, are equivalent to
\begin{align}
\label{eq: PVT2}
\la  V(p'',\e) | \bar{q}''_1  \sg^{\u\v} q_\v  \ q'_1 |P(p') \ra =&\ve^{\u\v\a\b}\e_{\v}^* P_\a q_\b T_1(q^2)\,,\\
\label{eq: PVT3}
\la  V(p'',\e) | \bar{q}''_1  \sg^{\u\v}\r_5 q_\v  \ q'_1 |P(p') \ra =
&-i\left[ (M'^2-M''^2)\e^{\u *} -\e^*\cdot q P^\u \right]T_2(q^2)\nonumber\\
&-i\e^*\cdot q\left[ q^\u-\frac{q^2}{M'^2-M''^2}P^\u \right]T_3(q^2)\,,
\end{align}
and
\begin{align}
\label{eq: PAT2}
\la \,^{i}\! A(p'',\e) | \bar{q}''_1  \sg^{\u\v}\r_5 q_\v  \ q'_1 |P(p') \ra =&-\ve^{\u\v\a\b}\e_{\v}^* P_\a q_\b T_1^{(i)}(q^2)\,,\\
\label{eq: PAT3}
\la  \,^{i}\!A(p'',\e) | \bar{q}''_1  \sg^{\u\v} q_\v  \ q'_1 |P(p') \ra =
&i\left[ (M'^2-M''^2)\e^{\u *} -\e^*\cdot q P^\u \right]T_2^{(i)}(q^2)\nonumber\\
&+i\e^*\cdot q\left[ q^\u-\frac{q^2}{M'^2-M''^2}P^\u \right]T_3^{(i)}(q^2)\,,
\end{align}
respectively.

The main work of LF approaches is to evaluate the current matrix element of $M'\to M''$ transition,
\begin{align}\label{eq:amp1}
{\cal B} \equiv \la  M''(p'') | \bar{q}''_1 (k_1'')\Gamma q'_1(k_1') |M'(p') \ra \,,\qquad \Gamma=\sigma_{\mu\nu},\,\sigma_{\mu\nu}\r_5,\,...
\end{align}
which will be further used to extract the form factors by matching to the definitions given above.

\subsection{ Theoretical results in the SLF QM}
The SLF and CLF QMs have been fully illustrated in, for instance, Refs.~\cite{Terentev:1976jk,Berestetsky:1977zk,Jaus:1989au,Jaus:1989av,Hwang:2010hw} and Refs.~\cite{Jaus:1999zv,Jaus:2002sv,Cheng:2003sm,Choi:2013mda}, respectively. One may refer to these literatures for detail. In this paper, we take the same notations and conventions as Refs.~\cite{Chang:2018zjq,Chang:2019mmh,Chang:2019obq}.

In the framework of the SLF QM, the matrix element, Eq.~\eqref{eq:amp1}, can be written as~\cite{Chang:2018zjq,Chang:2019mmh,Chang:2019obq}
\begin{eqnarray}
{\cal B}_{\rm  SLF}=\sum_{h'_1,h''_1,h_2} \int  \frac{\d x \,\d^2{ \mathbf{k}_\bot'}}{(2\pi)^3\,2x}  {\psi''}^{*}(x,\mathbf{k}_{\bot}''){\psi'}(x,\mathbf{k}_{\bot}')
S''^{\dagger}_{h''_1,h_2}(x,\mathbf{k}_{\bot}'')\,C_{h''_1,h'_1}(x,\mathbf{k}_{\bot}',\mathbf{k}_{\bot}'')\,S'_{h'_1,h_2}(x,\mathbf{k}_{\bot}')\,,
\label{eq:B}
\end{eqnarray}
where $C_{h''_1,h'_1}(x,\mathbf{k}_{\bot}',\mathbf{k}_{\bot}'') \equiv  \bar{u}_{h''_1}(x,\mathbf{k}_{\bot}'')  \Gamma   u_{h'_1}(x,\mathbf{k}_{\bot}')$ corresponds to the operator in Eq.~\eqref{eq:amp1}, $x$ and $\kb'$ are the  internal  LF relative momentum variables. The momenta of quark $q_1'$ and spectator anti-quark $\bar{q}_{2}$ in the initial state have been written in terms of $(x,\kb')$ as
\begin{align}\label{eq:momk1}
k_1'^+=xp'^+\,,\quad\, \mathbf{k}_{1\bot}'=x\mathbf{p}_{\bot}'+\mathbf{k}_{\bot}' \,;\qquad  k_2^+=\bar{x}p'^+ \,,\quad\, \mathbf{k}_{2\bot}=\bar{x}\mathbf{p}_{\bot}'-\mathbf{k}_{\bot}'\,,
\end{align}
where, $\bar{x}=1-x$. For convenience of calculation, it is usually assumed that the initial state moves along with $z$-direction, which implies that $\mathbf{p}_{\bot}'=0$. Taking the convenient Drell-Yan-West frame, $q^+=0$,  where $q\equiv p'-p''=k_1'-k_1''$ is the momentum  transfer, the momentum of quark $q_1''$ in the final state can be written as
\begin{align}\label{eq:momk1p}
k_1''^+=xp''^+=xp'^+\,,\quad\, \mathbf{k}_{1\bot}''=x\mathbf{p}_{\bot}''+\kb''=-x\qb+\kb''\,,
\end{align}
where $\kb''=\kb'-\bar{x}\qb$.

In Eq.~\eqref{eq:B},  $\psi(x,\kb)$ and $S_{h_1,h_2}(x,\kb)$ are  the radial and the  spin-orbital wavefunctions~(WFs). For the former, we adopt commonly used Gaussian-type WFs, which are written as 
\begin{align}
\psi_s(x,\kb) =&4\frac{\pi^{\frac{3}{4}}}{\beta^{\frac{3}{2}}} \sqrt{ \frac{\partial k_z}{\partial x}}\exp\left[ -\frac{k_z^2+\kb^2}{2\beta^2}\right]\,,\\
\psi_{p}(x,\kb) =&\frac{\sqrt{2}}{\beta}\psi_s(x,\kb) \,,
\end{align}
for s-wave and p-wave mesons, respectively. The Gaussian parameters $\beta$ can be determined by fitting to data, and $k_z$ is the relative momentum in the $z$-direction and can be written as
\begin{align}
 k_z=(x-\frac{1}{2})M_0+\frac{m_2^2-m_1^2}{2 M_0}\,,
\end{align}
with the invariant mass defined by
\begin{align}
M_0^2=\frac{m_1^2+\mathbf{k}_{\bot}^2}{x}+\frac{m_2^2+\mathbf{k}_{\bot}^2}{\bar{x}}\,.
\end{align}
For the later, $S_{h_1,h_2}(x,\mathbf{k}_{\bot}) $, it can be obtained by the interaction-independent Melosh transformation, and finally written as a covariant form~\cite{Jaus:1989av,Cheng:2003sm},
\begin{align}\label{eq:defS2}
S_{h_1,h_2}=\frac{\bar{u}(k_1,h_1)\Gamma_M v(k_2,h_2)}{\sqrt{2} \hat{M}_0}\,,
\end{align}
where $\hat{M}_0^2=M_0^2-(m_1-m_2)^2$. For the $P$, $S$, $V$ and $A$ states, $\Gamma_M$ has the form
\begin{align}
\Gamma_P=&\r_5\,,\\
 \Gamma_V=&-\not\!\hat{\e}+\frac{\hat{\e}\cdot (k_1-k_2)}{D_{V,{\rm LF}}}\,,\\
\Gamma_S=&\frac{\hat{M}_0^2}{2\sqrt{3}M_0}\,,\\
\Gamma_{ ^1\!A}=&-\frac{1}{D_{1,{\rm LF}}}\hat{\e}\cdot (k_1-k_2) \r_5\,,\\
 \Gamma_{ ^3\!A}=&-\frac{\hat{M}_0^2}{2\sqrt{2} M_0}\left[  \not\!\hat{\e}+\frac{\hat{\e}\cdot (k_1-k_2)}{D_{3,{\rm LF}}} \right]\r_5 \,,
\end{align}
where $D_{V,{\rm LF}}=M_0+m_1+m_2$, $D_{1,{\rm LF}}=2$, $D_{3,{\rm LF}}=\hat{M}_0^2/(m_1-m_2)$ and
\begin{align}
\hat{\epsilon}^{\mu}_{\lbd=0}=&\frac{1}{M_0}\left(p^+,\frac{-M_0^2+\mathbf{p}_{\bot}^2}{p^+},\mathbf{p}_{\bot}\right)\,,\\
\hat{\epsilon}^{\mu}_{\lbd=\pm}=&\left(0,\frac{2}{p^+}\boldsymbol{\epsilon}_{\bot}\cdot \mathbf{p}_{\bot}, \boldsymbol{\epsilon}_{\bot}\right)\,,
\quad \boldsymbol{\epsilon}_{\bot}\equiv \mp \frac{(1,\pm i)}{\sqrt{2}}\,.
\end{align}

Using the formulas given above, one can obtain the explicit expression of ${\cal B}_{\rm  SLF}$, which is further used to extract the form factors. The  form factor  in the SLF QM can be written as
\begin{equation}\label{eq:FSLF}
[{\cal F}(q^2)]_{\rm  SLF}=\int\frac{\d x\,\d^2{\bf k_\bot'}}{(2\pi)^3\,2x}\frac{{\psi''}^*(x,{\bf k''_\bot})\,{\psi'}(x,{\bf k_\bot'})}{2\hat {M}'_0\hat {M}''_0}\,{\cal \widetilde{F}}^{\rm SLF}(x,{\bf k}_\bot',q^2)\,.
\end{equation}
For the $P\to P$ and  $P\to S$ transitions, taking $\u=+$ and $\v=\bot$, we finally obtain
\begin{align}
\widetilde{ F}_T^{\rm SLF}=&-\frac{2(M'+M'')(m'_{1}{\bf k}_{\perp}''\cdot{\bf q}_{\perp}-m''_{1}{\bf k}_{\perp}'\cdot{\bf q}_{\perp}
-xm_{2}{\bf q}^{2}_{\bot})}{{\bf q}^{2}_{\bot}}\,,\\
\widetilde{ U}_T^{\rm{ SLF}}=&\frac{\hat M''^2_0}{2\sqrt3M''_0}\widetilde{F}_T^{\rm SLF}[m''_1\rightarrow -m''_1]\,,
\end{align}
where, ``$\widetilde{F}_T^{\rm SLF}[m''_1\rightarrow -m''_1]$'' means replacing $m''_1$ in $\widetilde{F}_T^{\rm SLF}$ by $-m''_1$.
For the $P\to V$ and  $P\to A$ transitions, we take $\lbd=+$ and  multiply  both sides of Eqs.~\eqref{eq:defPV} and~\eqref{eq:defPA} by $({\e}_\u q_\v\,,{\e}_\u P_\v\,,{\e}_{\u}{\e}^{*}_\v)$  for convenience of extracting  the form factors $T_{(1,2,3)}$. The final results are written as
\begin{align}
\label{eq:T1SLF}
\widetilde{ T}_1^{\rm SLF}=&\frac{1}{(M'^{2}-M''^{2}+{\bf q}^{2}_{\bot})}\frac{1}{x\bar{x}}
\bigg\{ 2x(xm_{2}+\bar{x}m'_{1})(xm_{2}+\bar{x}m''_{1})(M'^{2}-M''^{2})\nonumber\\
&+2x^{2}(M'^{2}-M''^{2}){\bf k}_{\bot}'\cdot{\bf k}_{\bot}''+(xm_{2}+\bar{x}m'_{1})\left[xm_{2}+\bar{x}(x-\bar{x})m'_{1}+2x\bar{x}m_1'' \right]{\bf q}^{2}_{\bot}\nonumber\\
&-[2x^{2}m^{2}_{2}+\bar{x}(x-\bar{x})m'^{2}_{1}-\bar{x}m''^{2}_{1}]{\bf k}_{\bot}'\cdot{\bf q}_{\bot}\nonumber\\
&+ 2(\bar{x}-x) {\bf k}_{\bot}''\cdot{\bf q}_{\bot} {\bf k}_{\bot}'\cdot{\bf k}_{\bot}''
+(1-2x\bar{x}){\bf k}_{\bot}'\cdot {\bf k}_{\bot}''{\bf q}^{2}_{\bot}\nonumber\\
&+\frac{2}{D''_{V}}\Big[
x\bar{x}(M'^{2}-M''^{2}) \left[(m'_{1}+m''_{1}){\bf k}_{\bot}'\cdot {\bf k}_{\bot}''   -(xm_2+\bar{x}m_1'){\bf k}_{\bot}''\cdot{\bf q}_{\bot} \right]\nonumber\\
&-(m'_{1}+m''_{1})(\bar{x}m_1'+xm_2)(\bar{x}m_1''-xm_2){\bf k}_{\bot}''\cdot{\bf q}_{\bot}\nonumber\\
&+\bar{x}(xm_{2} +\bar{x}m'_1) ({\bf k}_{\bot}''\cdot{\bf q}_{\bot})^2
-x\bar{x}\left(m'_{1}+m''_{1}\right)({\bf k}_{\bot}'\cdot{\bf q}_{\bot})^{2}
+x\bar{x}\left(m'_{1}+m''_{1}\right) {\bf k}_{\bot}'^{2}{\bf q}^{2}_{\bot}
 \nonumber\\
&+(x-\bar{x})(m'_{1}+m''_{1}) {\bf k}_{\bot}''\cdot{\bf q}_{\bot} {\bf k}_{\bot}'\cdot {\bf k}_{\bot}''
-\bar{x}(\bar{x}m''_{1}-xm_2) {\bf k}_{\bot}'\cdot{\bf q}_{\bot}{\bf k}_{\bot}''\cdot{\bf q}_{\bot}
\Big] \bigg\}\,, \nonumber\\
\widetilde{ T}_2^{\rm SLF}=&\widetilde{\cal T}_1^{\rm SLF}+\frac{q^2}{\left(M'^2-M''^2\right)\left(M'^{2}-M''^{2}+\qb^2\right)}\frac{1}{x^2\bar x}\Bigg\{ 4\bar x (\kb'\cdot\kb'')^2-x(1-2x\bar{x})\kb'\cdot\kb''\qb^2 \nonumber\\
 &+4\bar{x}\kb'\cdot\qb\kb''^2+2(x -2\bar{x})\kb'\cdot \kb'' \kb''\cdot\qb-x^2(\bar xm'_1+xm_2)(m_2-2\bar xm''_1)\qb^2  \nonumber\\
&+4\bar xm''_1\left(xm_2+m''_1\right)\kb'^2+(x^2+3x\bar x-4\bar x) m'_1(\bar xm'_1+xm_2)\kb''\cdot\qb\nonumber\\
&+\Big[3x\bar xm''^2_1-x^2m'_1(4\bar xm''_1-\bar xm_2+xm_2)+8\bar xx^2m''_1m_2+2x^3m^2_2\Big]\kb'\cdot\qb\nonumber\\
&-2x^3\left(M'^2+M''^2\right)\kb'\cdot\kb'' +4m_1'(\bar x^2m_1'+x^2m''_1+x\bar xm_2)\kb'\cdot\kb'' \nonumber\\
&-2(\bar xm'_1+xm_2)(\bar xm''_1+xm_2)\Big[x^2\left(M'^2+M''^2\right)-2m_1'm_1''\Big]\nonumber\\
&+\frac{2x}{D''_V}\bigg[4m'_1(\kb'\cdot\kb'')^2+\bar x\kb'\cdot\kb''\qb^2(xm''_1+2m_2-m'_1+3\bar xm'_1)\nonumber\\
&-2\kb'\cdot\kb''\kb'^2(m'_1-m''_1)-2x\bar xm_2\kb'\cdot\qb\kb''\cdot\qb-\bar x^2\kb''\cdot\qb\qb^2(\bar xm'_1+xm_2)\nonumber\\
&+\kb'^2\kb''\cdot\qb\Big(m'_1-m''_1-2\bar xm_2\Big)+2\kb'\cdot\kb''\kb''\cdot\qb(xm''_1-m'_1+2\bar xm_2)\nonumber\\
&-x\bar x\kb'\cdot\kb''\left(m'_1+m''_1\right)\left(M'^2+M''^2\right)+x\bar x\kb''\cdot\qb(\bar xm'_1+xm_2)(M'^2+M''^2)\nonumber\\
&+2\kb'\cdot\kb''\left(m'_1+m''_1\right)\left[\bar x(m'_1-m_2)(m''_1+m_2)+m^2_2\right]\nonumber\\
&+\kb''\cdot\qb\left(\bar xm'_1+xm_2\right)\left[\left(m'_1+m''_1\right)\left(xm_2-\bar xm''_1\right)-2m'_1m_2\right]
\bigg]\Bigg\}\,,\\
\label{eq:T3SLF}
\widetilde{ T}_3^{\rm{ SLF}}=&\frac{M'^{2}-M''^{2}}{q^2}\left[\widetilde{\cal T}_1^{\rm{ SLF}}-\widetilde{\cal T}_2^{\rm{ SLF}}\right]+\frac{2\left(M'^{2}-M''^{2}\right)}{x\bar x\left(M'^{2}-M''^{2}+\qb^2\right)\qb^2} \bigg\{ \bar xm''^2_1\kb'\cdot\qb-x^2m^2_2\qb^2\nonumber\\
&+\kb''\cdot\qb\left[(1-2x)\bar xm'^2_1-2x^2m^2_2\right]+ (1-2x)\kb'\cdot\kb''(2\kb''\cdot\qb+\qb^2)\nonumber\\
&+\frac{2}{D''_V}\kb''\cdot\qb\Big[(2x-1)(m'_1+m''_1)\kb'\cdot\kb''+(\bar x-x)(\bar xm'_1+xm_2)\kb''\cdot\qb\nonumber\\
&-\kb'\cdot\qb(\bar xm''_1-xm_2)+\left(\bar xm'_1+xm_2\right)(xm_2-\bar xm''_1)(m'_1+m''_1)
\Big]\bigg\}\,;\\
\widetilde{ T}_{1,2,3}^{(1)\,,\rm{ SLF}}=&\widetilde{T}_{1,2,3}^{\rm{ SLF}}\left[\text{$D''$-terms only}\,, D''_V\to D''_1\,, m''_1\to-m''_1\right]\,;\\
\widetilde{ T}_{1,2,3}^{(3)\,,\rm{ SLF}}=&\frac{\hat M''^2_0}{2\sqrt2M''_0}\widetilde{T}_{1,2,3}^{\rm{ SLF}}\left[ D''_V\to D''_3\,, m''_1\to-m''_1\right]\,.
\end{align}
It should be noted that only the $D''$-terms are kept in $\widetilde{ T}_{1,2,3}^{(1)\,,\rm{ SLF}}$, and the replacement $ m''_1\to-m''_1$ should not be applied to the $ m''_1$ in $D''$ factor.

\subsection{Theoretical results in the  CLF QM}
\begin{figure}[t]
\begin{center}
\includegraphics[scale=0.3]{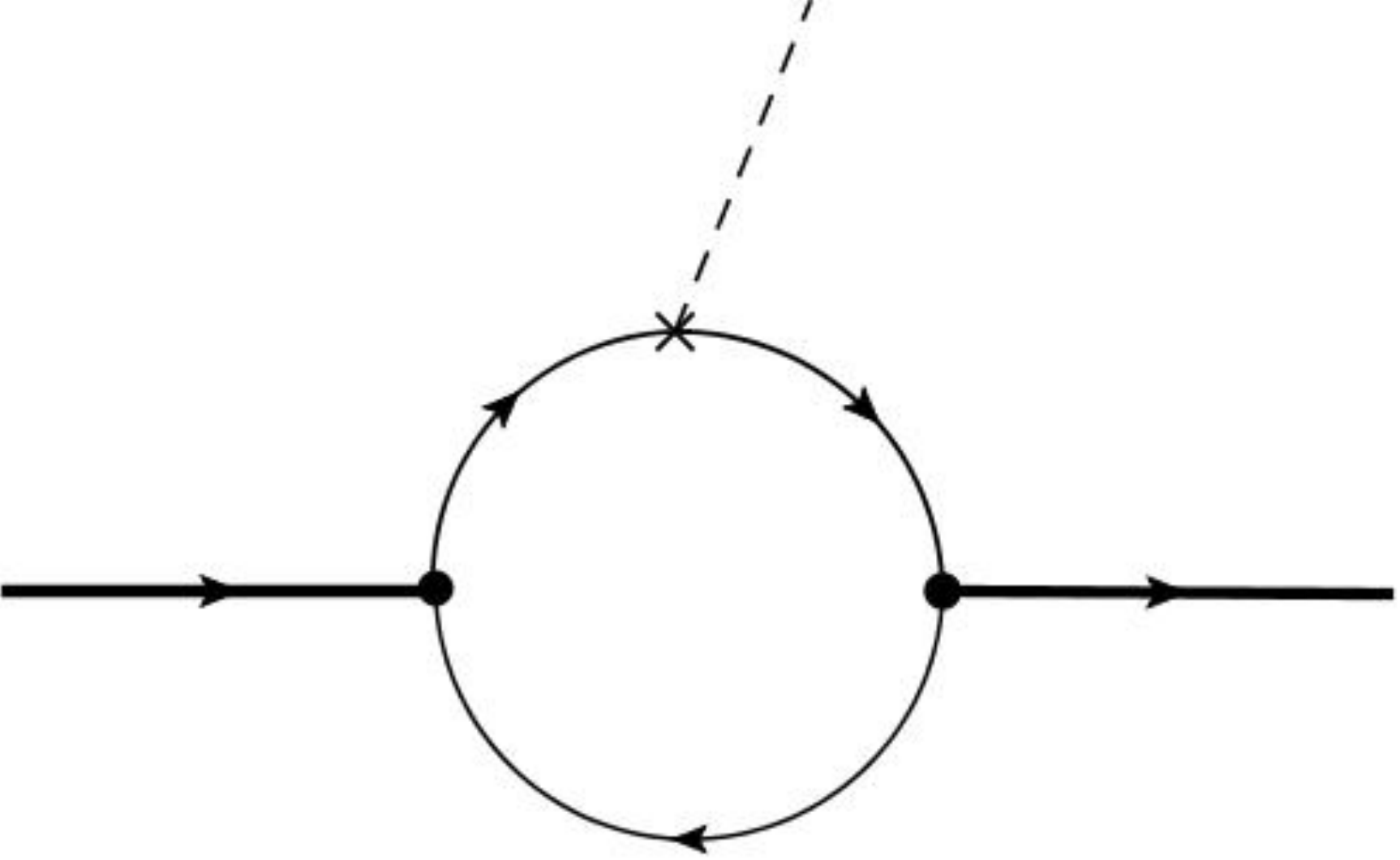}
\end{center}
\caption{\label{fig:fayn} The Feynman diagram for the matrix element $\cal B$ in the CLF QM.}
\end{figure}
In order to maintain manifest covariance and explore the zero-mode effects, a CLF approach is presented in Refs.~\cite{Jaus:1999zv,Jaus:2002sv,Cheng:2003sm} with the help of a manifestly covariant BS approach as a guide to the calculation. In the CLF QM, the matrix element for $M'\to M''$ transition is obtained by calculating the Feynman diagram shown in Fig.~\ref{fig:fayn}, and can be written  as a manifest covariant form,
\begin{eqnarray}\label{eq:Bclf1}
{\cal B}_{\rm CLF}=N_c \int \frac{\d^4 k_1'}{(2\pi)^4} \frac{H_{M'}H_{M''}}{N_1'\,N_1''\,N_2}iS\cdot (E_{M'}\, E_{M''}^*)\,,
\end{eqnarray}
where $\d^4 k_1'=\frac{1}{2} \d k_1'^- \d k_1'^+ \d^2 \mathbf{k}_{\bot}'$,  $E_{P,S}=1$ and $E_{V,A}=\e_\u$, the denominators $N_{1}^{(\prime,\prime\prime)}=k_{1}^{(\prime,\prime\prime)2}-m_1^{(\prime,\prime\prime)2}+i\e$ and $N_{2}=k_{2}^{2}-m_2^{2}+i\e$  come from the fermion propagators, and $H_{M', M''}$ are  vertex functions. The trace term  $S$  related to the fermion loop is written as
\begin{equation}\label{eq:Sterm}
S={\rm Tr}\left[\Gamma\, (\not\!k'_1+m'_1)\,(i\Gamma_{M'})\,(-\!\not\!k_2+m_2)\,(i\r^0{\Gamma}_{M''}^{\dag}\r^0) (\not\!k_1''+m_1'')\right]\,,
\end{equation}
where $\Gamma_{M^{(\prime,\prime\prime)}}$ is the vertex operator and can be written as~\cite{Cheng:2003sm,Choi:2013mda}
\begin{eqnarray}
i\Gamma_P&=&-i\gamma_5\,,\\
i\Gamma_S&=&-i\,,\\
i\Gamma_V&=&i\left[\gamma^\mu-\frac{ (k_1-k_2)^\mu}{D_{ V,{\rm con}}}\right]\,,\\
i\Gamma_{^1\!{A}}&=&i\frac{(k_1-k_2)^\mu}{D_{1,{\rm con}}}\gamma_5\,,\\
i\Gamma_{^3\!{A}}&=&i\left[\gamma^\mu+\frac{(k_1-k_2)^\mu}{D_{3,{\rm con}}}\right]\gamma_5\,.
\end{eqnarray}

Integrating out the minus component of loop momentum, one goes from the covariant calculation to the LF one. By closing the contour in the upper complex $k_1'^-$ plane and assuming that $H_{M', M''}$ are analytic within the contour, the integration picks up a residue at $k_2^2=\hat{k}_2^2=m_2^2$ corresponding to put the spectator antiquark on its mass-shell. Consequently, integrating out the minus component,  one has the following replacements~\cite{Jaus:1999zv,Cheng:2003sm}
\begin{eqnarray}
N_1 \to \hat{N}_1=x \left( M^2-M_0^2\right)
\end{eqnarray}
and
\begin{eqnarray}\label{eq:type1}
\chi_M \equiv H_M/N\to h_M/\hat{N}\,,\qquad  D_{\rm  con} \to D_{\rm  LF}\,,\qquad \text{(type-I)}
\end{eqnarray}
where the LF forms of vertex functions, $h_M$, for $P$, $S$, $V$ and $A$ mesons are given by
\begin{eqnarray}
h_P/\hat{N}&=&h_V/\hat{N}=\frac{1}{\sqrt{2N_c}}\sqrt{\frac{\bar{x}}{x}}\frac{\psi_s}{\hat{M}_0}\,,\\
\label{eq:vSV}
h_S/\hat{N}&=&\frac{1}{\sqrt{2N_c}}\sqrt{\frac{\bar{x}}{x}}\frac{\hat M'^2_0}{2\sqrt3M'_0}\frac{\psi_p}{\hat{M}_0}\,,\\
\label{eq:vPS}
h_{^1\!{A}}/\hat{N}&=&\frac{1}{\sqrt{2N_c}}\sqrt{\frac{\bar{x}}{x}}\frac{\psi_p}{\hat{M}_0}\,,\\
\label{eq:vP1A}
h_{^3\!{A}}/\hat{N}&=&\frac{1}{\sqrt{2N_c}}\sqrt{\frac{\bar{x}}{x}}\frac{\hat M'^2_0}{2\sqrt2M'_0}\frac{\psi_p}{\hat{M}_0}\,.
\label{eq:vP3A}
\end{eqnarray}
Eq.~\eqref{eq:type1} shows the correspondence between the manifestly covariant and the LF approaches.  In Eq.~\eqref{eq:type1},  the correspondence between $\chi$ and $\psi$  can be clearly derived by matching the CLF expressions to the SLF ones via some zero-mode independent  quantities, such as  $f_P$ and  $f_{+}^{P\to P}(q^2)$~\cite{Jaus:1999zv,Cheng:2003sm}, however, the validity of the correspondence for the $D$ factor appearing in the vertex operator, $D_{V, {\rm  con}} \to D_{V,{\rm  LF}}$, has not yet been clarified explicitly~\cite{Choi:2013mda}. Instead of the traditional type-I correspondence, a much more generalized correspondence,
\begin{eqnarray}\label{eq:type2}
\chi_M \equiv H_M/N\to h_M/\hat{N}\,,\qquad  M\to  M_0\,,\qquad \text{(type-II)}
\end{eqnarray}
is suggested by Choi {\it et al.} for  the purpose of  self-consistent results for $f_{V}$~\cite{Choi:2013mda}~(one may refer to Refs.~\cite{Choi:2013mda,Chang:2018zjq,Chang:2019mmh,Chang:2019obq} for more  discussions).

After integrating out $k_1'^-$, the matrix element, Eq.~\eqref{eq:Bclf1}, can be reduced to the LF form
\begin{eqnarray}
\label{eq:Bclf2}
\hat{\cal B}_{\rm CLF}=N_c \int \frac{\d x \d^2 \mathbf{k}_{\bot}'}{2(2\pi)^3}\frac{h_{M'}h_{M''}}{\bar{x} \hat{N}_1'\,\hat{N}_1''\,}\hat{S}\cdot (E_{M'}\, E_{M''}^*)\,.
\end{eqnarray}
It should be noted that ${\cal B}$ receives additional spurious contributions proportional to the light-like vector $\omega^\mu=(0,2,\mathbf{0}_\bot)$, and these undesired spurious contributions are expected to be cancelled out by the zero-mode contributions~\cite{Jaus:1999zv,Cheng:2003sm}. The inclusion of the zero-mode contribution in practice amounts to some proper replacements for $\hat{k}_1'$ and $\hat{N}_2$ in  $\hat{S}$ under integration~\cite{Jaus:1999zv}. In this work, we need
\begin{align}
\hat{k}_1'^{\mu} \to& P^\u A_1^{(1)}+q^\u A_2^{(1)} \,,
\label{eq:repk}\\
\hat{k}_1'^{\mu}\hat{k}_1'^{\nu} \to &g^{\u\v}A_1^{(2)}+P^\u P^\v A_2^{(2)}+(P^\u q^\v+q^\u P^\v)A_3^{(2)}+q^\u q^\v A_4^{(2)}\nonumber\\
&+\frac{P^\u\omega^\v+\omega^\u P^\v}{\omega\cdot P}B_1^{(2)}\,,
\label{eq:repkk}\\
\hat{k}_1'^{\mu}\hat{N}_2\to& q^\u\left(A_2^{(1)}Z_2+\frac{q\cdot P}{q^2}A_1^{(2)} \right) \,,
\label{eq:repkN}
\end{align}
where $A$ and $ B$ functions are written as
\begin{align}
A_1^{(1)}&=  \frac{x}{2}\,,\quad
A_2^{(1)}=\frac{x}{2} -\frac{\kb' \cdot \qb }{q^2}\,;\\
A_1^{(2)}&=-\kb'^2 -\frac{(\kb' \cdot \qb)^2}{q^2}\,,\quad
A_2^{(2)}=\left(A_1^{(1)}\right)^2\,,\quad
A_3^{(2)}=A_1^{(1)}A_2^{(1)}\,,\nonumber\\
A_4^{(2)}&=\left(A_2^{(1)}\right)^2-\frac{1}{q^2}A_1^{(2)}\,,\quad
B_1^{(2)}=\frac{x}{2}Z_2-A_1^{(2)}\,;\\
Z_2&=\hat{N}_1'+m_1'^2-m_2^2+(\bar{x}-x)M'^2+\left(q^2+q\cdot P\right)\frac{\kb' \cdot \qb}{q^2}\,.
\end{align}
In above formulas, the $\w$-dependent terms associated with the  $C$ functions are not given since they are eliminated exactly by the inclusion of the zero-mode contributions~\cite{Jaus:1999zv}.

In the CLF QM, the tensor form factors can be obtained directly by matching $\hat{\cal B}_{\rm CLF}$ to their definitions given by Eqs.~(\ref{eq:defPP}), (\ref{eq:defPS}) and (\ref{eq: PVT2}-\ref{eq: PAT3})\footnote{The definitions for the tensor form factors of $P\to V$ and $P\to A$ transitions given by Eqs.~(\ref{eq: PVT2}-\ref{eq: PAT3}) are used in this subsection because they are much more convenient  for the CLF  calculation. }. Our final CLF results for the tensor form factors can be written as
\begin{eqnarray}\label{eq:FCLF}
[\mathcal F(q^2)]_{\rm CLF}=N_c\int\frac{\d x\d^2{\bf k'_\bot}}{2(2\pi)^3}\frac{\chi_{M'}\chi_{M''}}{\bar x}{\cal \widetilde{\cal F}}^{\rm CLF}(x,{\bf k'_\bot},q^2)\,,
\end{eqnarray}
where, the integrands are
\begin{align}
\widetilde{F}_T^{\rm{CLF}}=&2(M'+M'')\left[m'_1-(m'_1+m''_1-2m_2)A^{(1)}_1-(m'_1-m''_1)A^{(1)}_2\right]\,;
\label{eq:CLFFT}\\
\widetilde{U}_T^{\rm CLF}=&\widetilde{F}_T^{\rm{CLF}}[m''_1\rightarrow-m''_1]\,;\\
\widetilde{T}_1^{\rm CLF}=&\left(2\bar x-1\right)\left(m'^2_1+\hat N'_1\right)+m''^2_1+\hat N''_1+\qb^2+2\left(\bar xm'_1m''_1+xm'_1m_2+xm''_1m_2\right)\nonumber\\
&-8A^{(2)}_1+2\left(M'^2-M''^2\right)\left(A^{(1)}_1+2A^{(2)}_2-2A^{(2)}_3\right)\nonumber\\
&+2\qb^2\left(A^{(1)}_1-2A^{(1)}_2-2A^{(2)}_3+2A^{(2)}_4\right)-\frac{4}{D''_{V,{\rm con}}}(m'_1+m''_1)A^{(2)}_1\,,
\label{eq:CLFT1}\\
\widetilde{T}_2^{\rm CLF}=&-(m'_1-m''_1)^2-\hat N'_1-\hat N''_1+\bar xq^2+x\left[M'^2+M''^2+2(m'_1-m_2)(m_2-m''_1)\right]\nonumber\\
&-\frac{q^2}{M'^2-M''^2}\bigg\{
2M'^2+(m''_1-m'_1)^2-2(m'_1-m_2)^2-\hat N'_1+\hat N''_1-q^2\nonumber\\
&-2Z_2+4Z_2A^{(1)}_2-4A^{(2)}_1-2\left[M'^2+M''^2-q^2+2(m'_1-m_2)(m_2-m''_1)\right]A^{(1)}_2\bigg\}\nonumber\\
&-\frac{4}{D''_{V,{\rm con}}}A^{(2)}_1\left[m'_1+m''_1+\frac{\qb^2}{M'^2-M''^2}(m'_1-m''_1-2m_2)\right]\,,
\label{eq:CLFT2}\\
\widetilde{T}_3^{\rm CLF}=&2M'^2-2(m'_1-m_2)^2+(m'_1-m''_1)^2-\hat N'_1+\hat N''_1-q^2-2Z_2\nonumber\\
&-4A^{(2)}_1+4\left(M'^2-M''^2\right)\left(A^{(1)}_1-A^{(2)}_2+A^{(2)}_4\right)+\frac{4\left(M'^2-M''^2\right)}{q^2}A^{(2)}_1\nonumber\\
&+2\left[M''^2-3M'^2+q^2+2Z_2+2(m'_1-m_2)(m''_1-m_2)\right]A^{(1)}_2\nonumber\\
&+\frac{4}{D''_{V,{\rm con}}}\bigg\{\left(M'^2-M''^2\right)\Big[m'_1\left(2A^{(1)}_1+2A^{(1)}_2-A^{(2)}_2-2A^{(2)}_3-A^{(2)}_4-1\right)\nonumber\\
&-m''_1\left(A^{(1)}_1-A^{(1)}_2-A^{(2)}_2+A^{(2)}_4\right)-2m_2\left(A^{(1)}_1-A^{(2)}_2-A^{(2)}_3\right)\Big]\nonumber\\
&+(m''_1-m'_1+2m_2)A^{(2)}_1\bigg\}\,;
\label{eq:CLFT3}\\
\widetilde{T}_{1,2,3}^{(1)\,,\rm{CLF}}=&\widetilde{T}_{1,2,3}^{\rm CLF}[\text{$D''$-terms only}\,, D''_{V,{\rm con}}\to D''_{1,{\rm con}}\,,m''_1\rightarrow-m''_1]\,;
\label{eq:CLFT1A}\\
\widetilde{T}_{1,2,3}^{(3)\,,\rm{CLF}}=&\widetilde{T}_{1,2,3}^{\rm CLF}[D''_{V,{\rm con}}\to D''_{3,{\rm con}}\,,m''_1\rightarrow-m''_1]\,.
\label{eq:CLFT3A}
\end{align}
Similar to the case of SLF results,  only the $D''$-terms are kept in $\widetilde{\cal T}_{1,2,3}^{(1)\,,\rm{ CLF}}$, and the replacement $ m''_1\to-m''_1$ should not be applied to the $ m''_1$ in $D''$ factors.  Our results given above are obtained with the traditional type-I correspondence scheme,  the ones with type-II scheme can be easily obtained by making an additional replacement $M\to  M_0$. It should be noted that the contributions related to the $B$ functions are not included in the results given above. These contributions would lead to the self-consistence and covariance problems, and will be given and analyzed  separately in the next section. Comparing our results for $P\to V~(A)$ transition, Eqs.~(\ref{eq:CLFT1}-\ref{eq:CLFT3}), with the ones obtained in the previous work~\cite{Cheng:2010yj,Cheng:2009ms}, Eqs.~(\ref{eq:chengT1}-\ref{eq:chengT3}), which are collected in the appendix A, we find that our result for $\widetilde{T}_1^{\rm CLF}$,  Eq.~\eqref{eq:CLFT1}, is exactly the same as the one in Refs.~\cite{Cheng:2010yj,Cheng:2009ms}, Eq.~\eqref{eq:chengT1}, however, the results for  $\widetilde{T}_{2,3}^{\rm CLF}$  are different. This inconsistence will be analyzed  in detail in the next section.

In the CLF QM,  for a given quantity ($\cal Q$), the CLF result (${\cal Q}^{\rm CLF}$) can be expressed as a sum of  valence (${\cal Q}^{\rm val.}$) and  zero-mode (${\cal Q}^{\rm z.m.}$) contributions~\cite{Choi:2013mda},  ${\cal Q}^{\rm CLF}={\cal Q}^{\rm val.}+{\cal Q}^{\rm z.m.}$, in which the CLF results for the tensor form factors has been given above. It has been found in Ref.~\cite{Choi:2013mda} and our previous works~\cite{Chang:2018zjq,Chang:2019mmh} that ${\cal Q}^{\rm CLF}\dot{=}{\cal Q}^{\rm val.}={\cal Q}^{\rm SLF}$ within type-II correspondence scheme, where  ``$\dot{=}$'' denotes that  two quantities are equal to each other only in numerical value, while ``$=$'' means that two quantities are exactly the same not only in numerical value but also in form. In order to  check the universality of such relation and clearly show the effects of zero-mode contributions, we have also calculated the valence contributions, which are written as
\begin{align}
\widetilde{F}_T^{\rm{val.}}=&\widetilde{F}_T^{\rm{CLF}}\,;\\
\widetilde{U}_T^{\rm val.}=&\widetilde{U}_T^{\rm{CLF}}\,;\\
\label{eq:T1val}
\widetilde{T}_1^{\rm val.}=&\frac{1}{\bar x\left(M'^2-M''^2+\qb^2\right)}\bigg\{-2\kb'\cdot\qb\kb''^2+\kb'\cdot\kb''\qb^2-\bar x(2x-1)\kb''\cdot\qb M'^2\nonumber\\
&+2x(M'^2-M''^2)(\kb'\cdot\kb''+m^2_2)+\kb'\cdot\qb(\bar xM''^2-2m^2_2)\nonumber\\
&+2\bar x\left[m'_1m''_1-x(m'_1-m_2)(m''_1-m_2)\right](M'^2-M''^2-\qb^2)+m^2_2\qb^2\nonumber\\
&+\frac{2}{D''_{V,{\rm con}}}(m'_1+m''_1)\Big[\kb'\cdot\qb\kb''^2+\kb''\cdot\qb(m^2_2-\bar x^2M'^2)+\bar x\kb'\cdot\kb''(M'^2-M''^2)\Big]\bigg\}\,,\\
\widetilde{T}_2^{\rm val.}=&\widetilde{T}_1^{\rm val.}-\frac{q^2}{\left(M'^2-M''^2\right)\left(M'^{2}-M''^{2}+\qb^2\right)}\frac{1}{\bar x}\Bigg\{2\kb'\cdot\kb''\kb'\cdot\qb-2\bar x\kb'\cdot\qb\kb''\cdot\qb\nonumber\\
&+\kb'\cdot\kb''\qb^2+2\kb'\cdot\kb''[(1+\bar x)M'^2+xM''^2+2(m'_1-m_2)(m_2-m''_1)]\nonumber\\
&+2\kb''\cdot\qb(m^2_2-\bar x^2M'^2)-2\bar x(\kb'\cdot\qb+\kb''\cdot\qb)(m'_1-m_2)(m_2-m''_1)\nonumber\\
&-\bar xM'^2\kb''\cdot\qb-3\bar xM''^2\kb'\cdot\qb+m^2_2\qb^2+2\bar xm_2(m'_1-m''_1)(M'^2-M''^2+\qb^2)\nonumber\\
&+2(M'^2+M''^2)\left[\bar x^2(m'_1-m_2)(m''_1-m_2)+m^2_2\right]-4\bar x^2M'^2M''^2+4m^2_2(m'_1-m_2)(m_2-m''_1)\nonumber\\
&+\frac{2}{D''_{V,{\rm con}}}(m'_1-m''_1-2m_2)\bigg[\kb'\cdot\qb\kb''^2+\bar x\kb'\cdot\kb''(M'^2-M''^2)+\kb''\cdot\qb(m^2_2-\bar x^2M'^2)
\bigg]\Bigg\}
\,,\\
\label{eq:T3val}
\widetilde{T}_3^{\rm val.}=&2\frac{M'^{2}-M''^{2}}{q^2}\left[\widetilde{T}_1^{\rm val.}-\widetilde{T}_2^{\rm val.}\right]+\frac{2\left(M'^{2}-M''^{2}\right)}{\bar x\left(M'^{2}-M''^{2}+\qb^2\right)\qb^2}\bigg\{(\bar x-x)\kb'\cdot\kb''\qb^2\nonumber\\
&-2\kb'\cdot\qb\cdot\kb''^2+\bar x(\bar x-x)M'^2\kb''\cdot\qb+\kb'\cdot\qb(\bar xM''^2-2m^2_2)+(\bar x-x)m^2_2\qb^2\nonumber\\
&+\frac{2}{D''_{V,{\rm con}}}\kb''\cdot\qb\Big[(m'_1+m''_1)(\kb'^2-2\bar x\kb'\cdot\qb+m^2_2)+\bar x(xm_2-\bar xm''_1)M'^2\nonumber\\
&-\bar x(xm_2+\bar xm'_1)(M''^2-\qb^2)
\Big]\bigg\}\,;
\\
\widetilde{T}_{1,2,3}^{(1)\,,\rm{val.}}=&\widetilde{T}_{1,2,3}^{\rm val.}[\text{$D''$-terms only}\,, D''_{V,{\rm con}}\to D''_{1,{\rm con}}, m''_1\to-m''_1]\,;\\
\widetilde{T}_{1,2,3}^{(3)\,,\rm{val.}}=&\widetilde{T}_{1,2,3}^{\rm val.}[D''_{V,{\rm con}}\to D''_{3,{\rm con}}, m''_1\to -m''_1]\,.
\end{align}
It can be easily found that the tensor form factors of $P\to (P,S)$  transitions are free from the zero-mode effects, while the ones of $P\to (V,A)$  transitions are  zero-mode dependent.

\section{Numerical results and discussions}
Using the theoretical results given in the last section and input parameters collected in appendix B, we then  present our numerical results and discussions in this section. It has been mentioned above that most of the  spurious $\w$-dependent contributions are neutralized by zero-mode contributions, but there are still some residuals associated with $B$ functions, which possibly  violate the self-consistence and covariance of  CLF QM, but are not taken into account in Eqs.~(\ref{eq:CLFFT}-\ref{eq:CLFT3A} and are not considered in the previous works~\cite{Cheng:2004yj,Cheng:2010yj,Cheng:2009ms} either. These residual $\w$-dependent contributions to the tensor matrix elements of $P\to V$ transition ( l.h.s. of Eqs.~\eqref{eq: PVT2} and \eqref{eq: PVT3}) can be written as
\begin{eqnarray}
[\mathcal B]_{ B}=N_c\int\frac{\d x\d^2{\bf k'_\bot}}{2(2\pi)^3}\frac{\chi_V'\chi_V''}{\bar x}{\cal \widetilde{\cal B}}_{ B}
\end{eqnarray}
where,
\begin{align}
\widetilde{\cal B}_B^\mu(\Gamma=\sigma^{\mu\nu}q_\v)=&4\frac{B_1^{(2)}}{\omega\cdot P}\bigg[
-\epsilon^{\delta\nu\alpha\beta}P_\delta q_\nu\omega_\alpha\epsilon^*_{\beta} P^\mu
+\epsilon^{\mu\nu\alpha\beta}P_\nu\omega_\alpha\epsilon^{*}_{\beta}\left(M'^2-M''^2\right)\nonumber\\
&+\epsilon^{\mu\nu\alpha\beta}q_\nu\omega_\alpha\epsilon^{*}_\beta\left(M'^2-M''^2\right)
-\epsilon^{\mu\nu\alpha\beta}P_\nu q_\alpha\omega_\beta(q\cdot \epsilon^*)\frac{m'_1+m''_1}{D_{V,{\rm con}}''}\bigg]\,,
\label{eq:BBT1}\\
\widetilde{\cal B}_B^\mu(\Gamma=\sigma^{\mu\nu}\gamma_5q_\v)=&4i\frac{B_1^{(2)}}{\omega\cdot P}\Bigg\{
-P^\mu(\omega\cdot\epsilon^*)q^2\left(1+\frac{m'_1-m''_1-2m_2}{D_{V,{\rm con}}''}\right)\nonumber\\
&+q^\mu(\omega\cdot\epsilon^*)\left(M'^2-M''^2\right)\left(1+\frac{m'_1-m''_1-2m_2}{D_{V,{\rm con}}''}\right)\nonumber\\
&-\omega^\mu(q\cdot\epsilon^*)\bigg[q^2\left(1+\frac{m'_1-m''_1-2m_2}{D_{V,{\rm con}}''}\right)\nonumber\\
&+\left(M'^2-M''^2\right)\left(1+\frac{m'_1-m''_1}{D_{V,{\rm con}}''}\right)\bigg]\Bigg\}\,.
\label{eq:BBT3}
\end{align}
 $\widetilde{B}_B^\mu=0$ for the $P\to (P,S)$ transitions,  and $\widetilde{B}_B^\mu$ for the $P\to A$ transitions can be obtained from above results by the replacements similar to Eqs.~\eqref{eq:CLFT1A}  and \eqref{eq:CLFT3A}. Taking the contributions  associated with $B$ functions into account, the full results for the tensor form factors in the CLF QM can be expressed as
\begin{align}
[{\cal F}]^{\rm full}=[{\cal F}]^{\rm CLF}+[{\cal F}]^{ B}\,.
\end{align}

\begin{table}[t]
\begin{center}
\caption{\label{tab:T1BD} Numerical results of $T_1({\bf{q}}^{2}_{\bot})$  for $B_{c}\to D^*$ transition at ${\bf q}_{\bot}^2=(0,4,9) \,{\rm GeV^2}$ and for  $D_{s}\to \phi$  transition at ${\bf q}_{\bot}^2=(0,0.5,1) \,{\rm GeV^2}$.}
\vspace{0.2cm}
\let\oldarraystretch=\arraystretch
\renewcommand*{\arraystretch}{0.9}
\setlength{\tabcolsep}{5pt}
\begin{tabular}{llccccccccccc}
\hline\hline
  $B_{c}\to D^*$ & &$ [T_1]^{\text{SLF}}$&$[T_1]^{\rm{val.}}$&$[T_1]^{\rm CLF}$&$[T_1]^{\rm full}_{\lambda=0}$&$[T_1]^{\rm full}_{\lambda=+}$&$[T_1]^{\rm full}_{\lambda=-}$
 \\\hline
\multirow{2}{*}{${\bf{q}}_\perp^2=0\,{\rm GeV^2}$}
&type-I    &$0.106$ & $0.106$ & $0.094$ & $0.118$ & $0.081$ & $0.094$\\
&type-II    &$0.106$ & $0.106$ & $0.106$ & $0.106$ & $0.106$ & $0.106$\\\hline
\multirow{2}{*}{${\bf{q}}_\perp^2=4\,{\rm GeV^2}$}
&type-I     &$0.072$ & $0.072$ & $0.063$ & $0.079$ & $0.055$ & $0.062$ \\
&type-II     &$0.073$ & $0.073$ & $0.073$ & $0.073$ & $0.073$ & $0.073$\\\hline
\multirow{2}{*}{${\bf{q}}_\perp^2=9\,{\rm GeV^2}$}
&type-I      &$0.046$ & $0.045$ & $0.040$ & $0.049$ & $0.035$ & $0.038$ \\
&type-II    &$0.047$ & $0.047$ & $0.047$ & $0.047$ & $0.047$ & $0.047$
\\\hline\hline
$D_{s}\to \phi$  & &$ [T_1]^{\text{SLF}}$&$[T_1]^{\rm{val.}}$&$[T_1]^{\rm CLF}$&$[T_1]^{\rm full}_{\lambda=0}$&$[T_1]^{\rm full}_{\lambda=+}$ &$[T_1]^{\rm full}_{\lambda=-}$\\\hline
\multirow{2}{*}{${\bf{q}}_\perp^2=0\,{\rm GeV^2}$}
&type-I    &$0.687$ & $0.687$ & $0.658$ & $0.681$ & $0.630$ & $0.658$ \\
&type-II    &$0.687$ & $0.687$ & $0.687$ & $0.687$ & $0.687$ & $0.687$\\\hline
\multirow{2}{*}{${\bf{q}}_\perp^2=0.5\,{\rm GeV^2}$}
&type-I     &$0.597$ & $0.593$ & $0.568$ & $0.589$ & $0.544$ & $0.564$ \\
&type-II     &$0.598$ & $0.598$ & $0.598$ & $0.598$ & $0.598$ & $0.598$\\\hline
\multirow{2}{*}{${\bf{q}}_\perp^2=1\,{\rm GeV^2}$}
&type-I      &$0.524$ & $0.517$ & $0.495$ & $0.513$ & $0.476$ & $0.488$\\
&type-II    &$0.526$ & $0.526$ & $0.526$ & $0.526$ & $0.526$ & $0.526$
\\\hline\hline
\end{tabular}
\end{center}
\end{table}

\begin{figure}[t]
\caption{The dependences of $\Delta_{\rm B}(x)$ on $x$ for  $B_{c}\to D^*$ transition at ${\bf q}_{\bot}^2=(0,4,9)\,{\rm GeV^2}$ and  for $D_{s}\to \phi$ transition at ${\bf q}_{\bot}^2=(0,0.5,1)\,{\rm GeV^2}$.}\label{fig:T1}
\vspace{0.32cm}
  \centering
  \subfigure[]{\includegraphics[width=0.30\textwidth]{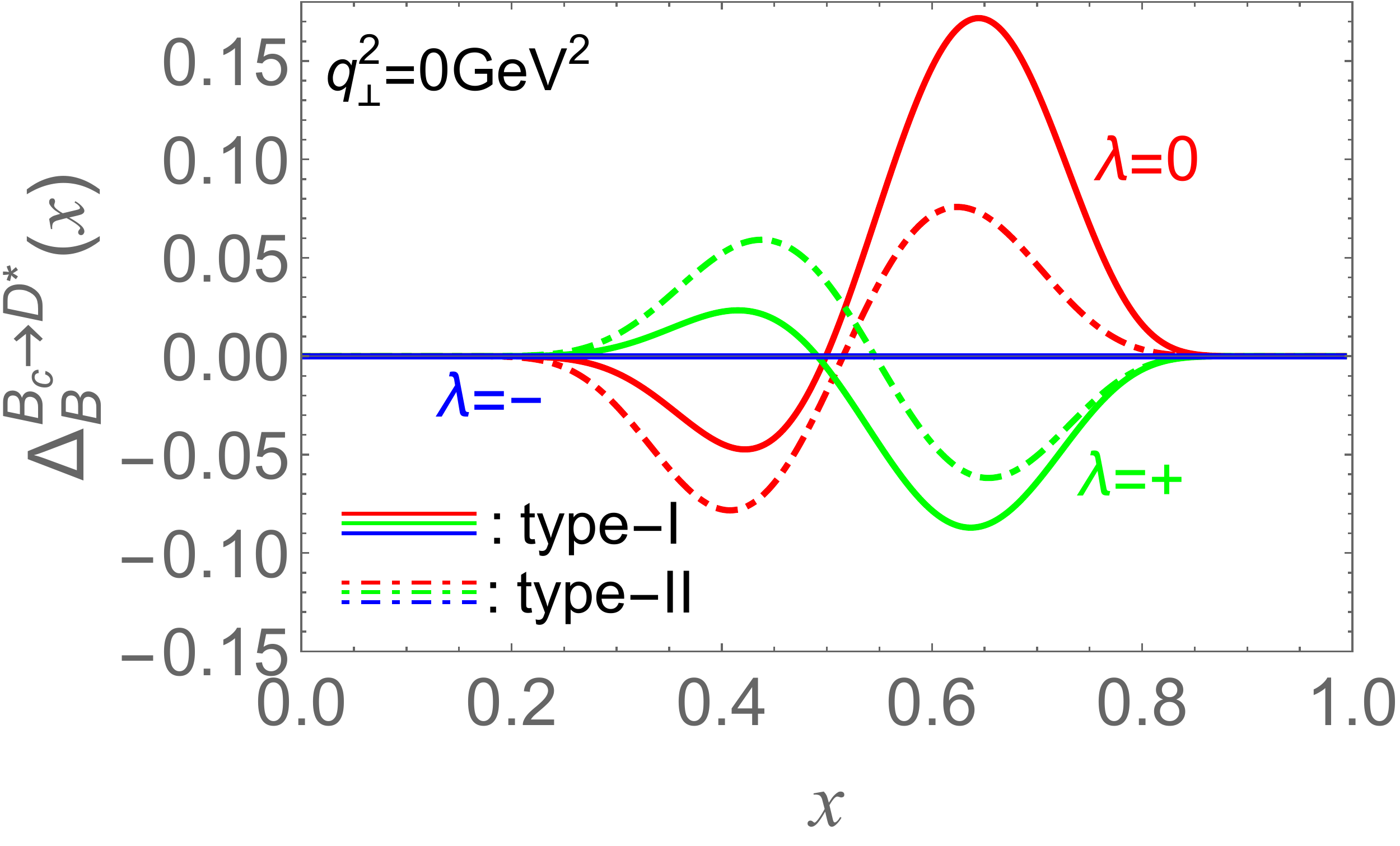}\quad}
  \subfigure[]{\includegraphics[width=0.30\textwidth]{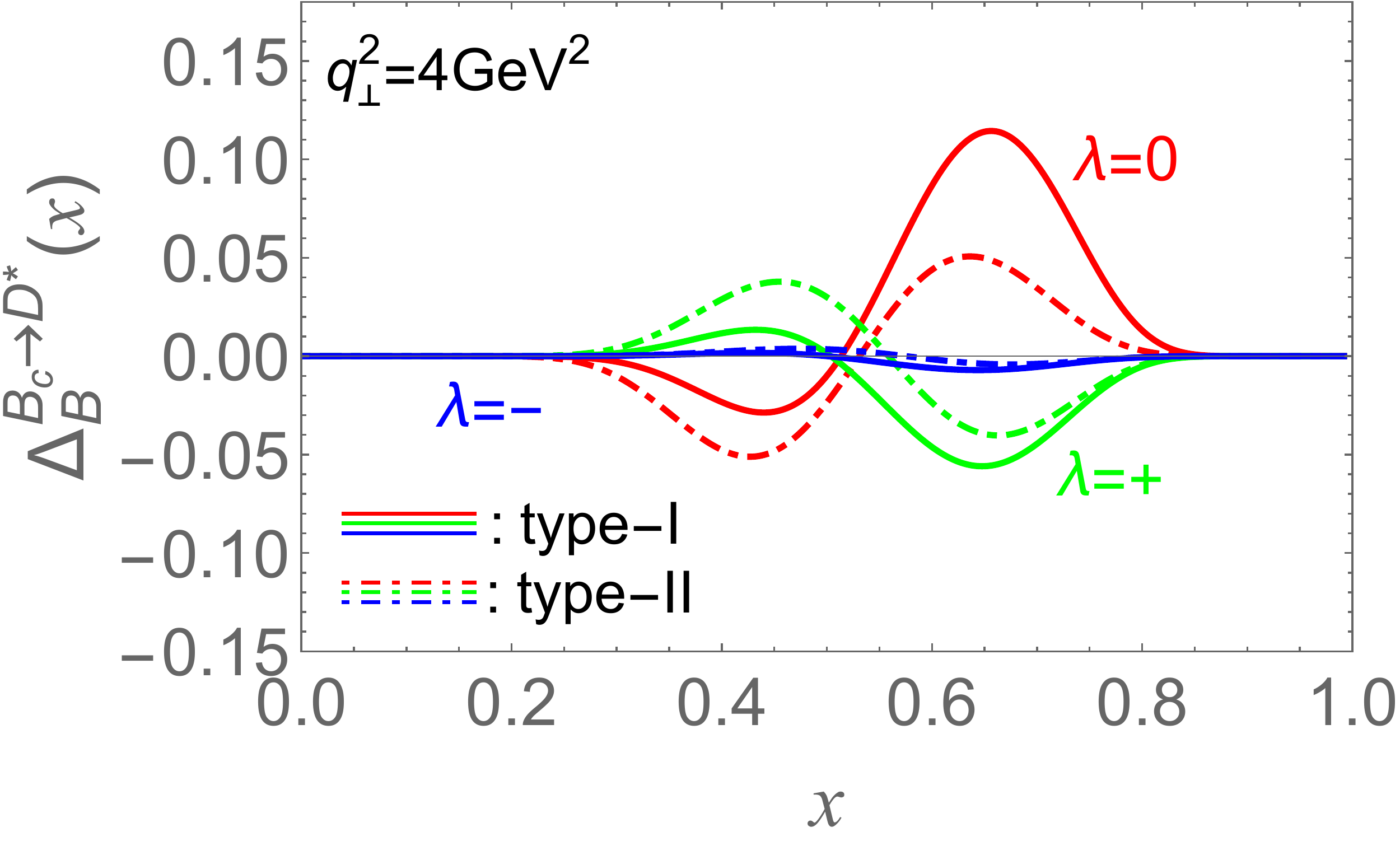}\quad}
  \subfigure[]{\includegraphics[width=0.30\textwidth]{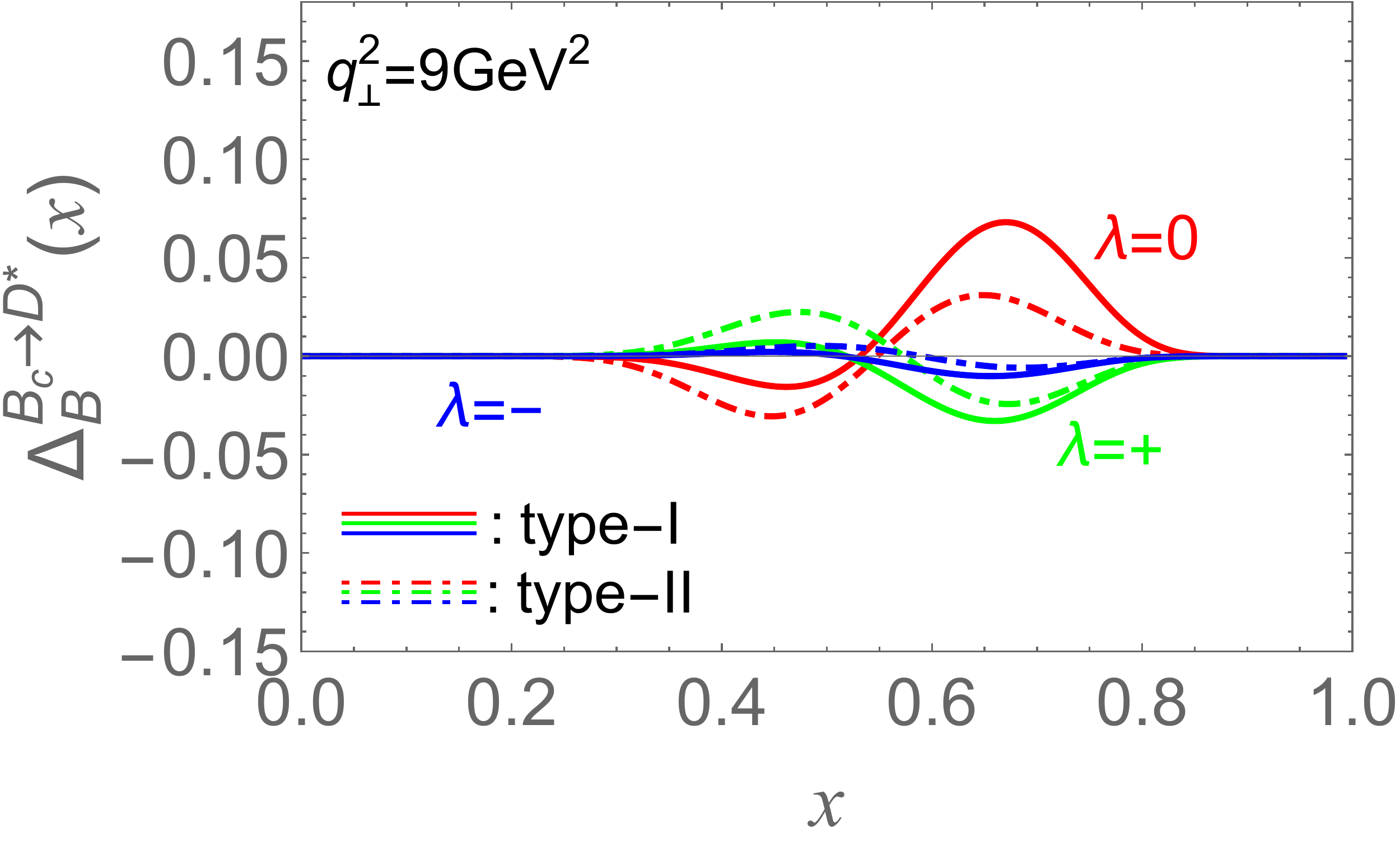}}\\
  \subfigure[]{\includegraphics[width=0.30\textwidth]{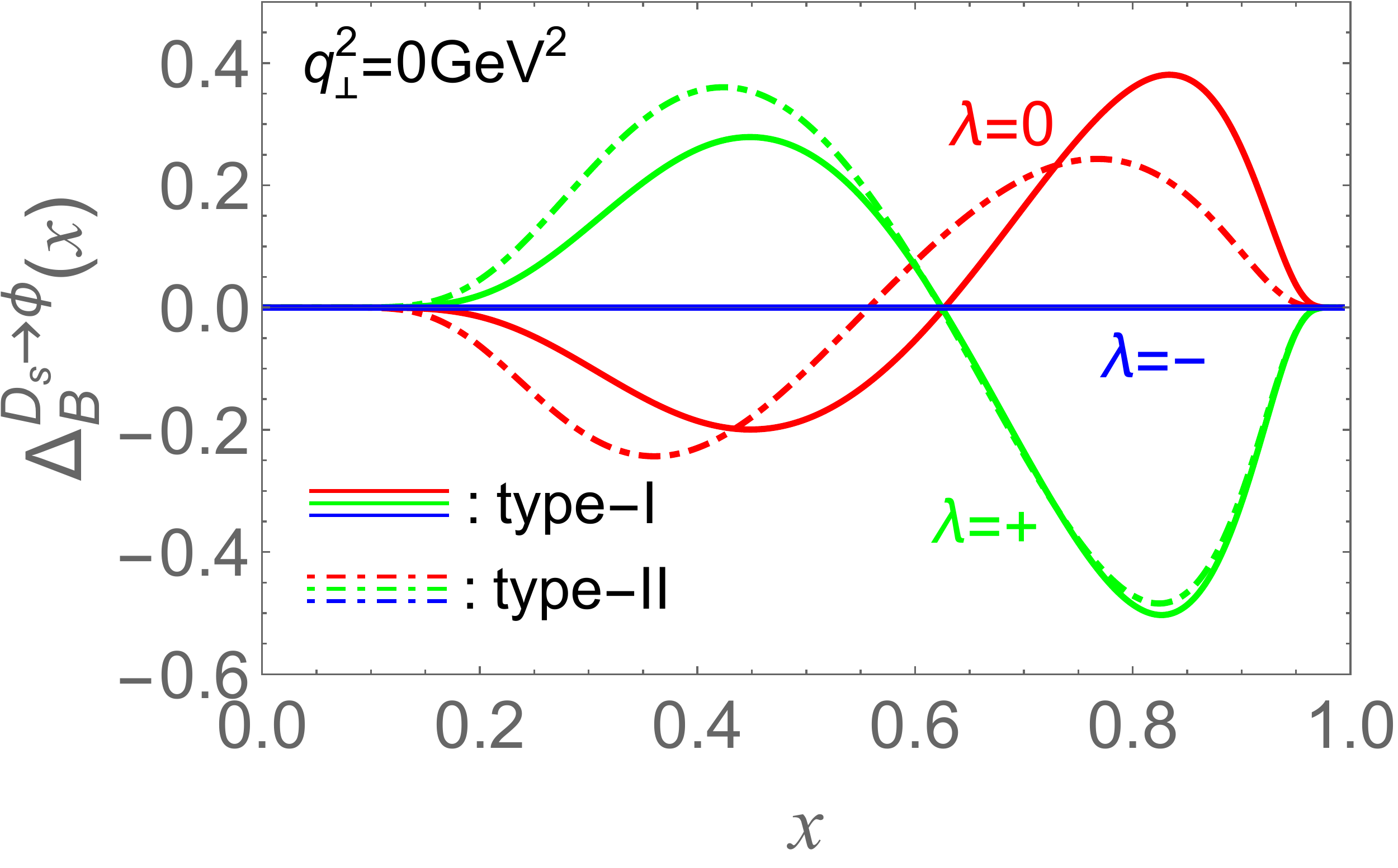}\quad}
  \subfigure[]{\includegraphics[width=0.30\textwidth]{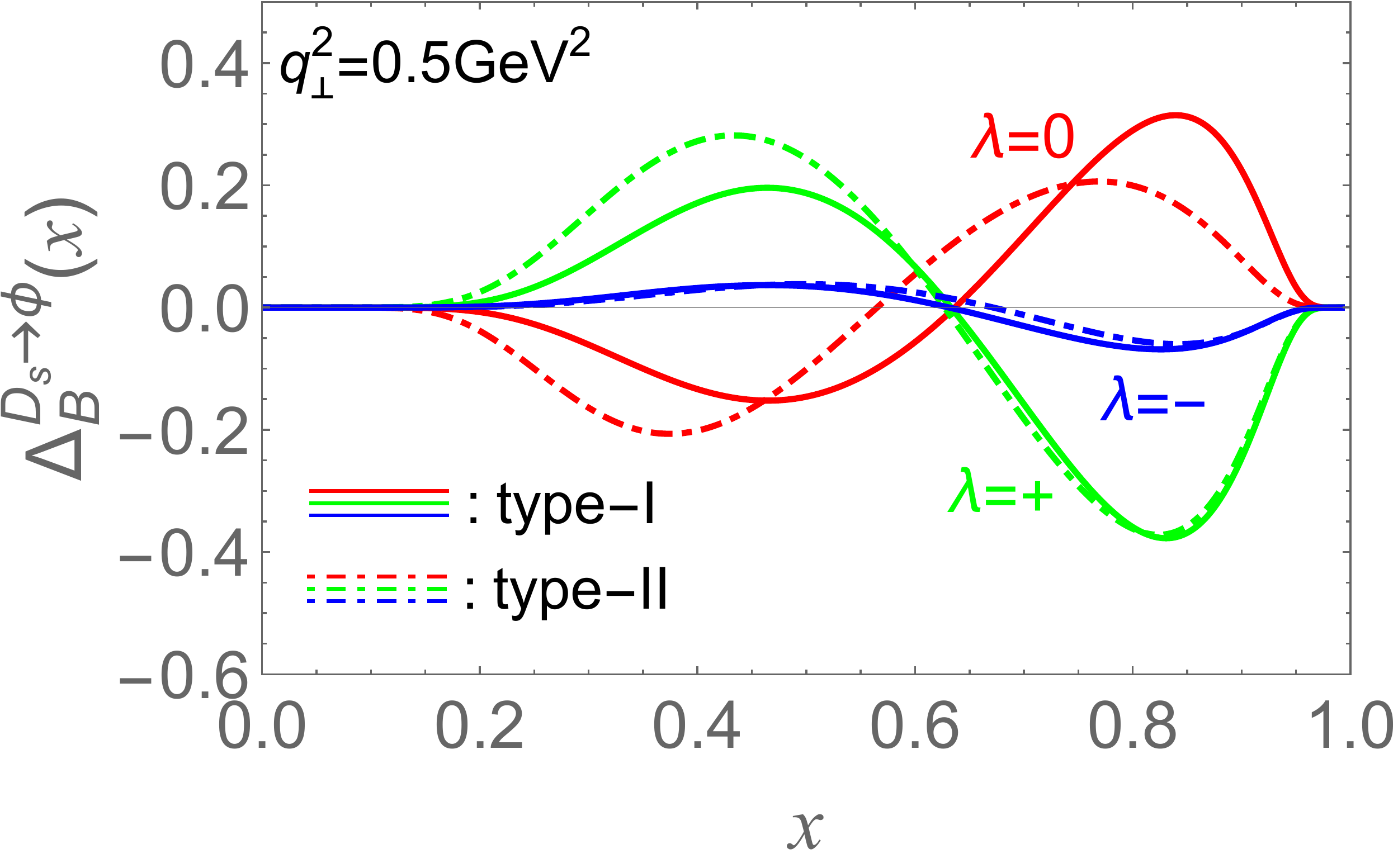}\quad}
  \subfigure[]{\includegraphics[width=0.30\textwidth]{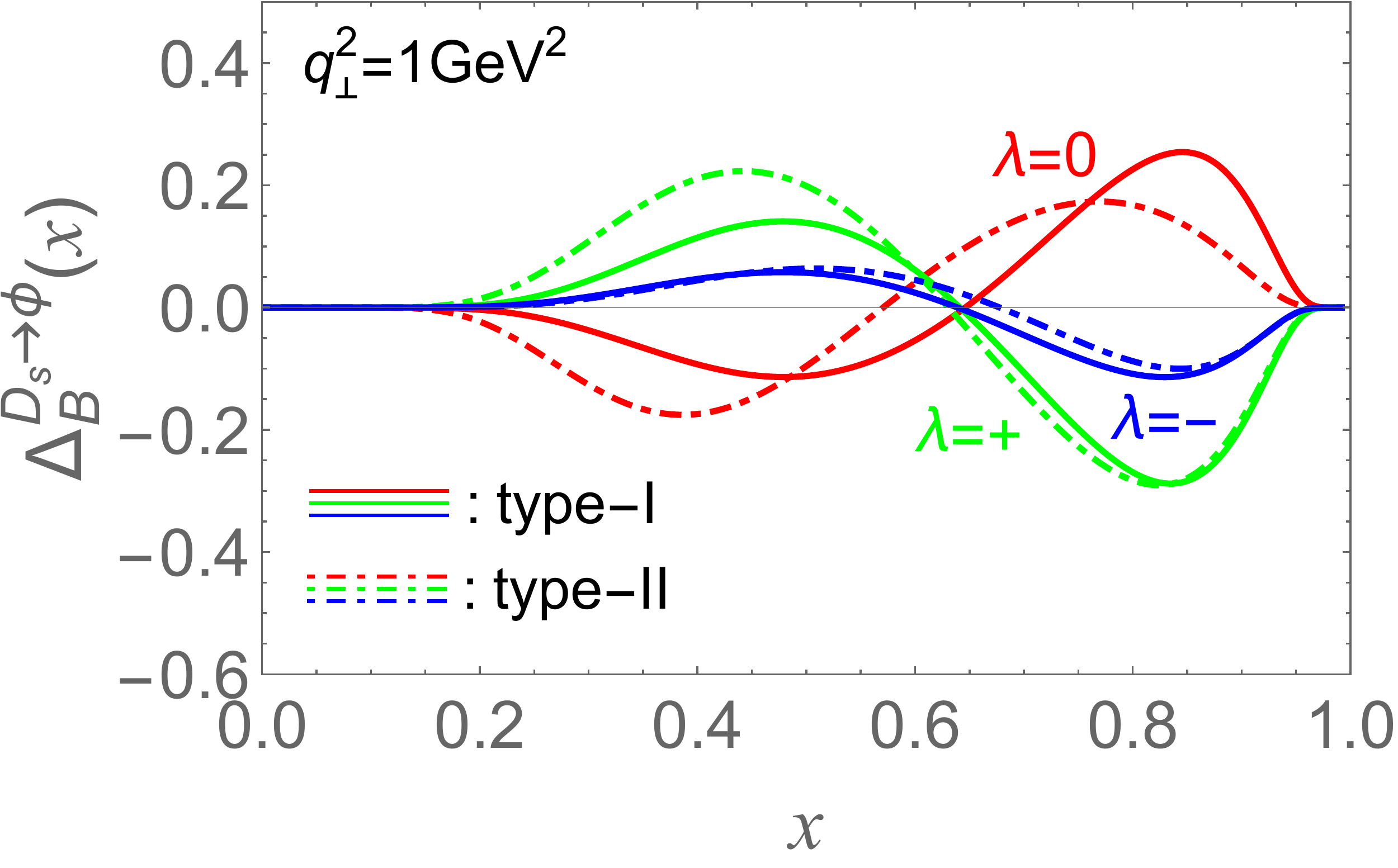}}
\end{figure}

\begin{figure}[t]
\caption{The dependences of $ \d [T_{1,2,3}]_{\rm z.m.}/ \d x$ on $x$ for  $B_{c}\to D^*$ transition at ${\bf q}_{\bot}^2=(0,4,9)\,{\rm GeV^2}$ and  for $D_{s}\to \phi$ transition at ${\bf q}_{\bot}^2=(0,0.5,1)\,{\rm GeV^2}$.}\label{fig:zm}
\vspace{0.32cm}
  \centering
  \subfigure[]{\includegraphics[width=0.30\textwidth]{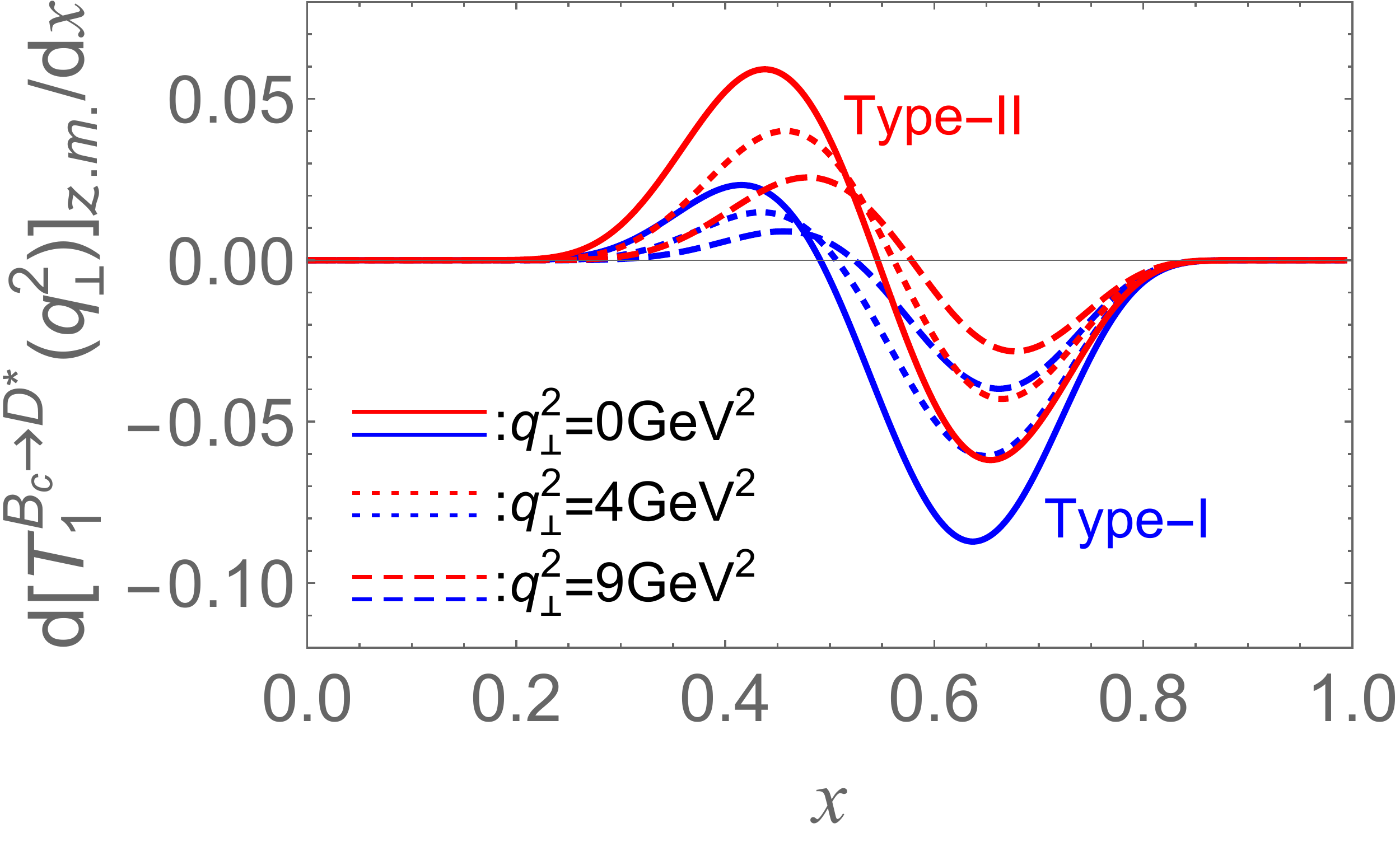}\quad}
  \subfigure[]{\includegraphics[width=0.30\textwidth]{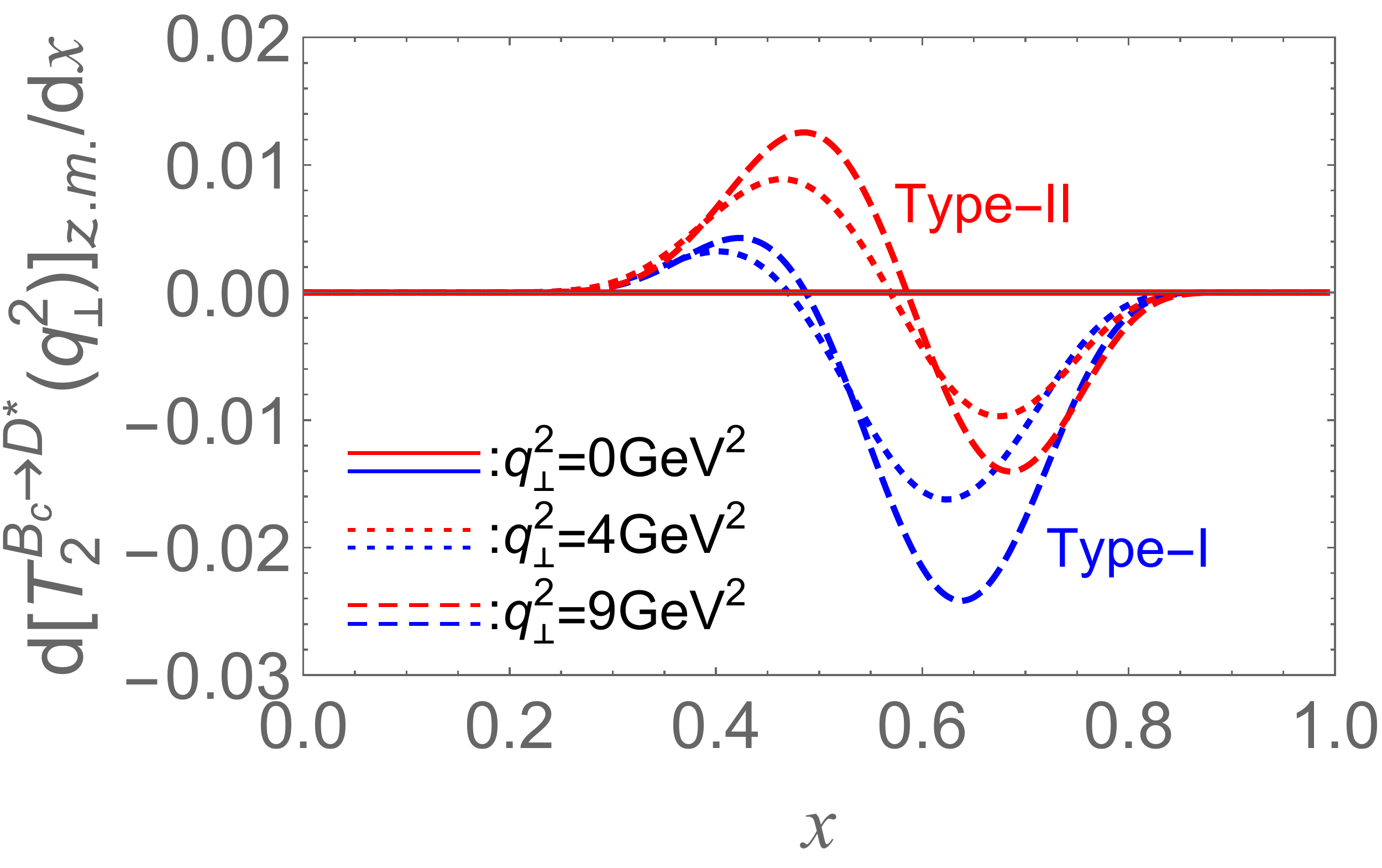}\quad}
  \subfigure[]{\includegraphics[width=0.30\textwidth]{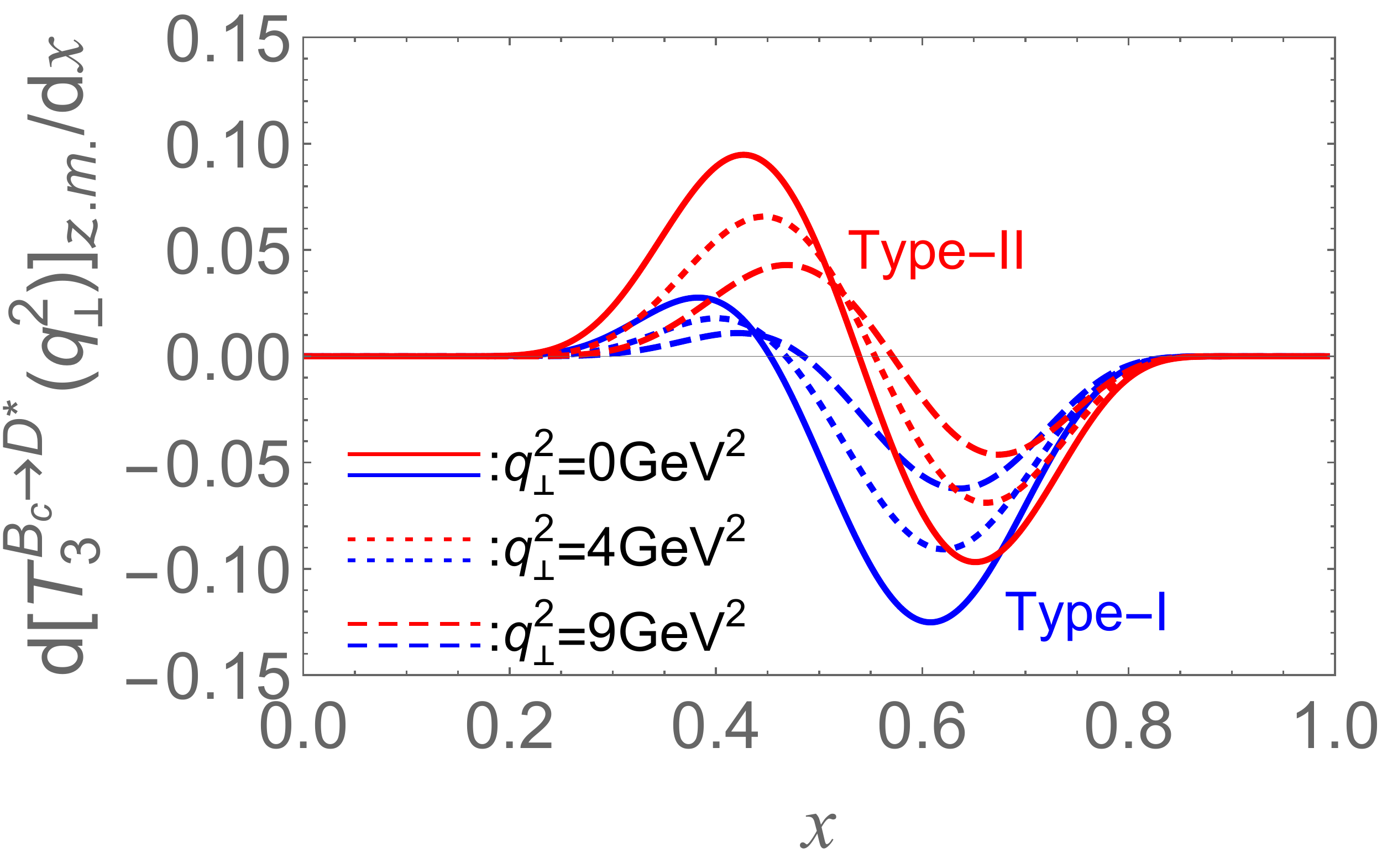}}\\
  \subfigure[]{\includegraphics[width=0.30\textwidth]{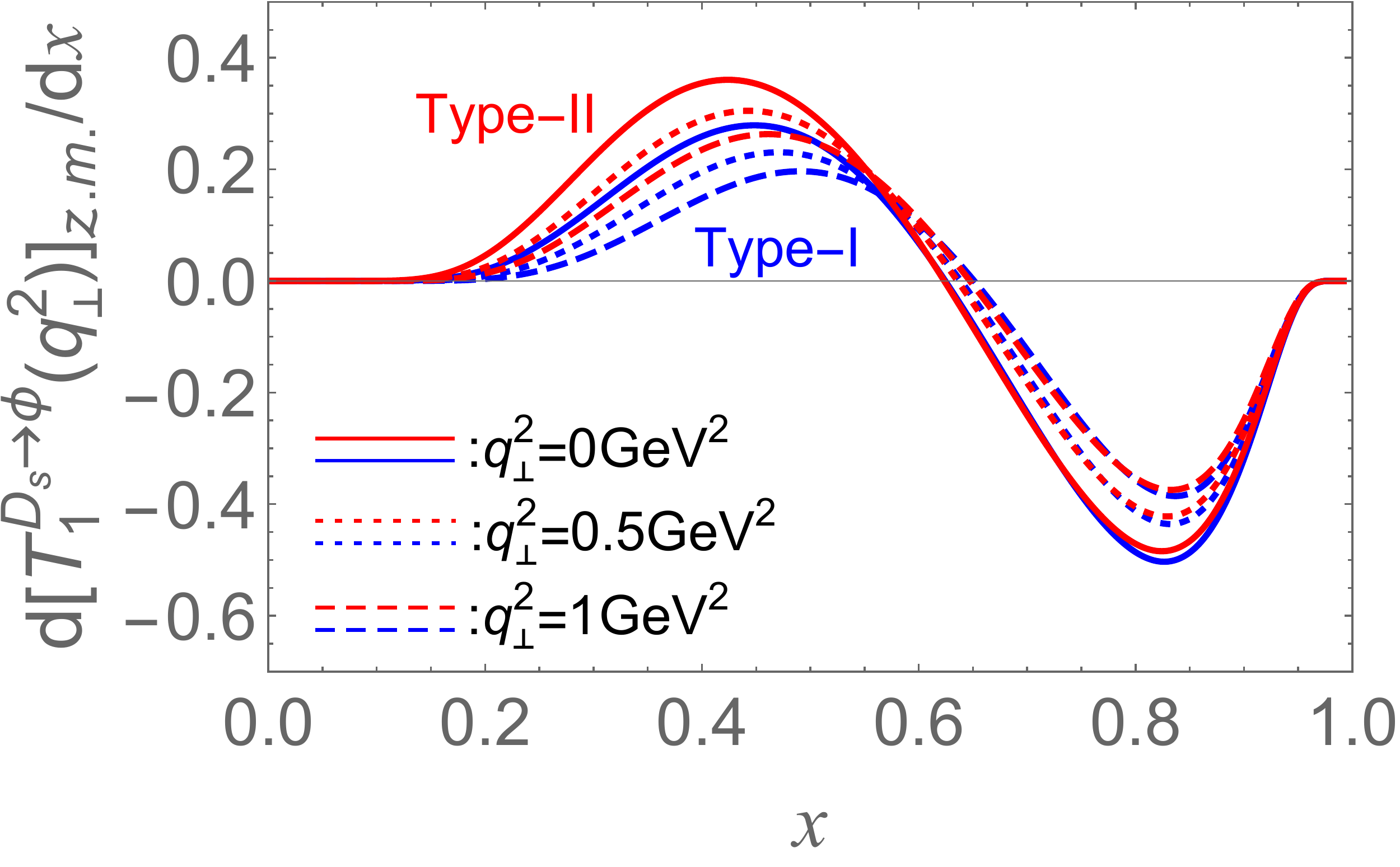}\quad}
  \subfigure[]{\includegraphics[width=0.30\textwidth]{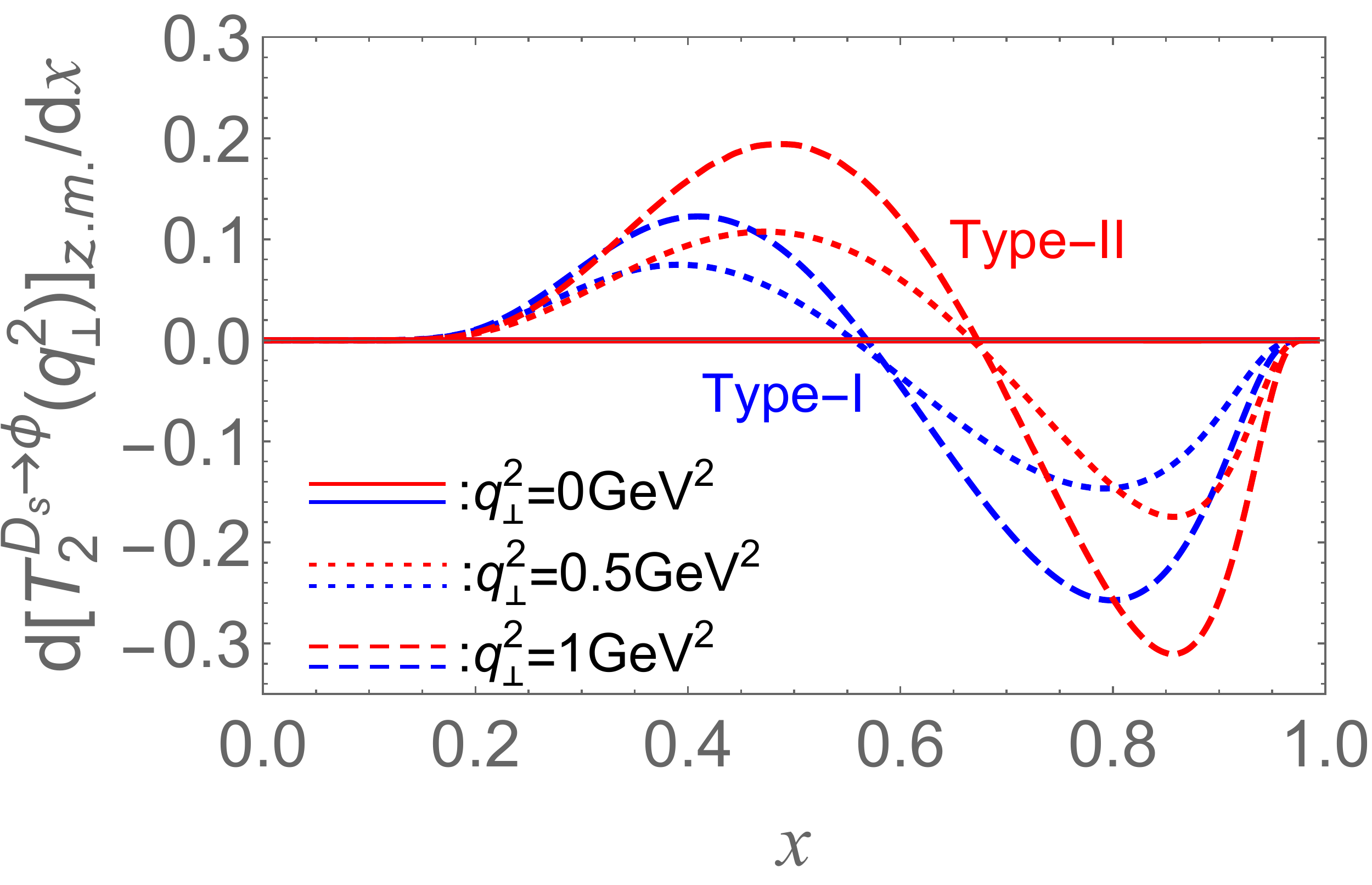}\quad}
  \subfigure[]{\includegraphics[width=0.30\textwidth]{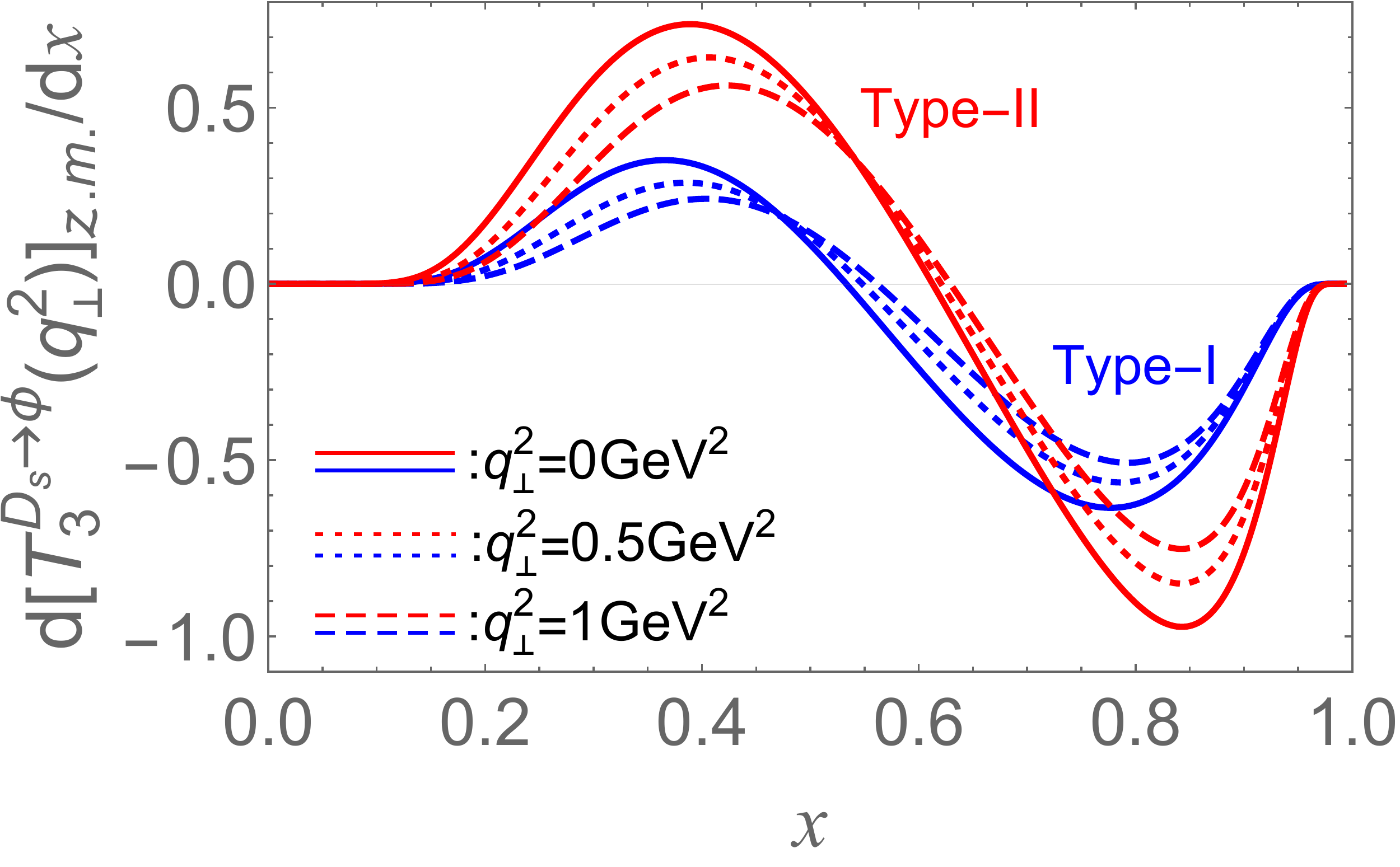}}
\end{figure}

Based on these formulas, we have following discussions and findings:
\begin{itemize}
\item In Eq.~\eqref{eq:BBT1}, the first term  would introduce a spurious unphysical form factor, and thus is expected to vanish.  Unfortunately, it is equal to zero for $\lbd=0$ but is nonzero for $\lbd=\pm$ within type-I scheme. The last three terms give additional contributions to $T_1$, which are however $\lbd$-dependent.  Explicitly, these contributions to $T_1$ can be written as
\begin{equation}
{ \widetilde{T}_1}^{ B}=\left\{
             \begin{array}{lr}
           \big[-2(M'^2-M''^2)+(M'^2-M''^2+\qb^2)\frac{m_1'+m_1''}{D_{V,{\rm con}}''}\big]\frac{1}{M''^2}B_1^{(2)}\,, &  \qquad \lambda=0 \\
            \big[2(M'^2-M''^2)-\qb^2\frac{m_1'+m_1''}{D_{V,{\rm con}}''}\big]\frac{2}{M'^2-M''^2+\qb^2}B_1^{(2)}\,,& \qquad \lambda=+  \\
             \big[1+\frac{m_1'+m_1''}{D_{V,{\rm con}}''}\big]\frac{2\qb^2}{M'^2-M''^2+\qb^2}B_1^{(2)}\,.& \qquad \lambda=-  \\
             \end{array}
\right.
\end{equation}
Further considering the fact that $[{\cal F}]^{\rm CLF}$ is independent of the choice of $\lambda$, it can be concluded that $T_1$ in the CLF QM would suffer from a problem of self-consistence, $[T_1]^{\rm full}_{\lambda=0}\neq [T_1]^{\rm full}_{\lambda=+}\neq [T_1]^{\rm full}_{\lambda=-}$, except that $[T_1]^{\rm B}$ vanishes numerically.
In order to  clearly show the contributions of $B$ function in type-I and II schemes, we take $B_{c}\to D^*$ and $D_{s}\to \phi$ transitions as examples, and list the numerical results of $ [T_1]^{\rm full}_{\lambda=0,\pm}$ in Table~\ref{tab:T1BD}; moreover, the dependence of $\Delta_{B}(x) $ defined as
\begin{align}
\Delta_{B}(x) \equiv \frac{\d [{\cal F}]^{\rm B}_{\lbd}}{ \d x}=N_c\int\frac{\d^2{\bf k'_\bot}}{2(2\pi)^3}\frac{\chi_V'\chi_V''}{\bar x} \widetilde {\cal F}^{ B}_{\lbd}\,,
\end{align}
where ${\cal F}=T_1$, on $x$ are shown in Fig.~\ref{fig:T1}. From these results, it can be easily find that the self-consistence is violated in the traditional type-I  scheme~({\it i.e.}, $[T_1]^{\rm full}_{\lambda=0}\neq [T_1]^{\rm full}_{\lambda=+}\neq [T_1]^{\rm full}_{\lambda=-}$ in type-I  scheme) due to the nonzero contributions of $B$ function, $[T_1]^{\rm full}_{\lambda=0,\pm}=\int_0^1\d x \Delta_{B}(x) \neq 0$~(type-I), but can be recovered by using the type-II scheme due to $[T_1]^{ B}_{\lbd=0,\pm}\dot{=}0$, {\it i.e.},
\begin{align}
[T_1]^{\rm full}_{\lambda=0}\dot{=} [T_1]^{\rm full}_{\lambda=+}\dot{=}  [T_1]^{\rm full}_{\lambda=-}\,.~\qquad \text{(type-II)}
\end{align}

\item  In Eq.~\eqref{eq:BBT3}, the first and the second terms give additional contributions to $T_2$ and $T_3$, the last term is proportional to $\w^\u$ and  corresponds to a unphysical form factor. We take $T_3$ as an example for convenience of discussion. The correction of $B$ function to $T_3$ is
\begin{align}
{\widetilde{T}_3}^{ B}=-4\frac{M'^2-M''^2}{ \epsilon^{*}\cdot q}\,\frac{\omega\cdot \epsilon^{*} }{\omega\cdot P}B_1^{(2)}\left(1+\frac{m_1'-m_1''-2m_2}{D_{V,{\rm con}}''}\right)\,,
\end{align}
which can be explicitly rewritten as $\lambda$-dependent form,
\begin{equation}
{ \widetilde{T}_3}^{\rm B}=\left\{
             \begin{array}{lr}
           -4\frac{M'^2-M''^2}{M'^2-M''^2+q_\bot^2}B_1^{(2)}\left(1+\frac{m_1'-m_1''-2m_2}{D_{V,{\rm con}}''}\right)\,,   \qquad &\lambda=0 \\
             0\,. \qquad &\lambda=\pm  \\
             \end{array}
\right.
\end{equation}
Comparing with the $B$ function contribution to $A_3$ for $V\to V$ transition,  $[{ \widetilde{A}_3}]^{\rm B}$,  given by Eqs.~(4.5) and (4.6) in Ref.~\cite{Chang:2019obq},  it can be found that $[{ \widetilde{T}_3}]^{\rm B}=-{ [\widetilde{A}_3}]^{\rm B}$. We have analyzed the effects of ${ [\widetilde{A}_3}]^{\rm B}$~($[{ \widetilde{T}_3}]^{\rm B}$) in Ref.~\cite{Chang:2019obq} in detail, and have obtained  the same conclusion as we have obtained in the last item via $T_1$.

\item The covariance of the matrix element of tensor operators in the type-I scheme is violated due to the non-zero $\w$-dependent contributions associated with $B$ function~(for instance, the last term  proportional to $\w^\u$  in  Eq.~\eqref{eq:BBT3}); while, the Lorentz covariance can be naturally recovered in the  type-II scheme because  all of the  contributions associated with $B$ function exist only in form but vanish numerically.

\item Taking  $B_c\to D^*$ and $D_s\to \phi$ transitions as examples again, the numerical results of valence contributions and CLF results for $T_1$  are also listed in Table~\ref{tab:T1BD}. The zero-mode contributions can be easily obtained by the relation that  $[{\cal F}]^{\rm CLF}=[{\cal F}]^{\rm val.}+[{\cal F}]^{\rm z.m.}$.
 In addition,  the dependences of $\d [{\cal F}]^{\rm z.m.}/\d x$ on $x$, where ${\cal F}=T_{1,2,3}$, are shown in  Fig.~\ref{fig:zm}.  From these results, it can be found that the  zero-mode effects are significant within the traditional type-I scheme; while, these contributions vanish numerically  in the  type-II scheme, {\it i.e.},  $[T_{1,2,3}(q^2)]_{\rm z.m.}\dot{=} 0$~(type-II). Here, we would like to  clarify that, the spurious $\w$-dependent contributions associated with $C$ functions have been neutralized by the zero-mode contributions~(one may refer to Ref.~\cite{Jaus:1999zv} for details),  therefore the zero-mode contribution, $[{\cal F}]^{\rm z.m.}$, discussed here is form the residual zero-mode contribution to the matrix element. It implies that the zero-mode contributions to the matrix element within the type-II scheme are only responsible for neutralizing spurious $\w$-dependent contributions associated with $C$ functions,  but do not contribute numerically to the form factors. 

\item Comparing $[T_{1,2,3}]^{\rm  SLF}$  with $[T_{1,2,3}]^{\rm  val.}$, which are given by Eqs.~(\ref{eq:T1SLF}-\ref{eq:T3SLF}) and Eqs.~(\ref{eq:T1val}-\ref{eq:T3val}), respectively,  it can be found that the SLF results for $T_{1,2,3}$ are exactly the same as the valence contributions in form after taking $M\to M_0$ replacement~(type-II), which can also be clearly seen from the numerical results for $T_1$ in Table~\ref{tab:T1BD}. It is exactly what we expect due to the following facts: (1) the CLF QM has employed the LF vertex functions which can only be extracted by mapping the CLF result to the SLF one; (ii) the zero-mode contributions are not taken into account in the SLF result, therefore the SLF result is in fact only corresponding to the valence contribution in the CLF QM. The findings in this and last items can be concluded as
\begin{equation}
[T_{1,2,3}]^{\rm SLF}=[T_{1,2,3}]^{\rm val.}\dot{=}[T_{1,2,3}]^{\rm CLF}\,.\qquad \text{(type-II)}
\end{equation}
This relation is also valid  for the form factors of $P\to A$  and $P\to (P,S)$ transitions, while,  for the later, the notation ``$\dot{=}$'' should be replaced by ``$=$'' because $F_T$ and $U_T$ are zero-mode independent.
\end{itemize}
The analyses and findings mentioned above confirm again the main conclusion obtained in our previous works~\cite{Chang:2018zjq,Chang:2019mmh} and Ref.~\cite{Choi:2013mda}. In addition to above-mentioned self-consistency problem of CLF QM caused by the contributions associated with $B$ function, we note a new inconsistence problem, which will be discussed in the following.

The tensor form factors $T_{1,2,3}$ have also been obtained by Cheng and Chua~(CC) in Ref.~\cite{Cheng:2009ms} within the CLF QM, these results are collected in the appendix A~( Eqs.~(\ref{eq:chengT1}-\ref{eq:chengT3}) ) for convenience of discussion. Comparing our results given by Eqs.~(\ref{eq:CLFT1}-\ref{eq:CLFT3}) with CC's results, one can easily find that the results  for $T_1$ are consistent with each other, but the ones for   $T_{2}$ and $T_{3}$ are obviously different. In addition,  for  $T_{2}$ and $T_{3}$, it is found that our and CC's numerical results are also inconsistent  with each other in the traditional type-I scheme. After carefully checking our and  CC's calculations,  we find that such new inconsistence problem is caused by the different ways to deal with the trace term $S^{\u\v\lbd}$ related to the fermion-loop in ${\cal B}_{\rm CLF}^{P\to V}[\Gamma=\sg^{\u\v}\r_5]$,
where ${\cal B}_{\rm CLF}$ and $S$ have been given by Eq.~\eqref{eq:Bclf1} and Eq.~\eqref{eq:Sterm}, respectively. Explicitly,  ${\cal B}_{\rm CLF}^{P\to V}[\Gamma=\sg^{\u\v}\r_5]$ is written as
\begin{eqnarray}
 {\cal B}_{\rm CLF}^{P\to V}[\Gamma=\sg^{\u\v}\r_5]=N_c \int \frac{\d^4 k_1'}{(2\pi)^4} \frac{H_{P}H_{V}}{N_1'\,N_1''\,N_2}iS^{\u\v\lbd}\e^*_\lbd \,.
\end{eqnarray}
The trace term, $S^{\u\v\lbd}$,  can be related to $S'^{\rho\sg\lbd}$ by using the identity $2\sg_{\u\v}\r_5 = i\ve_{\u\v\rho\sg}\sg^{\rho\sg}$, where $S'^{\rho\sg\lbd}$ is the trace term in ${\cal B}_{\rm CLF}^{P\to V}[\Gamma=\sg^{\rho\sg}]$ corresponding to $T_1$. Explicitly, it is written as
\begin{align}
S^{\u\v}_{\lbd}=&\frac{i}{2}\ve^{\u\v\rho\sg} S'_{\rho\sg\lbd} \label{eq:relaSS}\\
=&i\ve^{\u\v\rho\sg}\bigg\{
\ve_{\rho\sg\lbd\a}\left[2(m'_1m_2+m''_1m_2-m'_1m''_1)k'^\a_1+m'_1m''_1P^{\a}+(m'_1m''_1-2m'_1m_2)q^{\a}\right]\nonumber\\
&-\ve_{\rho\sigma\alpha\beta}\frac{(4k'_{1}-3q-P)_\lbd}{D''_V}\left[(m'_1+m''_1)k'^{\alpha}_1P^\beta+(m''_1-m'_1+2m_2)k'^{\alpha}_1q^\beta+m'_1P^\alpha q^\beta\right]\nonumber\\
&+\ve_{\rho\sg\lbd\a}\left[2(k'_1\cdot k_2-k''_1\cdot k_2-k'_1\cdot k''_1)k'^{\a}_1+k'_1\cdot k''_1P^\a+(k'_1\cdot k''_1-2k'_1\cdot k_2)q^\a\right]\nonumber\\
&+(g^{\sg}_{\lbd}\ve_{\rho\a\b\r}-g^{\rho}_{\lbd}\ve_{\sg\a\b\r})P^{\a}q^{\b}k'^{\r}_1+\ve_{\sg\rho\a\b}(P^\a q^\b k'_{1\lbd}+k'^\a_1P^\b q_{\lbd}+q^\a k'^\b_1P_{\lbd})\nonumber\\
&+\ve_{\rho\lbd\a\b}\left[ k'_{1\sg}P^\a q^\b+q_\sg P^\a k'^\b_1 +(P+2q)_\sg q^\a k'^\b_1+2k'_{1\sg} k'^\a_1 (P+q)^\b \right]\nonumber\\
&-\ve_{\sg\lbd\a\b} \left[k'_{1\rho}P^\a q^\b+q_\rho P^\a k'^\b_1+(P+2q)_\rho q^\a k'^\b_1+2k'_{1\rho}k'^\a_1(P+q)^\b\right]
\bigg\}\,.
\end{align}
For convenience of  discussion, we take the last term, $[S^{\u\v}_{\lbd}]_{\text{last term}}=-2i\ve^{\u\v\rho\sg}\ve_{\sg\lbd\a\b}\,k'_{1\rho}k'^\a_1(P+q)^\b$, as example.

\begin{figure}[t]
\caption{$\Delta^{T_{2,3}}_{\rm CLF}(x)$ for  $B_{c}\to D^*$ transition at ${\bf q}_{\bot}^2=(0,4,9)\,{\rm GeV^2}$ and  for $D_{s}\to \phi$ transition at ${\bf q}_{\bot}^2=(0,0.5,1)\,{\rm GeV^2}$.}\label{fig:diff}
\vspace{0.32cm}
  \centering
  \subfigure[]{\includegraphics[width=0.30\textwidth]{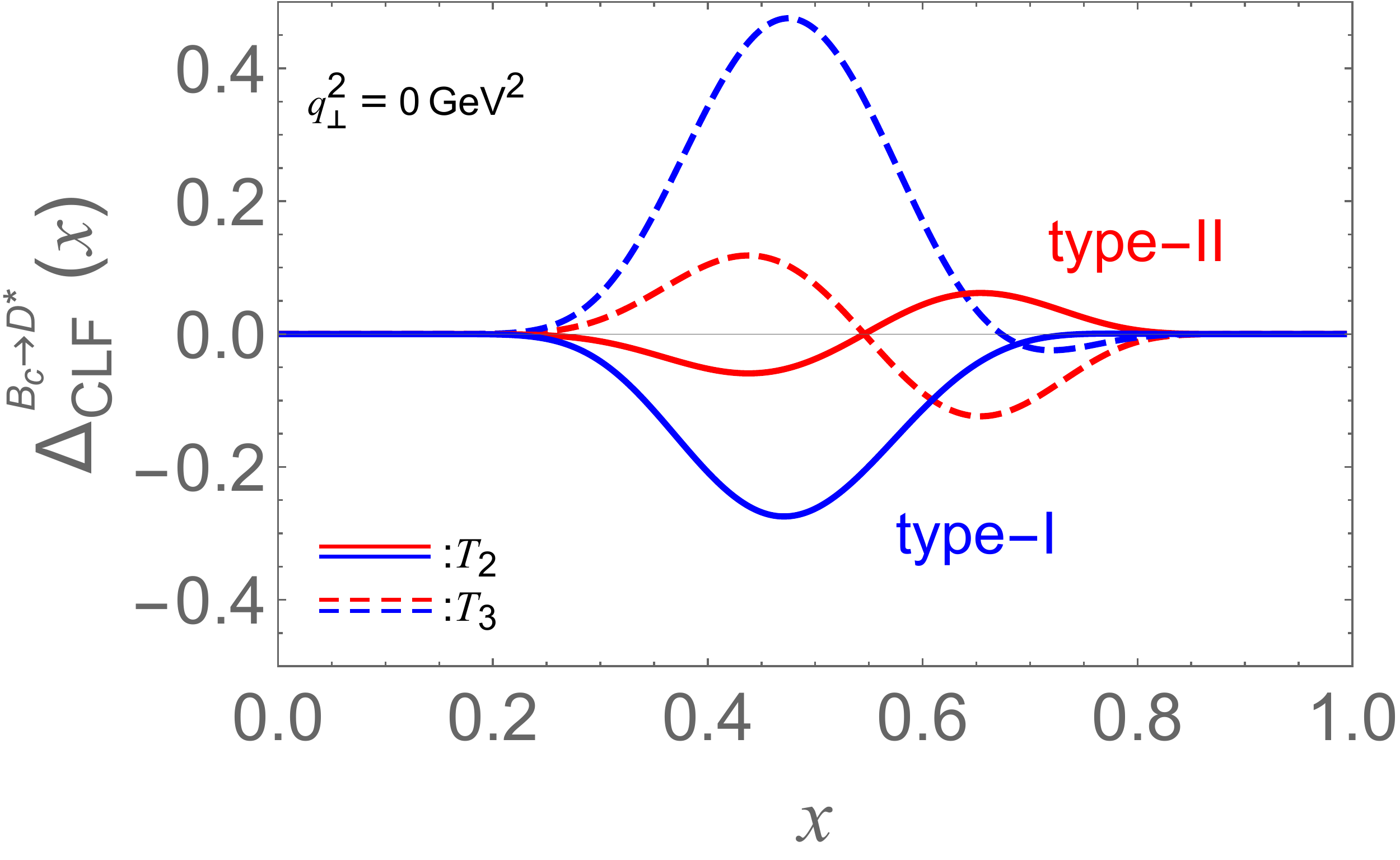}}\quad
  \subfigure[]{\includegraphics[width=0.30\textwidth]{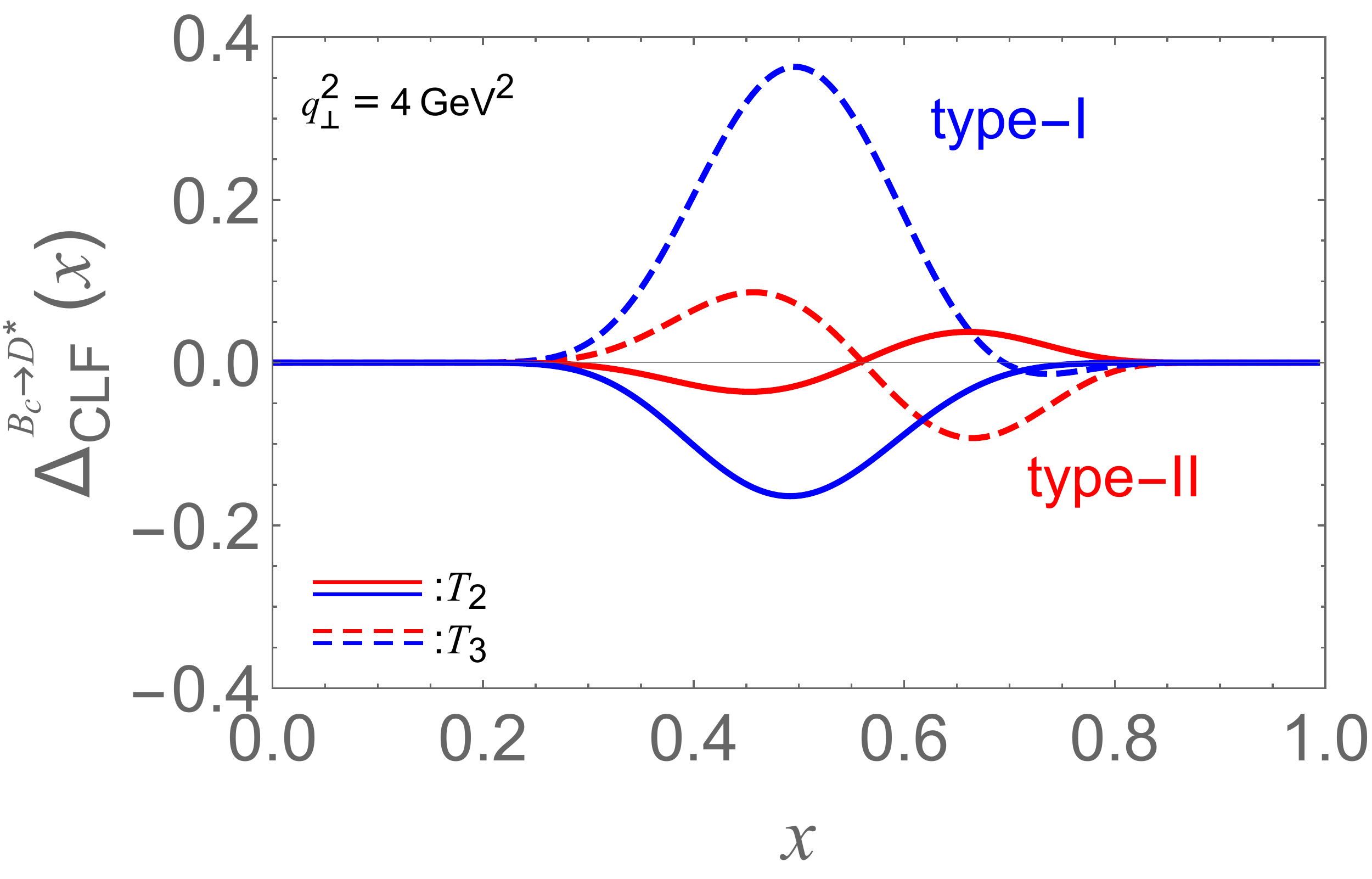}}\quad
  \subfigure[]{\includegraphics[width=0.30\textwidth]{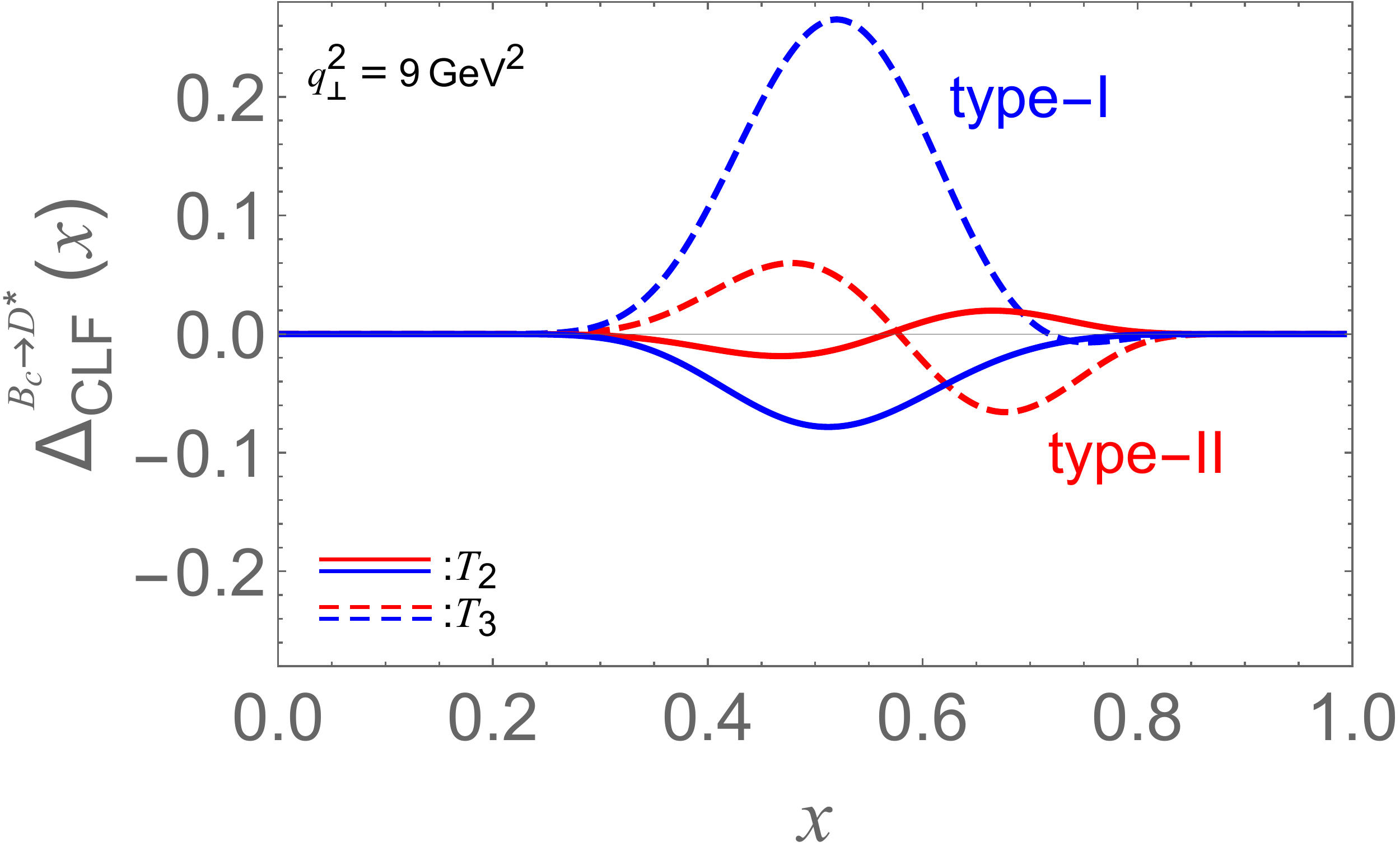}}\\
  \subfigure[]{\includegraphics[width=0.30\textwidth]{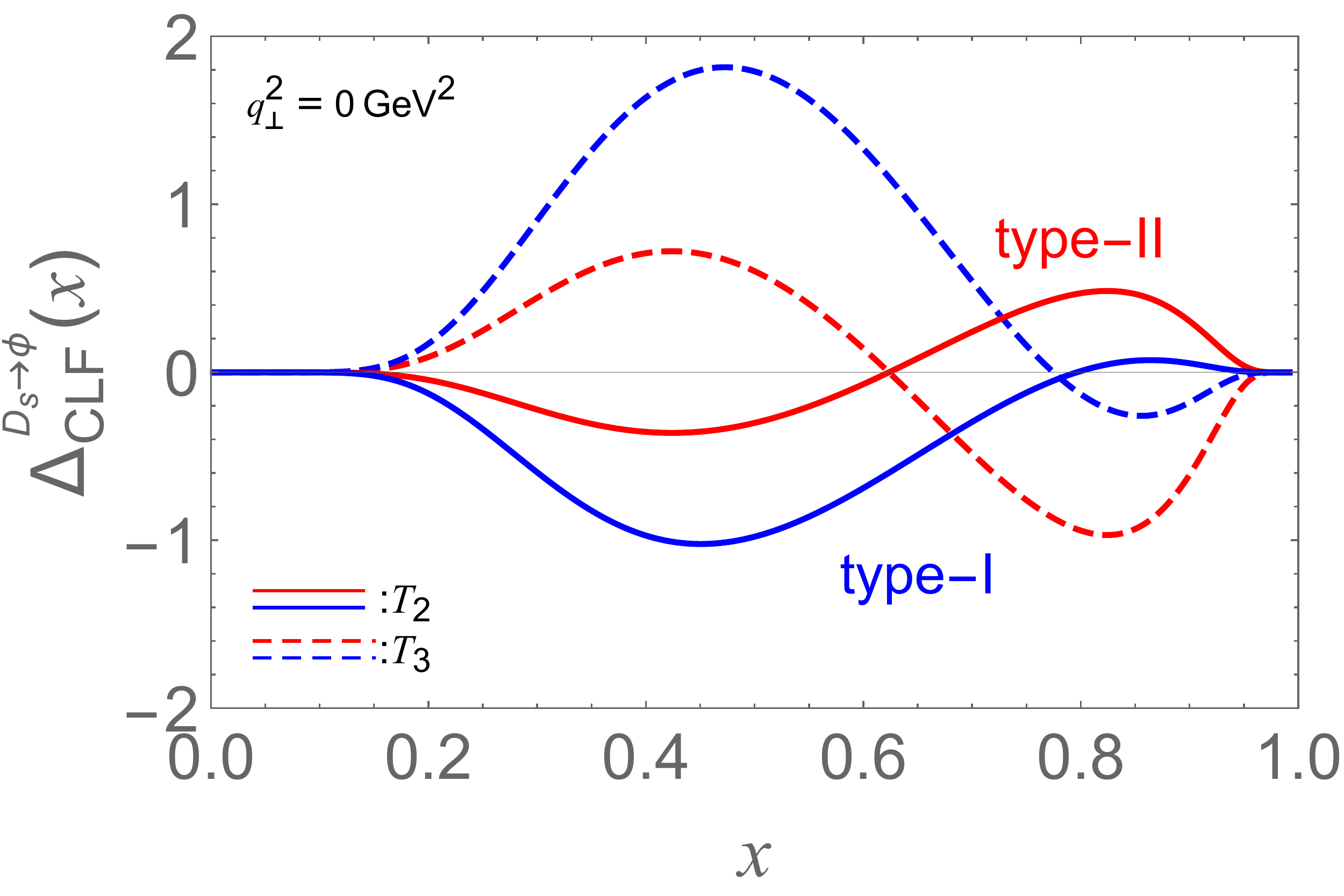}}\quad
   \subfigure[]{\includegraphics[width=0.30\textwidth]{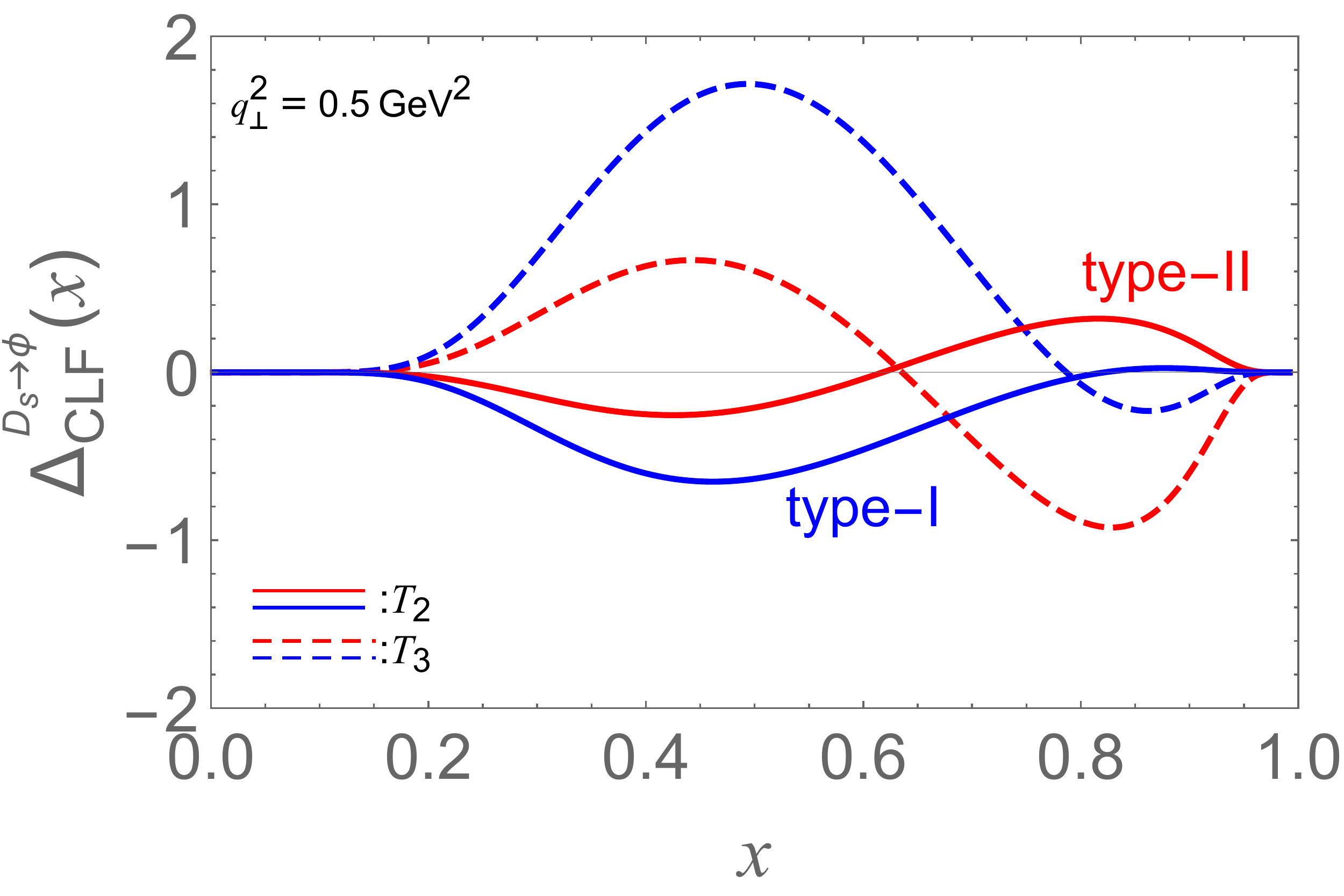}}\quad
    \subfigure[]{\includegraphics[width=0.30\textwidth]{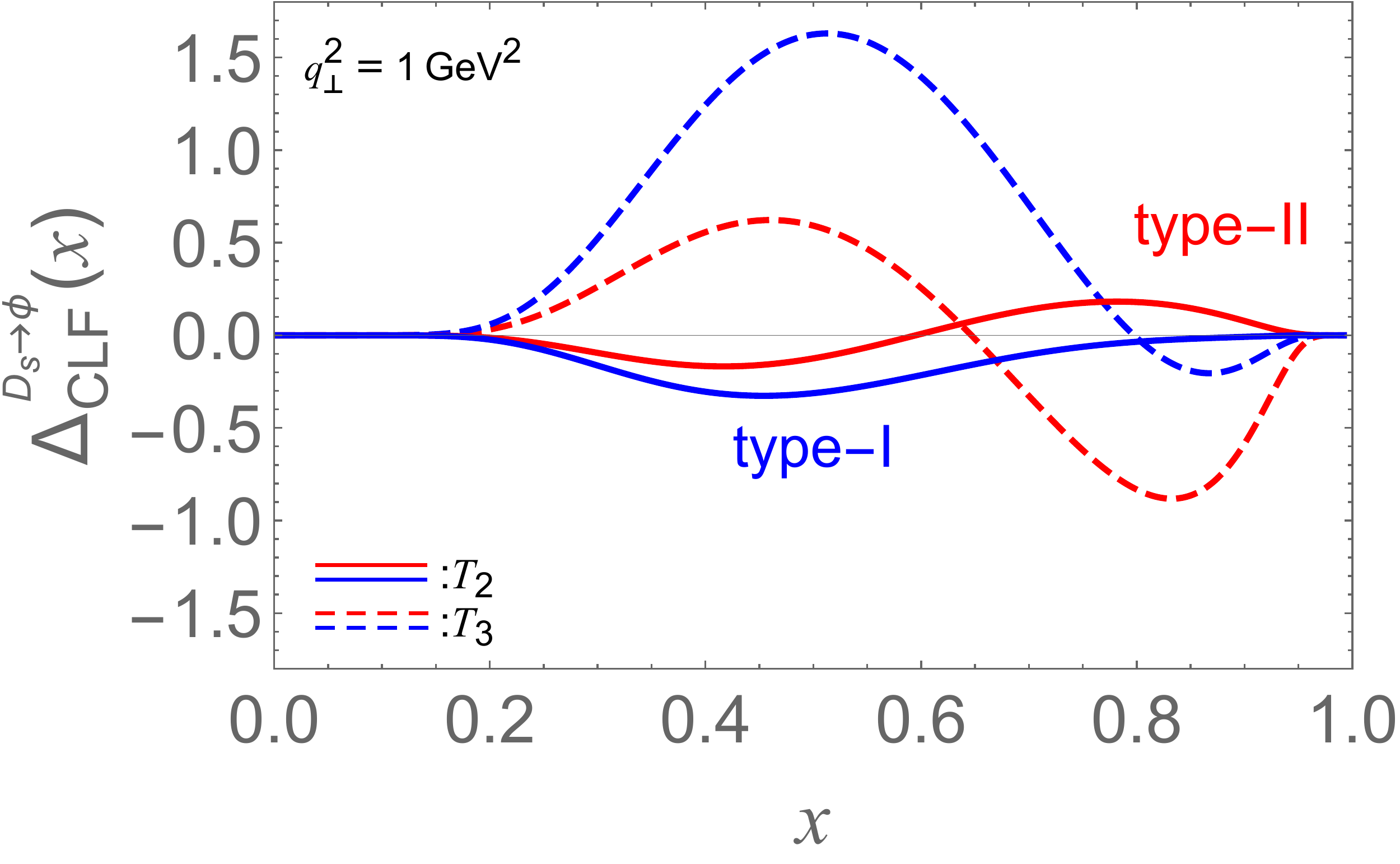}}
\end{figure}

In the CC's calculation~\cite{Cheng:2009ms}, the obtained results for  $\hat{S}'_{\rho\sg\lbd}$ is used directly to calculate $\hat{S}^{\u\v}_{\lbd}$ by using $\hat{S}^{\u\v}_{\lbd}=\frac{i}{2}\ve^{\u\v\rho\sg} \hat{S}'_{\rho\sg\lbd}$, which is formally similar to Eq.~\eqref{eq:relaSS}. It implies that,  after integrating out $k'^-_{1}$, the replacement for $\hat{k}'^{\rho}_1\hat{k}'^\a_1$ is made directly by using Eq.~\eqref{eq:repkk} even though $\rho$ and $\a$ are dummy indices; then, in the CC's way, using the identity
\begin{align}\label{eq:iden}
\ve^{\u\v\rho\sg}\ve_{\sg\lbd\a\b}=[g_\lbd^\u(g_\a^\v g_\b^\rho-g_\a^\rho g_\b^\v)+g_\lbd^\v(g_\a^\rho g_\b^\u-g_\a^\u g_\b^\rho)+g_\lbd^\rho(g_\a^\u g_\b^\v-g_\a^\v g_\b^\u)]\,,
\end{align}
it is obtained that
\begin{align}
[\hat{S}^{\u\v}_{\lbd}]_{\text{last term}}^{\text{CC}}=&2i g_\lbd^\v g_\a^\u g_\b^\rho[g^\a_\rho A^{(2)}_1+P^\a P_\rho A^{(2)}_2+(P_\rho q^\a+q_\rho P^\a)A^{(2)}_3+q^\a q_\rho A^{(2)}_4](P+q)^\b\,+\,...\nonumber\\
=&2ig_\lbd^\v\Big\{P^\u\left[A^{(2)}_1+(3M'^2-M''^2-q^2)A^{(2)}_2+(M'^2-M''^2+q^2)A^{(2)}_3\right]\nonumber\\
&+q^\u\left[A^{(2)}_1+\left(3M'^2-M''^2-q^2\right)A^{(2)}_3+\left(M'^2-M''^2+q^2\right)A^{(2)}_4\right]
\Big\}\,+\,...\,,\label{eq:CCresult}
\end{align}
where only the terms proportional to $g_\lbd^\v g_\a^\u g_\b^\rho$ are shown for convenience of comparison with our corresponding result given in the following.

In our calculation, we employ the standard procedure of CLF calculation instead of directly using the obtained result for  $\hat{S}'_{\rho\sg\lbd}$. Firstly, we write  $[S^{\u\v}_{\lbd}]_{\text{last term}}$ as
\begin{align}
[S^{\u\v}_{\lbd}]_{\text{last term}}&=-2i g_\lbd^\v k'^\u_{1} k'_1\cdot(P+q)+...\nonumber\\
&=-2i g_\lbd^\v k'^\u_{1} \left(M'^2+m'^2_1+m'^2_1-m^2_2- N_2 + N'_1 \right)+...\,,
\end{align}
where only the terms proportional to $g_\lbd^\v g_\a^\u g_\b^\rho$ corresponding to the CC' result, Eq.~\eqref{eq:CCresult}, are shown, by using Eq.~\eqref{eq:iden} and
\begin{align}
k'_1\cdot q =& \frac{1}{2}\left( N'_1+m'^2_1- N''_1-m''^2_1-\qb^2\right)\,,\\
k'_1\cdot P=&\frac{1}{2}\left(2M'^2+ N'_1+m'^2_1+ N''_1+m''^2_1-2 N_2-2m^2_2+\qb^2\right)\,.
\end{align}
Then, after integrating out $k'^-_{1}$, we further make replacements for  $\hat{k}_1'^\u$ and $\hat{k}_1'^\u\hat{N}_2 $~(note that $\u$ is free index) by using Eqs.~\eqref{eq:repk} and \eqref{eq:repkN}. Finally, we arrive at
\begin{align}
[\hat{S}^{\u\v}_{\lbd}]_{\text{last term}}^{\text{ours}}=&2ig_\lbd^\v\bigg\{P^\u\left(M'^2+m'^2_1-m_2^2{+}\hat N'_1\right)A^{(1)}_1\nonumber\\
&+q^\u\left[\left(M'^2+m'^2_1-m_2^2{+}\hat N'_1-Z_2\right)A^{(1)}_2-\frac{M'^2-M''^2}{q^2}A^{(2)}_1\right]
\bigg\}+...\,.\label{eq:ours}
\end{align}

Comparing CC's calculation with ours, it can be found that different replacements are needed due to the different strategies for dealing with $S$ term, which further results in the different theoretical  results  for $\hat{S}$,  as well as for $[T_2]^{\rm CLF}$ and $[T_3]^{\rm CLF}$. In order to clearly show the divergence between  CC's results and ours,  we take $B_c\to D^*$ and $D_s\to \phi$ transitions as examples, and plot the difference defined by
\begin{align}
\Delta^{\cal F}_{\rm CLF}(x,\qb^2)\equiv \frac{\d [{\cal F}]^{\rm CLF}_{\rm ours}}{\d x}-\frac{\d [{\cal F}]^{\rm CLF}_{\rm CC}}{\d x}\,, \qquad {\cal F}=T_2\text{ and } T_3
\end{align}
in Fig.~\ref{fig:diff}. In can be easily found from  Fig.~\ref{fig:diff} that: our and  CC's numerical results for  $[T_{2,3}]^{\rm CLF}$ are inconsistent with each other within the traditional type-I scheme because $[T_{2,3}]^{\rm CLF}_{\rm ours} - [T_{2,3}]^{\rm CLF}_{\rm CC}=\int_0^1 \d x\,\Delta_{\rm CLF}^{T_{2,3}}(x)\neq 0$; however, it is interesting that the consistence can be achieved numerically  within the  type-II scheme because $\int_0^1 \d x\,\Delta_{\rm CLF}^{T_{2,3}}(x)= 0$. The case of $P\to A$ transition is similar to the one of $P\to V$ transition.

\begin{table}[t]
\scriptsize
\begin{center}\renewcommand{\baselinestretch}{1}
\caption{\label{tab:PP} The numerical results of the tensor form factors for  $c\to q\,,s$~($q=u\,,d$) induced $D_{q,s}\to P $ and $b\to q\,,s\,,c$ induced $B_{q,s,c}\to P $ transitions. The theoretical errors are caused by the uncertainties  of input parameters~($\beta$ and $m_{q,s,c,b}$).}
\vspace{0.2cm}
\let\oldarraystretch=\arraystretch
\renewcommand*{\arraystretch}{1}
\setlength{\tabcolsep}{3pt}
\begin{tabular}{lccc|lccccccc}
\hline\hline
 &$\mathcal{F}(0)$  &$a$   &$b$ & &$\mathcal{F}(0)$  &$a$   &$b$
\\\hline
$F_T^{D\to \pi}$ &$0.84^{+0.16}_{-0.13}$&$-0.03^{+0.26}_{-0.18}$ &$0.07^{+0.06}_{-0.09}$
&$F_T^{D_{s}\to K}$ &$0.93^{+0.20}_{-0.15}$&$-0.01^{+0.29}_{-0.20}$ &$0.10^{+0.07}_{-0.10}$
\\\hline
$F_T^{D\to K}$ &$0.96^{+0.17}_{-0.15}$&$-0.02^{+0.23}_{-0.23}$ &$0.10^{+0.08}_{-0.10}$
&$F_T^{D_{s}\to \eta_{s}}$ &$1.05^{+0.21}_{-0.17}$&$0.01^{+0.26}_{-0.20}$ &$0.14^{+0.09}_{-0.11}$\\\hline
$F_T^{B\to \pi}$ &$0.32^{+0.04}_{-0.04}$&$0.42^{+0.14}_{-0.10}$ &$0.03^{+0.05}_{-0.06}$
&$F_T^{B_{s}\to K}$ &$0.30^{+0.06}_{-0.06}$&$0.66^{+0.10}_{-0.08}$ &$0.17^{+0.05}_{-0.06}$
\\\hline
$F_T^{B_{c}\to D}$ &$0.24^{+0.12}_{-0.10}$&$1.38^{+0.10}_{-0.11}$ &$1.23^{+0.43}_{-0.42}$
&$F_T^{B\to K}$ &$0.39^{+0.06}_{-0.05}$&$0.43^{+0.14}_{-0.10}$ &$0.02^{+0.05}_{-0.06}$
\\\hline
$F_T^{B_{s}\to \eta_{s}}$ &$0.36^{+0.08}_{-0.07}$&$0.66^{+0.07}_{-0.05}$ &$0.16^{+0.06}_{-0.07}$
&$F_T^{B_{c}\to D_{s}}$ &$0.36^{+0.14}_{-0.12}$&$1.20^{+0.08}_{-0.09}$ &$0.87^{+0.29}_{-0.30}$
\\\hline
$F_T^{B\to D}$ &$0.78^{+0.12}_{-0.12}$ &$0.49^{+0.10}_{-0.10}$ &$-0.02^{+0.01}_{-0.01}$
&$F_T^{B_{s}\to D_{s}}$ &$0.81^{+0.11}_{-0.14}$&$0.57^{+0.03}_{-0.02}$ &$0.08^{+0.06}_{-0.06}$\\\hline
$F_T^{B_{c}\to \eta_{c}}$ &$0.90^{+0.17}_{-0.22}$ &$1.09^{+0.15}_{-0.20}$ &$0.57^{+0.23}_{-0.23}$
\\\hline\hline
\end{tabular}
\end{center}
\end{table}

\begin{table}[t]
\scriptsize
\begin{center}
\caption{\label{tab:PS} Same as Table~\ref{tab:PP} except for $D_{q,s}\to S $ and $B_{q,s,c}\to S $ transitions. }
\vspace{0.2cm}
\let\oldarraystretch=\arraystretch
\renewcommand*{\arraystretch}{1}
\setlength{\tabcolsep}{3pt}
\begin{tabular}{lccc|lccccccc}
\hline\hline
 &$\mathcal{F}(0)$  &$a$   &$b$ & &$\mathcal{F}(0)$  &$a$   &$b$
\\\hline
$U_T^{D\to S_{(q,\bar{q})}}$ &$0.77^{+0.13}_{-0.12}$&$-0.17^{+0.26}_{-0.17}$ &$0.18^{+0.11}_{-0.18}$
&$U_T^{D_{s}\to S_{(q,\bar{s})}}$ &$0.94^{+0.17}_{-0.14}$&$-0.16^{+0.26}_{-0.18}$ &$0.17^{+0.09}_{-0.14}$
\\\hline
$U_T^{D\to S_{(s,\bar{q})}}$ &$0.74^{+0.13}_{-0.12}$&$-0.17^{+0.25}_{-0.17}$ &$0.16^{+0.11}_{-0.16}$
&$U_T^{D_{s}\to S_{(s,\bar{s})}}$ &$0.93^{+0.16}_{-0.15}$&$-0.18^{+0.26}_{-0.18}$ &$0.20^{+0.11}_{-0.16}$
\\\hline
$U_T^{B\to S_{(q,\bar{q})}}$ &$0.35^{+0.05}_{-0.04}$&$0.28^{+0.21}_{-0.17}$ &$0.02^{+0.04}_{-0.03}$
&$U_T^{B_{s}\to S_{(q,\bar{s})}}$ &$0.38^{+0.04}_{-0.05}$&$0.47^{+0.08}_{-0.05}$ &$0.12^{+0.05}_{-0.07}$
\\\hline
$U_T^{B_{c}\to S_{(q,\bar{c})}}$ &$0.40^{+0.13}_{-0.13}$&$1.16^{+0.18}_{-0.17}$ &$0.88^{+0.38}_{-0.38}$
&$U_T^{B\to S_{(s,\bar{q})}}$ &$0.37^{+0.05}_{-0.05}$&$0.25^{+0.19}_{-0.15}$ &$0.01^{+0.03}_{-0.03}$
\\\hline
$U_T^{B_{s}\to S_{(s,\bar{s})}}$ &$0.43^{+0.06}_{-0.05}$&$0.44^{+0.26}_{-0.21}$ &$0.11^{+0.05}_{-0.01}$
&$U_T^{B_{c}\to S_{(s,\bar{c})}}$ &$0.56^{+0.12}_{-0.14}$&$0.98^{+0.10}_{-0.11}$ &$0.62^{+0.24}_{-0.24}$
\\\hline
$U_T^{B\to S_{(c,\bar{q})}}$ &$0.51^{+0.09}_{-0.08}$&$0.33^{+0.12}_{-0.07}$ &$-0.08^{+0.04}_{-0.05}$
&$U_T^{B_{s}\to S_{(c,\bar{s})}}$ &$0.71^{+0.12}_{-0.11}$&$0.36^{+0.11}_{-0.11}$ &$0.03^{+0.01}_{-0.02}$
\\\hline
$U_T^{B_{c}\to S_{(c,\bar{c})}}$ &$1.21^{+0.32}_{-0.25}$&$0.83^{+0.35}_{-0.31}$ &$0.40^{+0.30}_{-0.16}$
\\\hline\hline
\end{tabular}
\end{center}
\end{table}

\begin{table}[t]
\scriptsize
\begin{center}
\caption{\label{tab:PV} \small Same as Table~\ref{tab:PP} except for $D_{q,s}\to V $ and $B_{q,s,c}\to V $ transitions.}
\vspace{0.2cm}
\let\oldarraystretch=\arraystretch
\renewcommand*{\arraystretch}{1}
\setlength{\tabcolsep}{4.8pt}
\begin{tabular}{lccc|lccccccc}
\hline\hline
 &$\mathcal{F}(0)$  &$a$   &$b$ & &$\mathcal{F}(0)$  &$a$   &$b$
\\\hline
$T_1^{D\to\rho}$ &$0.62_{-0.08}^{+0.08}$&$0.05_{-0.18}^{+0.26}$ &$0.10_{-0.15}^{+0.08}$
&$T_1^{D_{s}\to K^*}$ &$0.56_{-0.09}^{+0.09}$&$0.15^{+0.23}_{-0.16}$ &$0.11^{+0.09}_{-0.13}$\\
$T_2^{D\to\rho}$ & $0.62_{-0.08}^{+0.08}$ &$-0.81^{+0.05}_{-0.05}$ & $0.59^{+0.03}_{-0.03}$
&$T_2^{D_{s}\to K^*}$ &$0.56_{-0.09}^{+0.09}$&$-0.64^{+0.05}_{-0.15}$ &$0.47^{+0.03}_{-0.03}$\\
$T_3^{D\to\rho}$ & $0.30^{+0.02}_{-0.02}$ &$-0.11^{+0.26}_{-0.16}$ & $0.15^{+0.11}_{-0.18}$
&$T_3^{D_{s}\to K^*}$ &$0.23_{-0.03}^{+0.02}$&$-0.02^{+0.33}_{-0.12}$ &$0.15^{+0.04}_{-0.15}$
\\\hline
$T_1^{D\to K^*}$ &$0.71_{-0.09}^{+0.08}$&$0.07^{+0.20}_{-0.15}$ &$0.11^{+0.09}_{-0.12}$
&$T_1^{D_{s}\to \phi}$ &$0.69_{-0.10}^{+0.08}$&$0.11^{+0.17}_{-0.13}$ &$0.13^{+0.08}_{-0.08}$\\
$T_2^{D\to K^*}$ &$0.71_{-0.09}^{+0.08}$&$-0.74^{+0.06}_{-0.06}$ &$0.55^{+0.16}_{-0.16}$
&$T_2^{D_{s}\to \phi}$ &$0.69_{-0.10}^{+0.08}$ &$-0.69^{+0.07}_{-0.07}$ &$0.48^{+0.02}_{-0.02}$\\
$T_3^{D\to K^*}$ &$0.28_{-0.06}^{+0.03}$&$-0.08^{+0.19}_{-0.18}$ &$0.15^{+0.11}_{-0.10}$
&$T_3^{D_{s}\to \phi}$ &$0.23_{-0.07}^{+0.03}$ &$-0.03^{+0.21}_{-0.20}$ &$0.17^{+0.10}_{-0.09}$
\\\hline
$T_1^{B\to \rho}$ &$0.27_{-0.04}^{+0.05}$ &$0.55^{+0.14}_{-0.09}$ &$0.06^{+0.05}_{-0.06}$
&$T_1^{B_{s}\to K^{*}}$ &$0.19_{-0.05}^{+0.06}$ &$0.90^{+0.11}_{-0.09}$ &$0.33^{+0.07}_{-0.08}$\\
$T_2^{B\to \rho}$ &$0.27_{-0.04}^{+0.05}$ &$-0.29^{+0.01}_{-0.01}$ &$0.19^{+0.02}_{-0.03}$
&$T_2^{B_{s}\to K^{*}}$ &$0.19_{-0.05}^{+0.06}$ &$0.10^{+0.06}_{-0.06}$ &$0.18^{+0.03}_{-0.04}$\\
$T_3^{B\to \rho}$ &$0.18_{-0.03}^{+0.03}$ &$0.35^{+0.11}_{-0.07}$ &$0.09^{+0.03}_{-0.06}$
&$T_3^{B_{s}\to K^{*}}$ &$0.12_{-0.03}^{+0.03}$ &$0.69^{+0.08}_{-0.07}$ &$0.28^{+0.07}_{-0.07}$\\
\hline
$T_1^{B_{c}\to D^{*}}$ &$0.11_{-0.04}^{+0.06}$ &$1.68^{+0.12}_{-0.12}$ &$1.85^{+0.57}_{-0.57}$
&$T_1^{B\to K^{*}}$ &$0.32_{-0.06}^{+0.06}$ &$0.56^{+0.12}_{-0.09}$ &$0.06^{+0.05}_{-0.06}$\\
$T_2^{B_{c}\to D^{*}}$ &$0.11_{-0.04}^{+0.06}$ &$1.03^{+0.19}_{-0.23}$ &$0.94^{+0.43}_{-0.38}$
&$T_2^{B\to K^{*}}$ &$0.32_{-0.06}^{+0.06}$ &$-0.24^{+0.01}_{-0.01}$ &$0.16^{+0.02}_{-0.03}$\\
$T_3^{B_{c}\to D^{*}}$ &$0.05_{-0.02}^{+0.03}$ &$1.46^{+0.27}_{-0.17}$ &$1.51^{+0.35}_{-0.52}$
&$T_3^{B\to K^{*}}$ &$0.20_{-0.03}^{+0.03}$ &$0.40^{+0.10}_{-0.07}$ &$0.08^{+0.04}_{-0.06}$
\\\hline
$T_1^{B_{s}\to\phi}$ &$0.27_{-0.06}^{+0.07}$ &$0.82^{+0.09}_{-0.07}$ &$0.26^{+0.06}_{-0.07}$
&$T_1^{B_{c}\to D^{*}_{s}}$ &$0.20_{-0.07}^{+0.09}$ &$1.34^{+0.11}_{-0.11}$ &$1.06^{+0.34}_{-0.34}$\\
$T_2^{B_{s}\to\phi}$ &$0.27_{-0.06}^{+0.07}$ &$0.05^{+0.04}_{-0.04}$ &$0.16^{+0.03}_{-0.03}$
&$T_2^{B_{c}\to D^{*}_{s}}$ &$0.20_{-0.07}^{+0.09}$ &$0.63^{+0.17}_{-0.20}$ &$0.49^{+0.22}_{-0.18}$\\
$T_3^{B_{s}\to\phi}$ &$0.16_{-0.03}^{+0.04}$ &$0.64^{+0.07}_{-0.06}$ &$0.23^{+0.05}_{-0.06}$
&$T_3^{B_{c}\to D^{*}_{s}}$ &$0.10_{-0.03}^{+0.04}$ &$1.16^{+0.10}_{-0.11}$ &$0.87^{+0.30}_{-0.29}$\\
\hline
$T_1^{B\to D^{*}}$ &$0.70_{-0.11}^{+0.10}$ &$0.55^{+0.04}_{-0.03}$ &$0.00^{+0.05}_{-0.04}$
&$T_1^{B_{s}\to D^{*}_{s}}$ &$0.70_{-0.12}^{+0.11}$ &$0.63^{+0.10}_{-0.10}$ &$0.10^{+0.06}_{-0.07}$\\
$T_2^{B\to D^{*}}$ &$0.70_{-0.11}^{+0.10}$ &$-0.28^{+0.01}_{-0.02}$ &$0.19^{+0.03}_{-0.02}$
&$T_2^{B_{s}\to D^{*}_{s}}$ &$0.70_{-0.12}^{+0.11}$ &$-0.22^{+0.07}_{-0.07}$ &$0.24^{+0.02}_{-0.02}$\\
$T_3^{B\to D^{*}}$ &$0.30_{-0.03}^{+0.01}$ &$0.42^{+0.08}_{-0.10}$ &$0.02^{+0.02}_{-0.02}$
&$T_3^{B_{s}\to D^{*}_{s}}$ &$0.30_{-0.02}^{+0.02}$ &$0.54^{+0.01}_{-0.01}$ &$0.06^{+0.01}_{-0.01}$
\\\hline
$T_1^{B_{c}\to J/ \Psi}$ &$0.56_{-0.17}^{+0.16}$ &$1.30^{+0.17}_{-0.23}$ &$0.80^{+0.31}_{-0.31}$\\
$T_2^{B_{c}\to J/ \Psi}$ &$0.56_{-0.17}^{+0.16}$ &$0.54^{+0.20}_{-0.29}$ &$0.34^{+0.16}_{-0.12}$\\
$T_3^{B_{c}\to J/ \Psi}$ &$0.19_{-0.03}^{+0.03}$ &$1.17^{+0.03}_{-0.02}$ &$0.68^{+0.17}_{-0.21}$\\
\hline\hline
\end{tabular}
\end{center}
\end{table}

\begin{table}[t]
\scriptsize
\begin{center}
\caption{\label{tab:PA1} \small Same as Table~\ref{tab:PP} except for $D_{q,s}$$\to$$^1\!A $  and $B_{q,s,c}$$\to$$^1\!A $  transitions.}
\vspace{0.2cm}
\let\oldarraystretch=\arraystretch
\renewcommand*{\arraystretch}{0.9}
\setlength{\tabcolsep}{3pt}
\begin{tabular}{lccc|lccccccc}
\hline\hline
 &$\mathcal{F}(0)$  &$a$   &$b$ & &$\mathcal{F}(0)$  &$a$   &$b$
 \\\hline
$T_1^{D\to {^1\!A}_{(q,\bar{q})}}$ &$0.23^{+0.02}_{-0.03}$ &$0.02^{+0.23}_{-0.14}$ &$0.18^{+0.10}_{-0.15}$
&$T_1^{D_{s}\to {^1\!A}_{(q,\bar{s})}}$ &$0.19^{+0.02}_{-0.03}$ &$0.13^{+0.20}_{-0.20}$ &$0.18^{+0.08}_{-0.14}$\\
$T_2^{D\to {^1\!A}_{(q,\bar{q})}}$ &$0.23^{+0.02}_{-0.03}$ &$-0.82^{+0.07}_{-0.13}$ &$0.72^{+0.09}_{-0.06}$
&$T_2^{D_{s}\to {^1\!A}_{(q,\bar{s})}}$ &$0.19^{+0.02}_{-0.03}$ &$-0.40^{+0.04}_{-0.04}$ &$0.35^{+0.14}_{-0.14}$\\
$T_3^{D\to {^1\!A}_{(q,\bar{q})}}$ &$-0.02^{+0.12}_{-0.09}$ &$3.76^{+0.25}_{-0.25}$ &$5.25^{+0.47}_{-0.47}$
&$T_3^{D_{s}\to {^1\!A}_{(q,\bar{s})}}$ &$-0.13^{+0.12}_{-0.06}$ &$1.16^{+0.07}_{-0.07}$ &$1.01^{+0.23}_{-0.23}$
\\\hline
$T_1^{D\to {^1\!A}_{(s,\bar{q})}}$ &$0.19^{+0.03}_{-0.04}$ &$0.04^{+0.21}_{-0.21}$ &$0.20^{+0.09}_{-0.09}$
&$T_1^{D_{s}\to {^1\!A}_{(s,\bar{s})}}$ &$0.16^{+0.03}_{-0.03}$ &$0.09^{+0.28}_{-0.22}$ &$0.20^{+0.09}_{-0.08}$\\
$T_2^{D\to {^1\!A}_{(s,\bar{q})}}$ &$0.19^{+0.03}_{-0.04}$ &$-1.32^{+0.26}_{-0.40}$ &$1.69^{+0.59}_{-0.59}$
&$T_2^{D_{s}\to {^1\!A}_{(s,\bar{s})}}$ &$0.16^{+0.03}_{-0.03}$ &$-0.83^{+0.04}_{-0.20}$ &$0.79^{+0.18}_{-0.06}$\\
$T_3^{D\to {^1\!A}_{(s,\bar{q})}}$ &$0.01^{+0.17}_{-0.11}$ &$-4.77^{+0.40}_{-0.40}$ &$-9.34^{+0.50}_{-0.49}$
&$T_3^{D_{s}\to {^1\!A}_{(s,\bar{s})}}$ &$-0.10^{+0.11}_{-0.11}$ &$1.10^{+0.38}_{-0.38}$ &$0.93^{+0.26}_{-0.26}$
\\\hline
$T_1^{B\to {^1\!A}_{(q,\bar{q})}}$ &$0.12^{+0.03}_{-0.03}$ &$0.72^{+0.11}_{-0.08}$ &$0.17^{+0.05}_{-0.05}$
&$T_1^{B_{s}\to {^1\!A}_{(q,\bar{s})}}$ &$0.08^{+0.03}_{-0.03}$ &$1.07^{+0.05}_{-0.05}$ &$0.53^{+0.11}_{-0.10}$\\
$T_2^{B\to {^1\!A}_{(q,\bar{q})}}$ &$0.12^{+0.03}_{-0.03}$ &$-0.15^{+0.06}_{-0.08}$ &$0.19^{+0.03}_{-0.03}$
&$T_2^{B_{s}\to {^1\!A}_{(q,\bar{s})}}$ &$0.08^{+0.03}_{-0.03}$ &$0.33^{+0.11}_{-0.13}$ &$0.20^{+0.07}_{-0.03}$\\
$T_3^{B\to {^1\!A}_{(q,\bar{q})}}$ &$-0.01^{+0.02}_{-0.03}$ &$3.37^{+0.14}_{-0.14}$ &$3.09^{+0.24}_{-0.24}$
&$T_3^{B_{s}\to {^1\!A}_{(q,\bar{s})}}$ &$-0.07^{+0.03}_{-0.03}$ &$1.87^{+0.27}_{-0.22}$ &$1.30^{+0.23}_{-0.23}$
\\\hline
$T_1^{B_{c}\to {^1\!A}_{(q,\bar{c})}}$ &$0.03^{+0.03}_{-0.01}$ &$1.78^{+0.16}_{-0.16}$ &$2.05^{+0.67}_{-0.67}$
&$T_1^{B\to {^1\!A}_{(s,\bar{q})}}$ &$0.13^{+0.03}_{-0.03}$ &$0.72^{+0.09}_{-0.06}$ &$0.18^{+0.05}_{-0.06}$\\
$T_2^{B_{c}\to {^1\!A}_{(q,\bar{c})}}$ &$0.03^{+0.03}_{-0.01}$ &$1.55^{+0.24}_{-0.28}$ &$1.51^{+0.69}_{-0.69}$
&$T_2^{B\to {^1\!A}_{(s,\bar{q})}}$ &$0.13^{+0.03}_{-0.03}$ &$-0.27^{+0.13}_{-0.13}$ &$0.28^{+0.06}_{-0.05}$\\
$T_3^{B_{c}\to {^1\!A}_{(q,\bar{c})}}$ &$-0.17^{+0.05}_{-0.03}$ &$2.41^{+0.13}_{-0.13}$ &$2.02^{+0.15}_{-0.16}$
&$T_3^{B\to {^1\!A}_{(s,\bar{q})}}$ &$-0.02^{+0.03}_{-0.04}$ &$2.88^{+0.51}_{-0.51}$ &$2.55^{+0.58}_{-0.58}$
\\\hline
$T_1^{B_{s}\to {^1\!A}_{(s,\bar{s})}}$ &$0.09^{+0.03}_{-0.02}$ &$0.97^{+0.05}_{-0.03}$ &$0.42^{+0.07}_{-0.07}$
&$T_1^{B_{c}\to {^1\!A}_{(s,\bar{c})}}$ &$0.06^{+0.03}_{-0.03}$ &$1.41^{+0.15}_{-0.14}$ &$1.19^{+0.49}_{-0.40}$\\
$T_2^{B_{s}\to {^1\!A}_{(s,\bar{s})}}$ &$0.09^{+0.03}_{-0.02}$ &$0.11^{+0.13}_{-0.16}$ &$0.22^{+0.02}_{-0.02}$
&$T_2^{B_{c}\to {^1\!A}_{(s,\bar{c})}}$ &$0.06^{+0.03}_{-0.03}$ &$1.11^{+0.24}_{-0.27}$ &$0.74^{+0.49}_{-0.38}$\\
$T_3^{B_{s}\to {^1\!A}_{(s,\bar{s})}}$ &$-0.10^{+0.04}_{-0.04}$ &$1.93^{+0.24}_{-0.18}$ &$1.43^{+0.28}_{-0.21}$
&$T_3^{B_{c}\to {^1\!A}_{(s,\bar{c})}}$ &$-0.28^{+0.07}_{-0.10}$ &$2.22^{+0.39}_{-0.37}$ &$1.82^{+0.54}_{-0.47}$
\\\hline
$T_1^{B\to {^1\!A}_{(c,\bar{q})}}$ &$0.15^{+0.02}_{-0.02}$ &$0.64^{+0.01}_{-0.01}$ &$0.11^{+0.04}_{-0.03}$
&$T_1^{B_{s}\to {^1\!A}_{(c,\bar{s})}}$ &$0.12^{+0.02}_{-0.01}$ &$0.73^{+0.17}_{-0.13}$ &$0.22^{+0.04}_{-0.05}$\\
$T_2^{B\to {^1\!A}_{(c,\bar{q})}}$ &$0.15^{+0.02}_{-0.02}$ &$-1.89^{+0.44}_{-0.65}$ &$2.90^{+0.94}_{-0.94}$
&$T_2^{B_{s}\to {^1\!A}_{(c,\bar{s})}}$ &$0.12^{+0.02}_{-0.01}$ &$-1.57^{+0.05}_{-0.05}$ &$2.31^{+0.06}_{-0.06}$\\
$T_3^{B\to {^1\!A}_{(c,\bar{q})}}$ &$-0.07^{+0.10}_{-0.06}$ &$3.56^{+0.92}_{-0.92}$ &$4.69^{+0.89}_{-0.89}$
&$T_3^{B_{s}\to {^1\!A}_{(c,\bar{s})}}$ &$-0.20^{+0.10}_{-0.06}$ &$2.42^{+0.20}_{-0.20}$ &$2.82^{+0.39}_{-0.39}$
\\\hline
$T_1^{B_{c}\to {^1\!A}_{(c,\bar{c})}}$ &$0.08^{+0.02}_{-0.02}$ &$1.34^{+0.05}_{-0.04}$ &$0.88^{+0.21}_{-0.21}$\\
$T_2^{B_{c}\to {^1\!A}_{(c,\bar{c})}}$ &$0.08^{+0.02}_{-0.02}$ &$0.23^{+0.20}_{-0.21}$ &$0.47^{+0.04}_{-0.04}$\\
$T_3^{B_{c}\to {^1\!A}_{(c,\bar{c})}}$ &$-0.48^{+0.12}_{-0.11}$ &$2.45^{+0.60}_{-0.60}$ &$2.90^{+0.52}_{-0.52}$\\
\hline\hline
\end{tabular}
\end{center}
\end{table}

\begin{table}[t]
\scriptsize
\begin{center}
\caption{\label{tab:PA2} \small Same as Table~\ref{tab:PP} except for $D_{q,s}$$\to$$^3\!A $ and $B_{q,s,c}$$\to$$^3\!A $ transitions.}
\vspace{0.2cm}
\let\oldarraystretch=\arraystretch
\renewcommand*{\arraystretch}{0.9}
\setlength{\tabcolsep}{3.0pt}
\begin{tabular}{lccc|cccccccc}
\hline\hline
 &$\mathcal{F}(0)$  &$a$   &$b$ & &$\mathcal{F}(0)$  &$a$   &$b$
 \\\hline
$T_1^{D\to {^3\!A}_{(q,\bar{q})}}$ &$0.49^{+0.05}_{-0.04}$ &$-0.09^{+0.29}_{-0.21}$ &$0.17^{+0.14}_{-0.18}$
&$T_1^{D_{s}\to {^3\!A}_{(q,\bar{s})}}$ &$0.49^{+0.05}_{-0.05}$ &$-0.04^{+0.32}_{-0.22}$ &$0.19^{+0.12}_{-0.17}$\\
$T_2^{D\to {^3\!A}_{(q,\bar{q})}}$ &$0.49^{+0.05}_{-0.04}$ &$-2.06^{+0.66}_{-0.66}$ &$3.24^{+0.41}_{-0.41}$
&$T_2^{D_{s}\to {^3\!A}_{(q,\bar{s})}}$ &$0.49^{+0.05}_{-0.05}$  &$-1.93^{+0.43}_{-0.66}$ &$3.06^{+0.53}_{-0.53}$\\
$T_3^{D\to {^3\!A}_{(q,\bar{q})}}$ &$0.50^{+0.12}_{-0.10}$ &$-0.21^{+0.29}_{-0.21}$ &$0.20^{+0.19}_{-0.18}$
&$T_3^{D_{s}\to {^3\!A}_{(q,\bar{s})}}$ &$0.54^{+0.14}_{-0.11}$ &$-0.14^{+0.34}_{-0.24}$ &$0.22^{+0.19}_{-0.19}$\\\hline
$T_1^{D\to {^3\!A}_{(s,\bar{q})}}$ &$0.43^{+0.05}_{-0.07}$ &$-0.09^{+0.14}_{-0.13}$ &$0.18^{+0.11}_{-0.11}$
&$T_1^{D_{s}\to {^3\!A}_{(s,\bar{s})}}$ &$0.47^{+0.04}_{-0.06}$ &$-0.07^{+0.11}_{-0.10}$ &$0.22^{+0.08}_{-0.10}$\\
$T_2^{D\to {^3\!A}_{(s,\bar{q})}}$ &$0.43^{+0.05}_{-0.07}$  &$-3.05^{+0.72}_{-0.72}$ &$7.09^{+0.51}_{-0.51}$
&$T_2^{D_{s}\to {^3\!A}_{(s,\bar{s})}}$ &$0.47^{+0.04}_{-0.06}$ &$-3.10^{+0.79}_{-0.79}$ &$7.20^{+0.14}_{-0.14}$\\
$T_3^{D\to {^3\!A}_{(s,\bar{q})}}$ &$0.51^{+0.14}_{-0.13}$ &$-0.26^{+0.39}_{-0.27}$ &$0.28^{+0.22}_{-0.29}$
&$T_3^{D_{s}\to {^3\!A}_{(s,\bar{s})}}$ &$0.66^{+0.16}_{-0.13}$ &$-0.14^{+0.28}_{-0.22}$ &$0.19^{+0.20}_{-0.11}$\\\hline
$T_1^{B\to {^3\!A}_{(q,\bar{q})}}$ &$0.29^{+0.03}_{-0.03}$ &$0.35^{+0.21}_{-0.20}$ &$0.03^{+0.03}_{-0.02}$
&$T_1^{B_{s}\to {^3\!A}_{(q,\bar{s})}}$ &$0.26^{+0.04}_{-0.06}$ &$0.65^{+0.06}_{-0.04}$ &$0.21^{+0.07}_{-0.09}$\\
$T_2^{B\to {^3\!A}_{(q,\bar{q})}}$ &$0.29^{+0.03}_{-0.03}$ &$-0.71^{+0.01}_{-0.01}$ &$0.46^{+0.05}_{-0.02}$
&$T_2^{B_{s}\to {^3\!A}_{(q,\bar{s})}}$ &$0.26^{+0.04}_{-0.06}$ &$-0.38^{+0.14}_{-0.16}$ &$0.37^{+0.07}_{-0.05}$\\
$T_3^{B\to {^3\!A}_{(q,\bar{q})}}$ &$0.25^{+0.04}_{-0.03}$ &$0.12^{+0.23}_{-0.18}$ &$0.12^{+0.03}_{-0.02}$
&$T_3^{B_{s}\to {^3\!A}_{(q,\bar{s})}}$ &$0.25^{+0.04}_{-0.05}$ &$0.49^{+0.01}_{-0.01}$ &$0.23^{+0.06}_{-0.07}$\\\hline
$T_1^{B_{c}\to {^3\!A}_{(q,\bar{c})}}$ &$0.18^{+0.09}_{-0.07}$ &$1.43^{+0.19}_{-0.21}$ &$1.28^{+0.51}_{-0.51}$
&$T_1^{B\to {^3\!A}_{(s,\bar{q})}}$ &$0.29^{+0.04}_{-0.03}$ &$0.34^{+0.23}_{-0.19}$ &$0.04^{+0.02}_{-0.02}$\\
$T_2^{B_{c}\to {^3\!A}_{(q,\bar{c})}}$ &$0.18^{+0.09}_{-0.07}$ &$0.46^{+0.39}_{-0.44}$ &$0.71^{+0.13}_{-0.13}$
&$T_2^{B\to {^3\!A}_{(s,\bar{q})}}$ &$0.29^{+0.04}_{-0.03}$ &$-0.85^{+0.02}_{-0.04}$ &$0.61^{+0.10}_{-0.07}$\\
$T_3^{B_{c}\to {^3\!A}_{(q,\bar{c})}}$ &$0.23^{+0.09}_{-0.08}$ &$1.49^{+0.18}_{-0.23}$ &$1.34^{+0.51}_{-0.51}$
&$T_3^{B\to {^3\!A}_{(s,\bar{q})}}$ &$0.26^{+0.05}_{-0.04}$ &$0.03^{+0.21}_{-0.16}$ &$0.16^{+0.03}_{-0.02}$
\\\hline
$T_1^{B_{s}\to {^3\!A}_{(s,\bar{s})}}$ &$0.31^{+0.04}_{-0.04}$ &$0.57^{+0.07}_{-0.04}$ &$0.16^{+0.06}_{-0.07}$
&$T_1^{B_{c}\to {^3\!A}_{(s,\bar{c})}}$ &$0.30^{+0.09}_{-0.09}$ &$1.09^{+0.18}_{-0.17}$ &$0.75^{+0.40}_{-0.32}$\\
$T_2^{B_{s}\to {^3\!A}_{(s,\bar{s})}}$ &$0.31^{+0.04}_{-0.04}$ &$-0.61^{+0.16}_{-0.21}$ &$0.53^{+0.16}_{-0.10}$
&$T_2^{B_{c}\to {^3\!A}_{(s,\bar{c})}}$ &$0.30^{+0.09}_{-0.09}$ &$-0.12^{+0.41}_{-0.45}$ &$0.62^{+0.14}_{-0.14}$\\
$T_3^{B_{s}\to {^3\!A}_{(s,\bar{s})}}$ &$0.31^{+0.03}_{-0.04}$ &$0.38^{+0.20}_{-0.16}$ &$0.20^{+0.02}_{-0.02}$
&$T_3^{B_{c}\to {^3\!A}_{(s,\bar{c})}}$ &$0.43^{+0.11}_{-0.12}$ &$1.16^{+0.17}_{-0.19}$ &$0.78^{+0.39}_{-0.32}$\\\hline
$T_1^{B\to {^3\!A}_{(c,\bar{q})}}$ &$0.34^{+0.05}_{-0.06}$ &$0.39^{+0.03}_{-0.03}$ &$-0.03^{+0.04}_{-0.04}$
&$T_1^{B_{s}\to {^3\!A}_{(c,\bar{s})}}$ &$0.43^{+0.04}_{-0.07}$ &$0.45^{+0.02}_{-0.02}$ &$0.05^{+0.07}_{-0.07}$\\
$T_2^{B\to {^3\!A}_{(c,\bar{q})}}$ &$0.34^{+0.05}_{-0.06}$ &$-2.73^{+0.55}_{-0.55}$ &$4.69^{+0.29}_{-0.29}$
&$T_2^{B_{s}\to {^3\!A}_{(c,\bar{s})}}$ &$0.43^{+0.04}_{-0.07}$ &$-2.73^{+0.55}_{-0.55}$ &$4.72^{+0.34}_{-0.34}$\\
$T_3^{B\to {^3\!A}_{(c,\bar{q})}}$ &$0.44^{+0.12}_{-0.04}$ &$0.02^{+0.21}_{-0.21}$ &$0.14^{+0.22}_{-0.22}$
&$T_3^{B_{s}\to {^3\!A}_{(c,\bar{s})}}$ &$0.67^{+0.17}_{-0.14}$ &$0.28^{+0.17}_{-0.15}$ &$0.09^{+0.04}_{-0.02}$
\\\hline
$T_1^{B_{c}\to {^3\!A}_{(c,\bar{c})}}$ &$0.50^{+0.01}_{-0.04}$ &$1.05^{+0.38}_{-0.38}$ &$0.56^{+0.30}_{-0.26}$\\
$T_2^{B_{c}\to {^3\!A}_{(c,\bar{c})}}$ &$0.50^{+0.01}_{-0.04}$  &$-2.29^{+0.14}_{-0.23}$ &$4.79^{+0.16}_{-0.23}$\\
$T_3^{B_{c}\to {^3\!A}_{(c,\bar{c})}}$ &$1.06^{+0.37}_{-0.25}$ &$1.06^{+0.44}_{-0.41}$ &$0.57^{+0.28}_{-0.27}$
\\\hline\hline
\end{tabular}
\end{center}
\end{table}

\begin{figure}[t]
  \centering
{\includegraphics[width=0.25\textwidth]{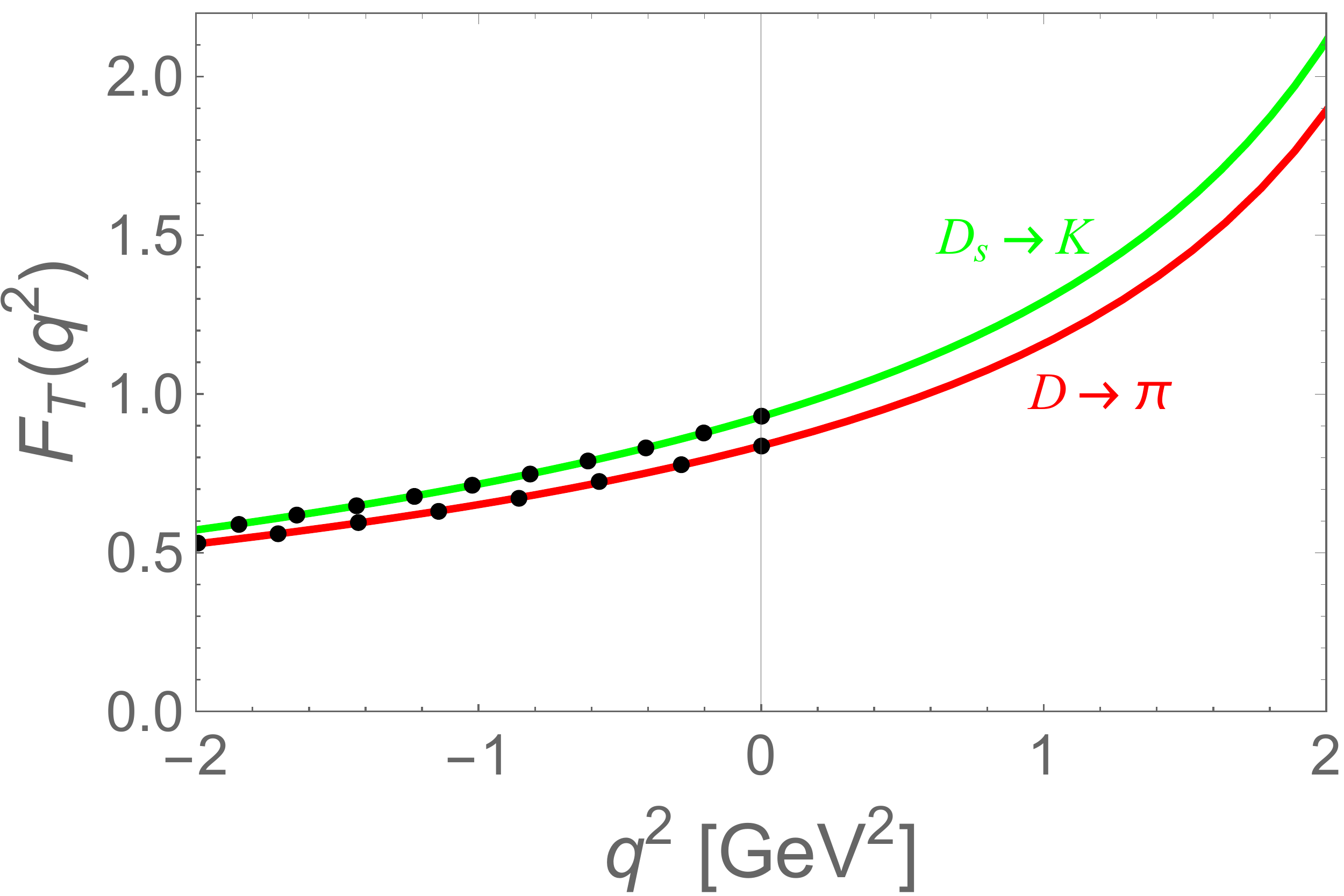}\quad}
{\includegraphics[width=0.25\textwidth]{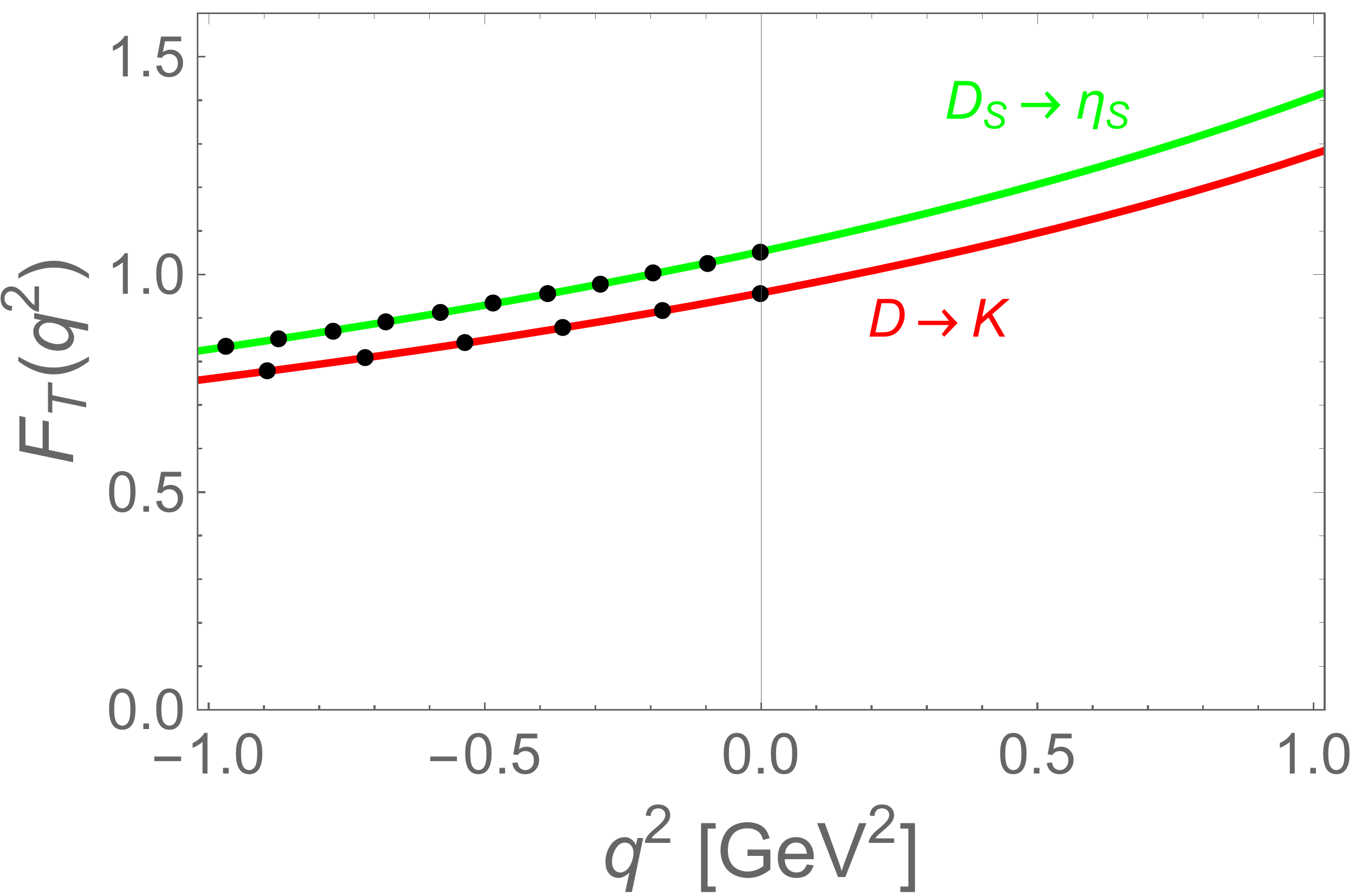}}\\
\vspace{0.3cm}
{\includegraphics[width=0.25\textwidth]{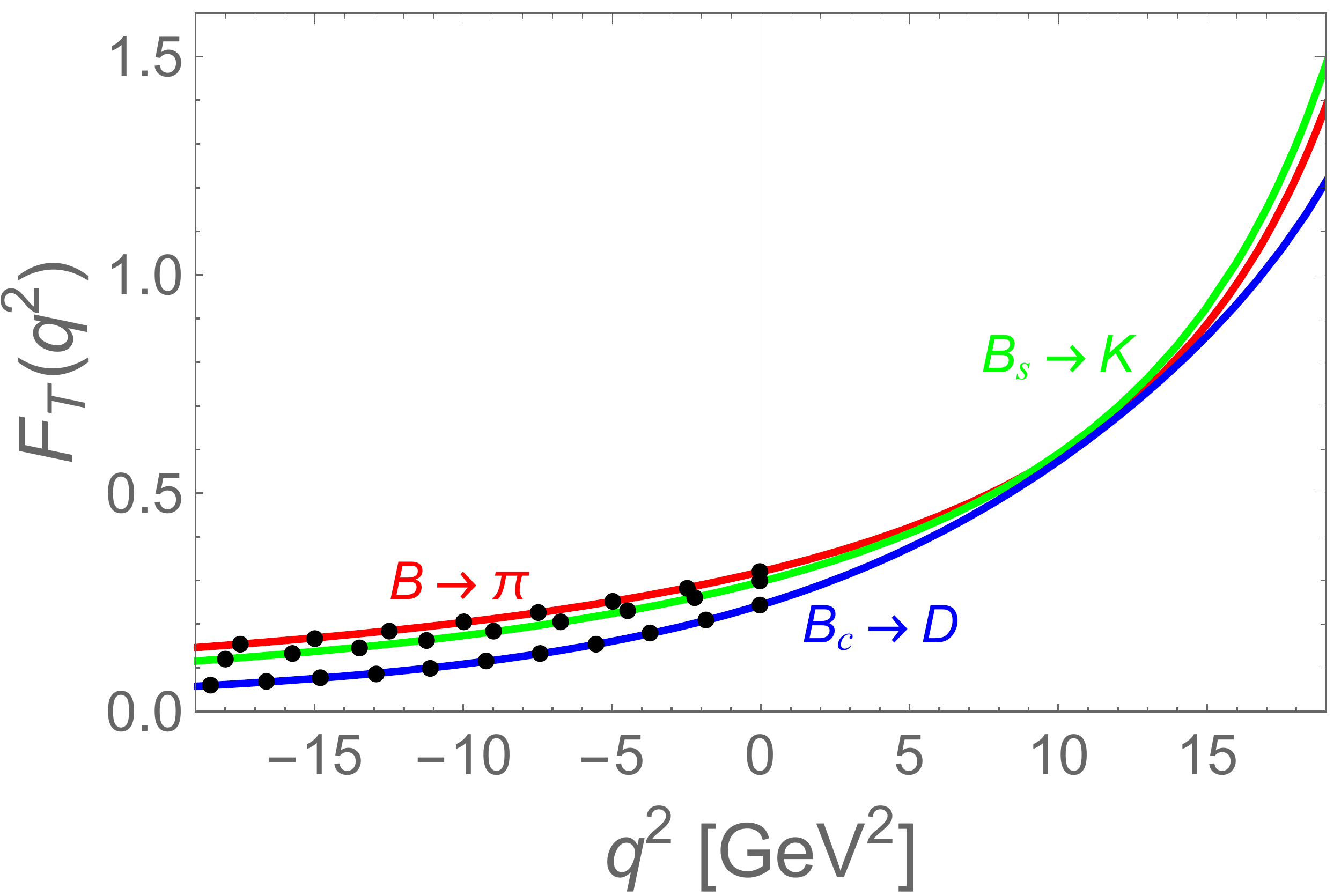}\quad}
{\includegraphics[width=0.25\textwidth]{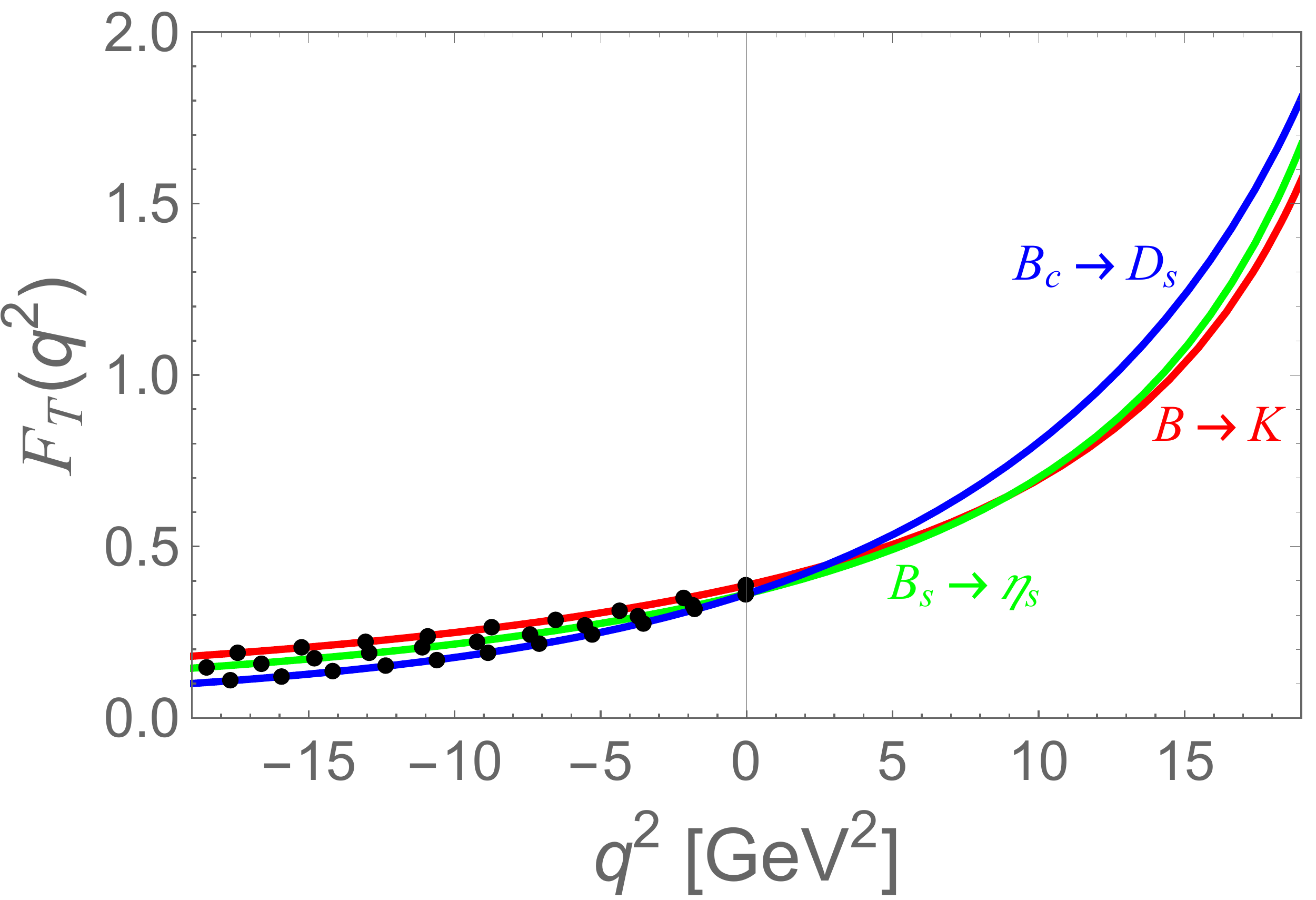}\quad}
{\includegraphics[width=0.25\textwidth]{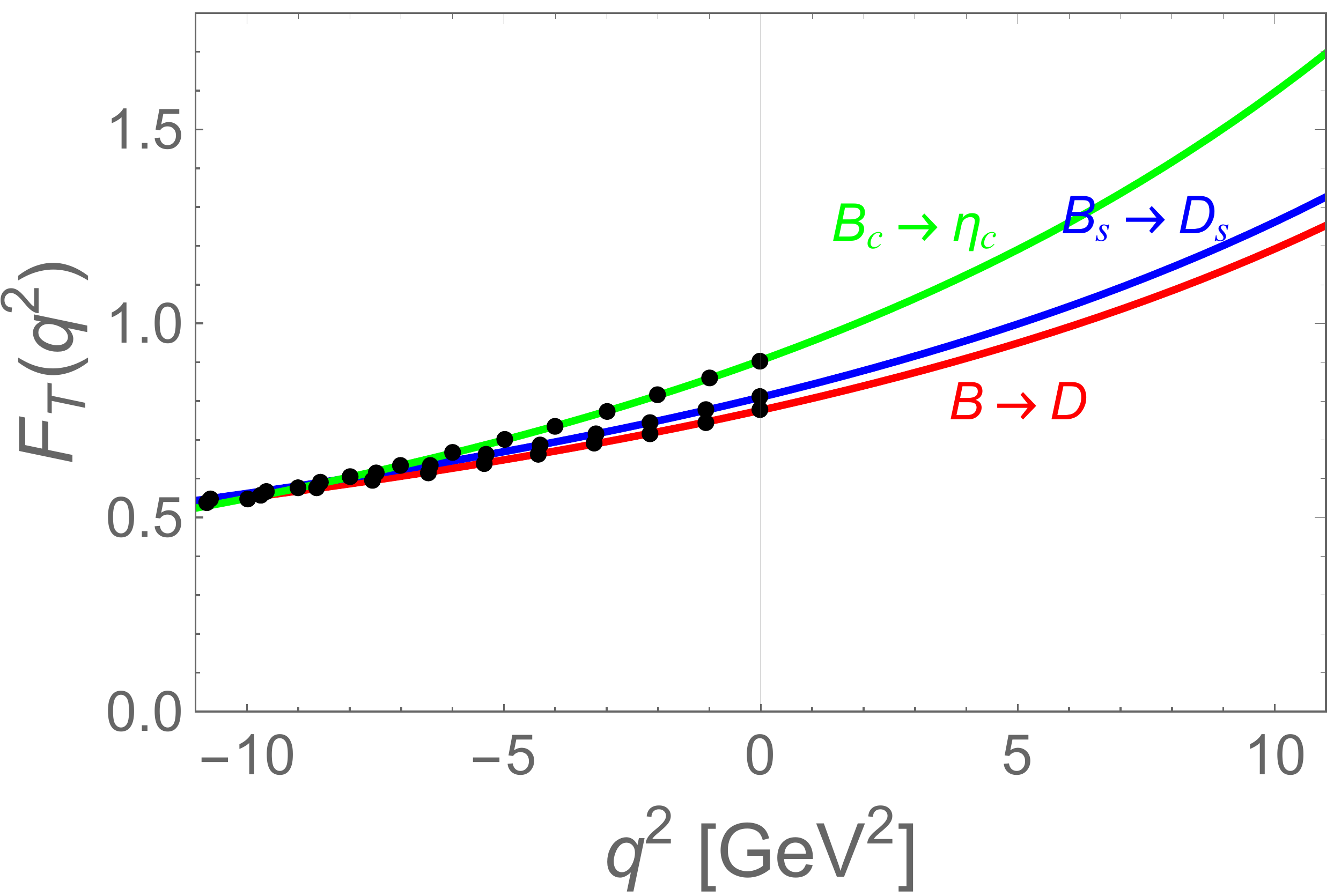}}
\renewcommand{\baselinestretch}{1}
\caption{\label{fig:PP}$q^2$ dependences of tensor form factors for $c\to q\,,s$~($q=u\,,d$) induced $D_{q,s}\to P $ and $b\to q\,,s\,,c$ induced $B_{q,s,c}\to P $ transitions. The dots in the space-like region are the results obtained directly via the CLF QM, and the lines are fitting results.}
\end{figure}

\begin{figure}[t]
  \centering
{\includegraphics[width=0.25\textwidth]{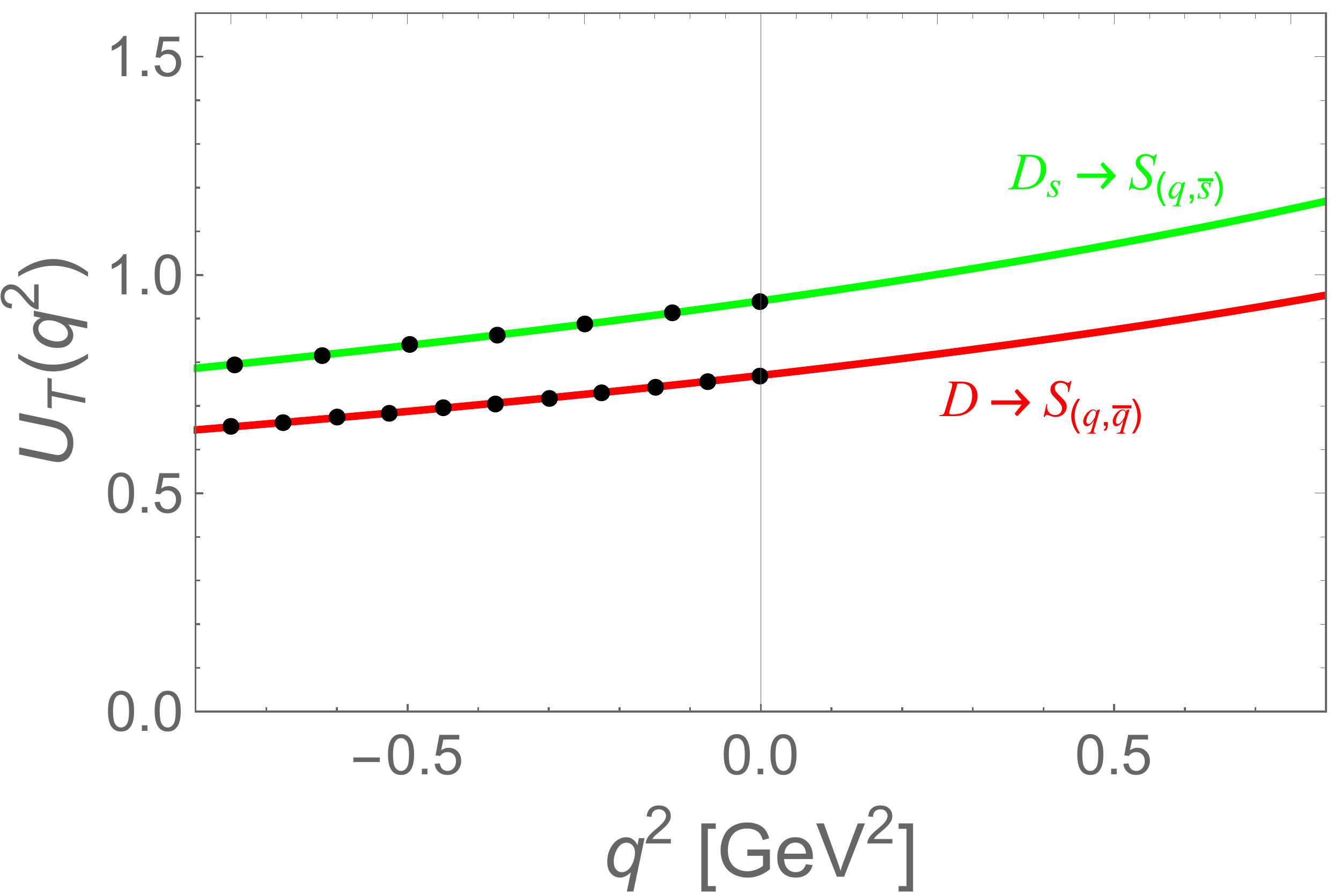}\quad}
{\includegraphics[width=0.25\textwidth]{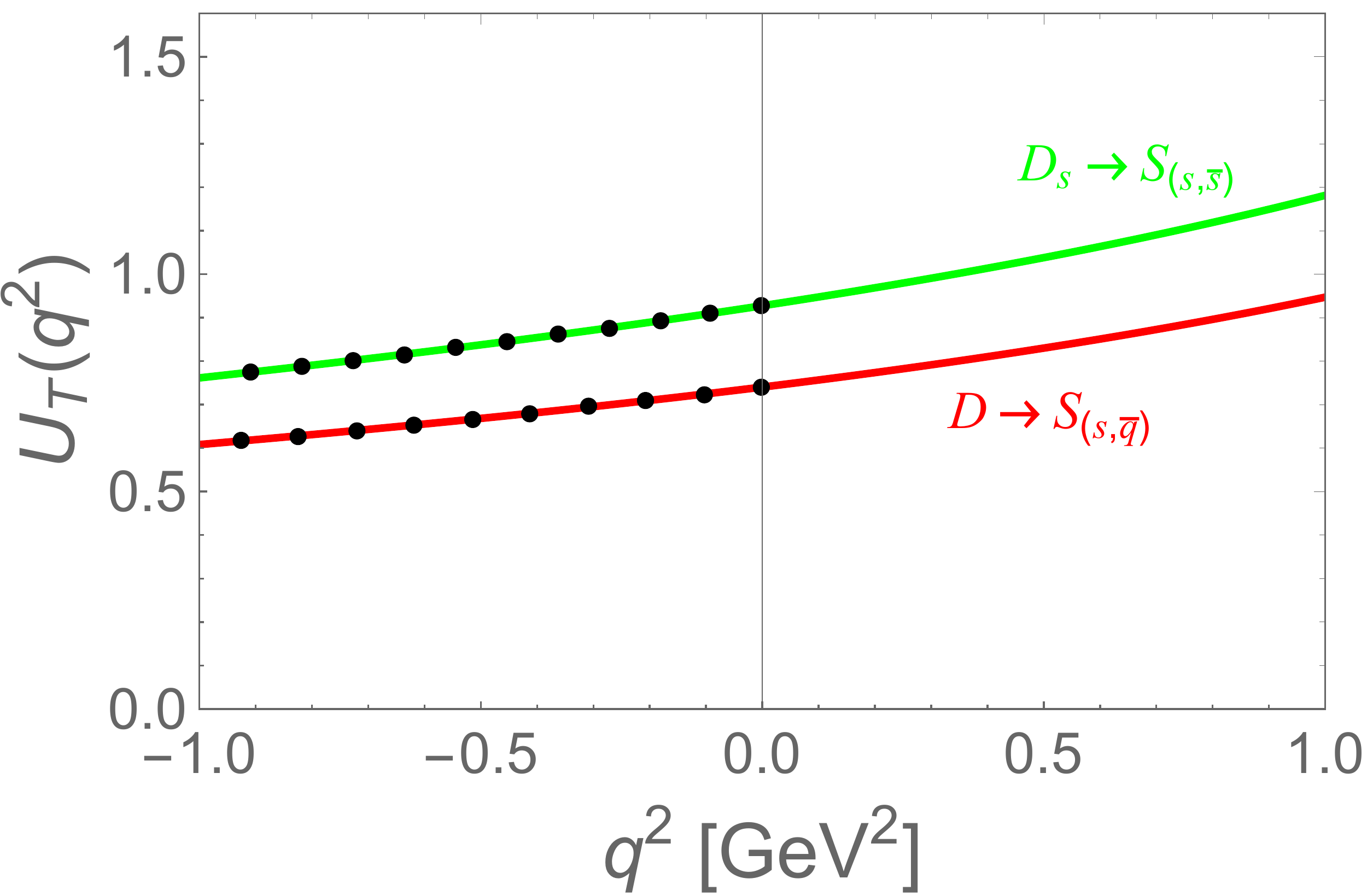}}\\
\vspace{0.3cm}
{\includegraphics[width=0.25\textwidth]{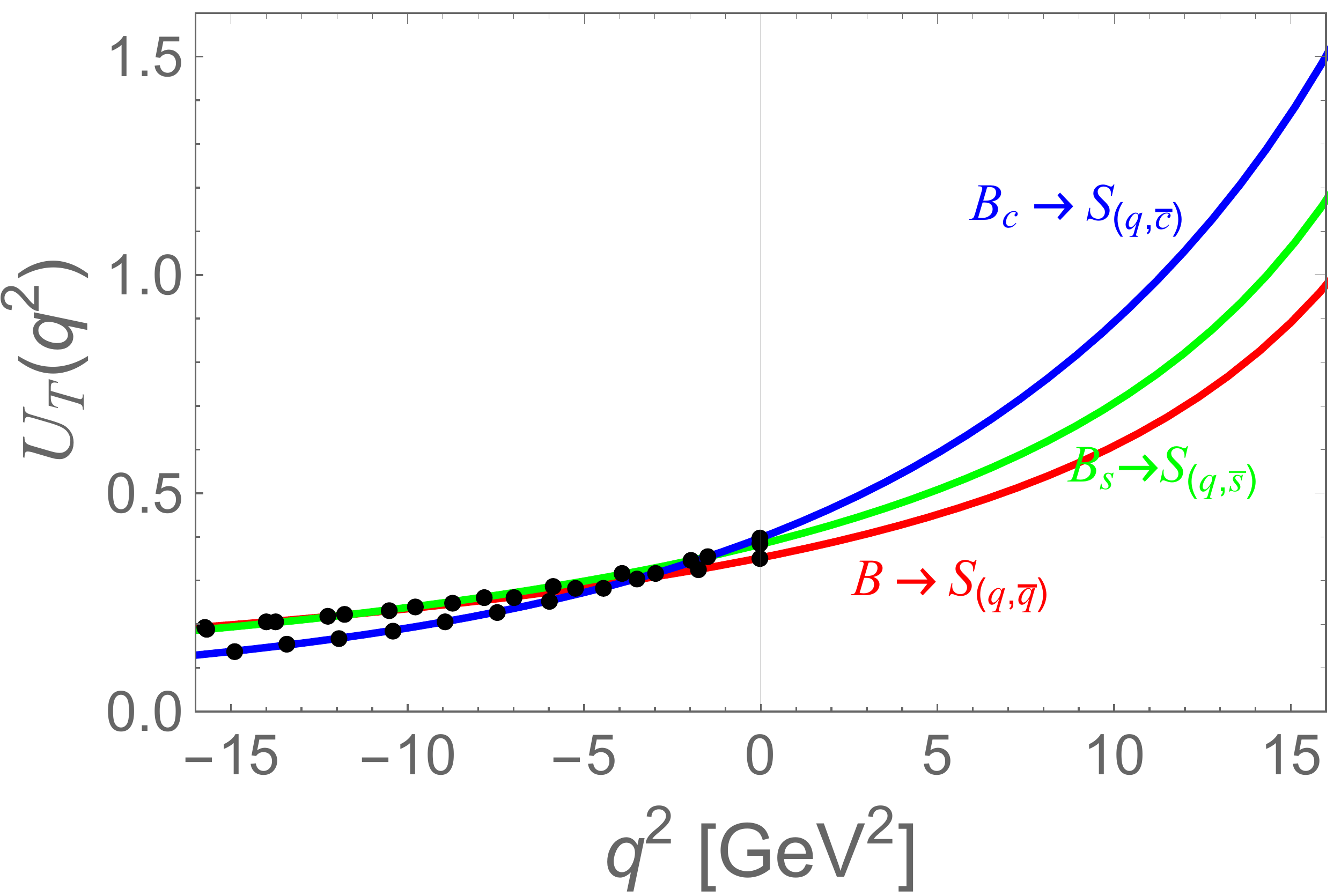}\quad}
{\includegraphics[width=0.25\textwidth]{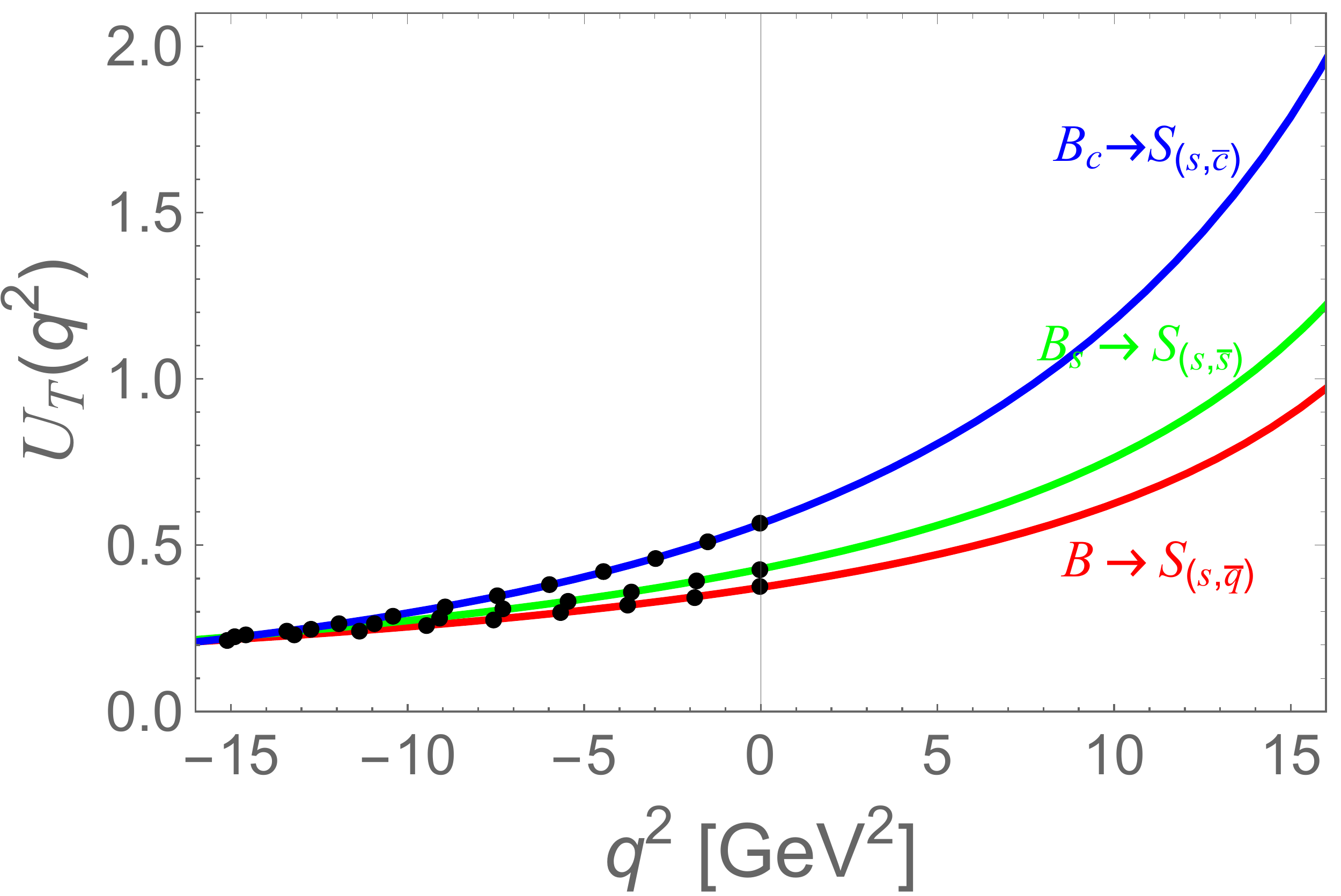}\quad}
{\includegraphics[width=0.25\textwidth]{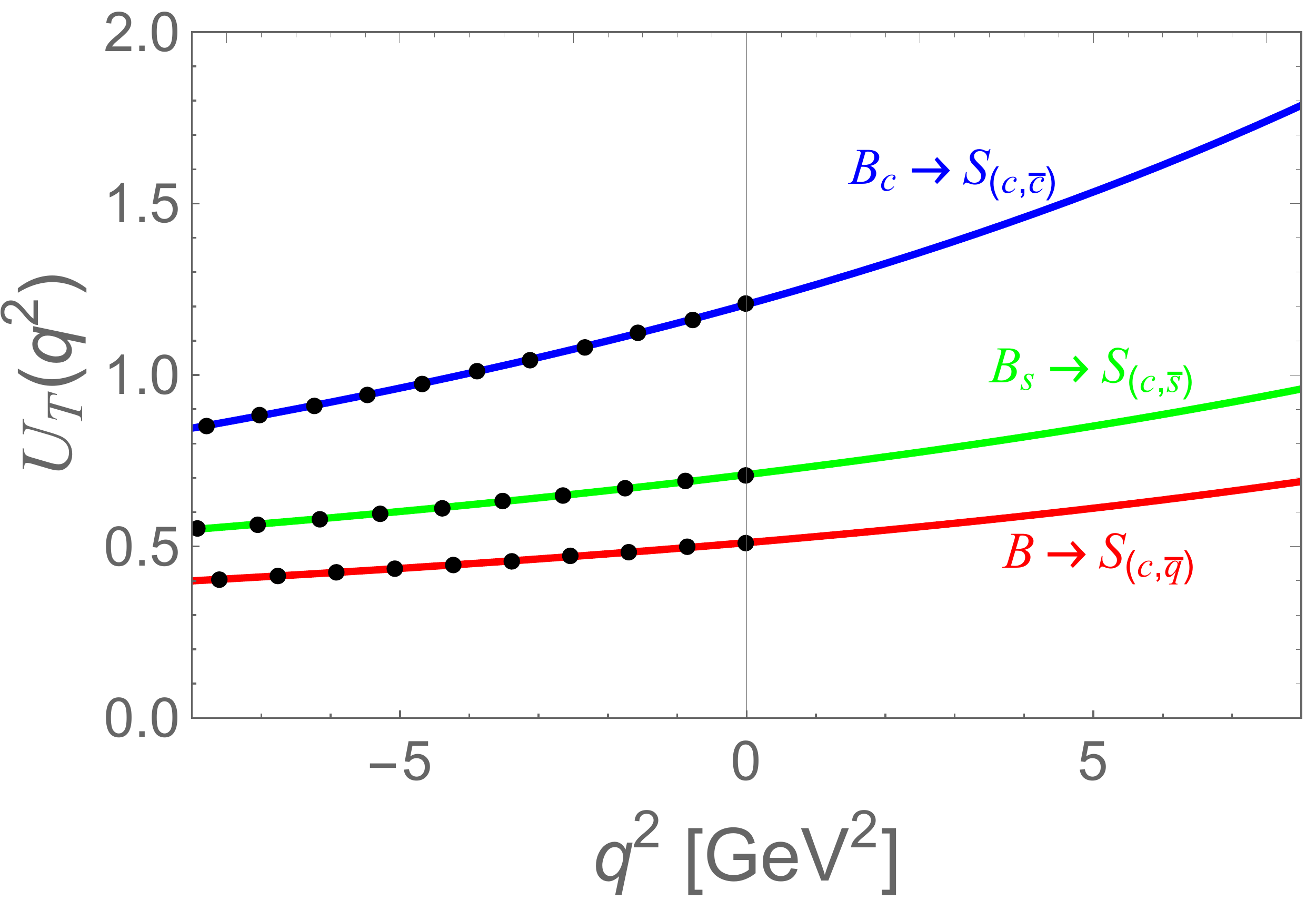}}
\caption{\label{fig:PS} Same as Fig. \ref{fig:PP} except for $D_{q,s}\to S $ and $B_{q,s,c}\to S $ transitions.}
\end{figure}

\begin{figure}[t]
  \centering
{\includegraphics[width=0.24\textwidth]{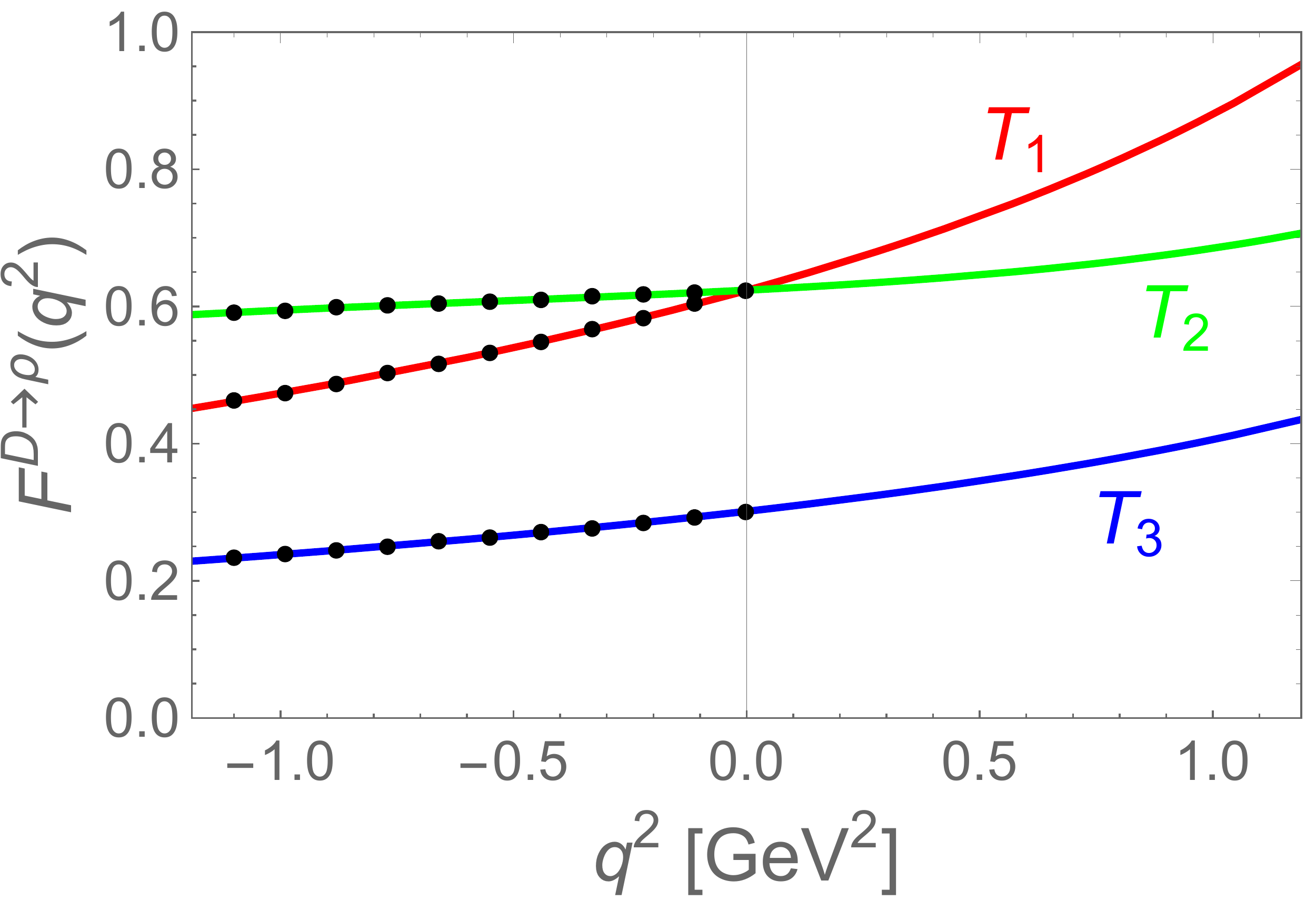}}
{\includegraphics[width=0.24\textwidth]{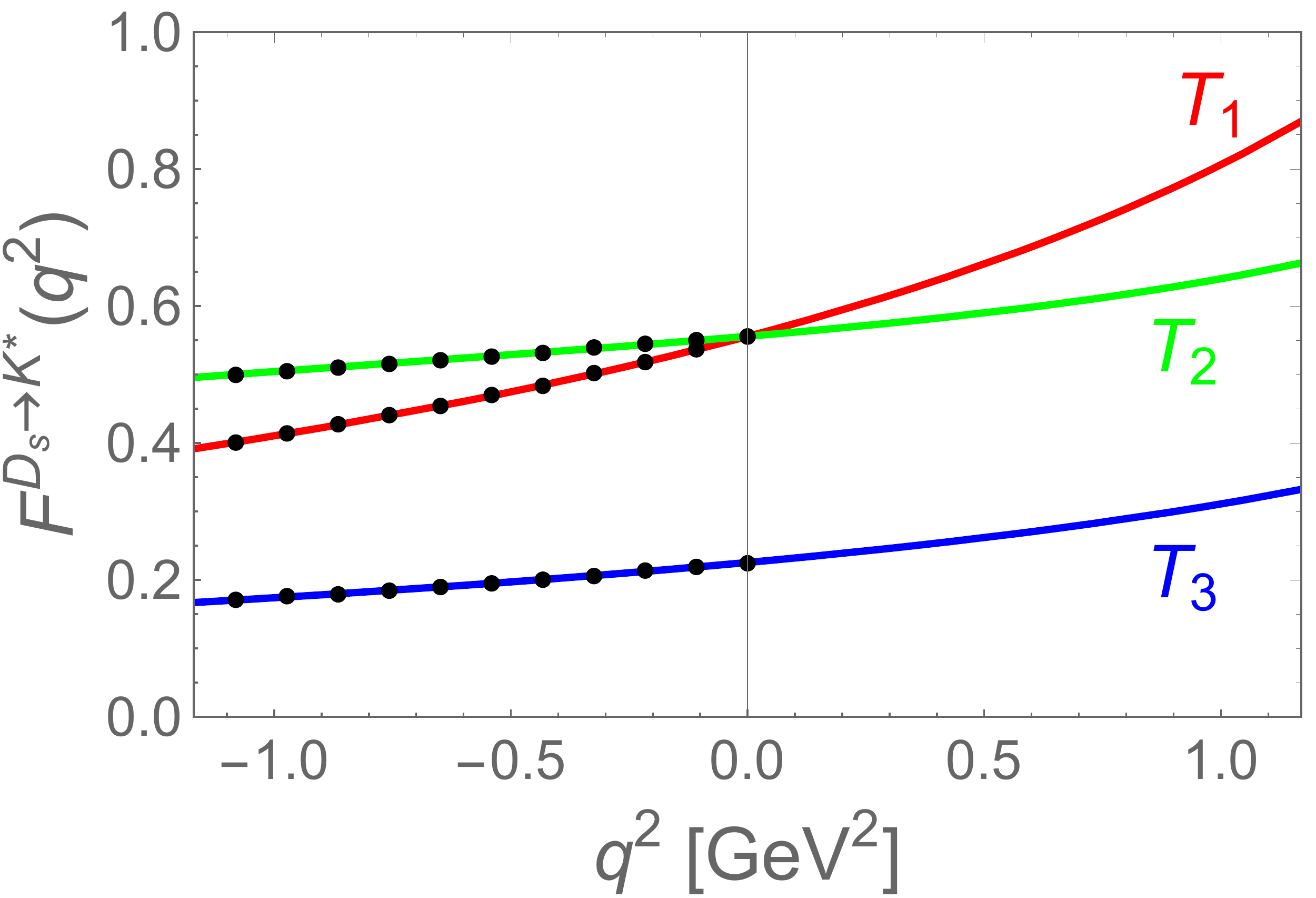}}
{\includegraphics[width=0.24\textwidth]{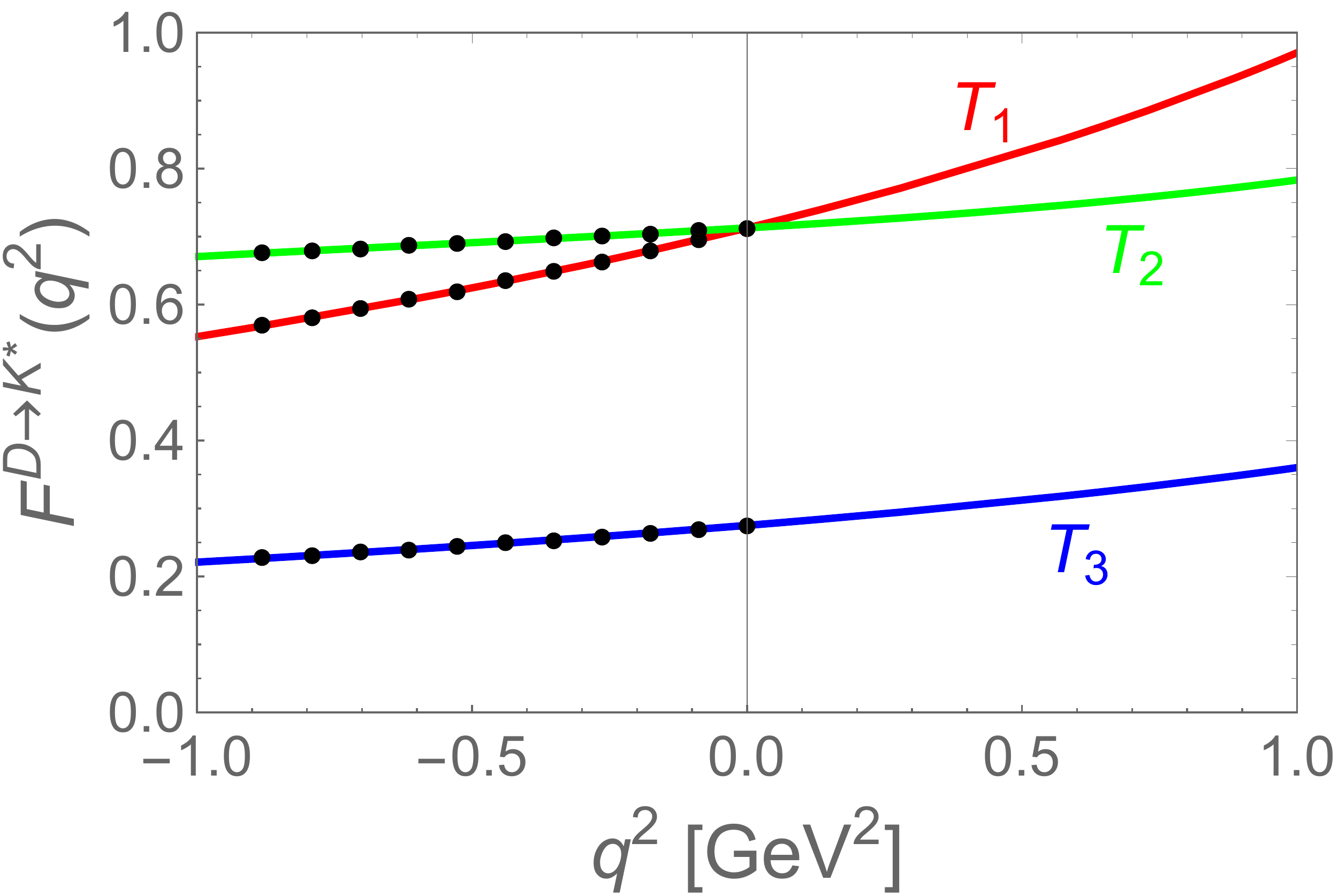}}
{\includegraphics[width=0.24\textwidth]{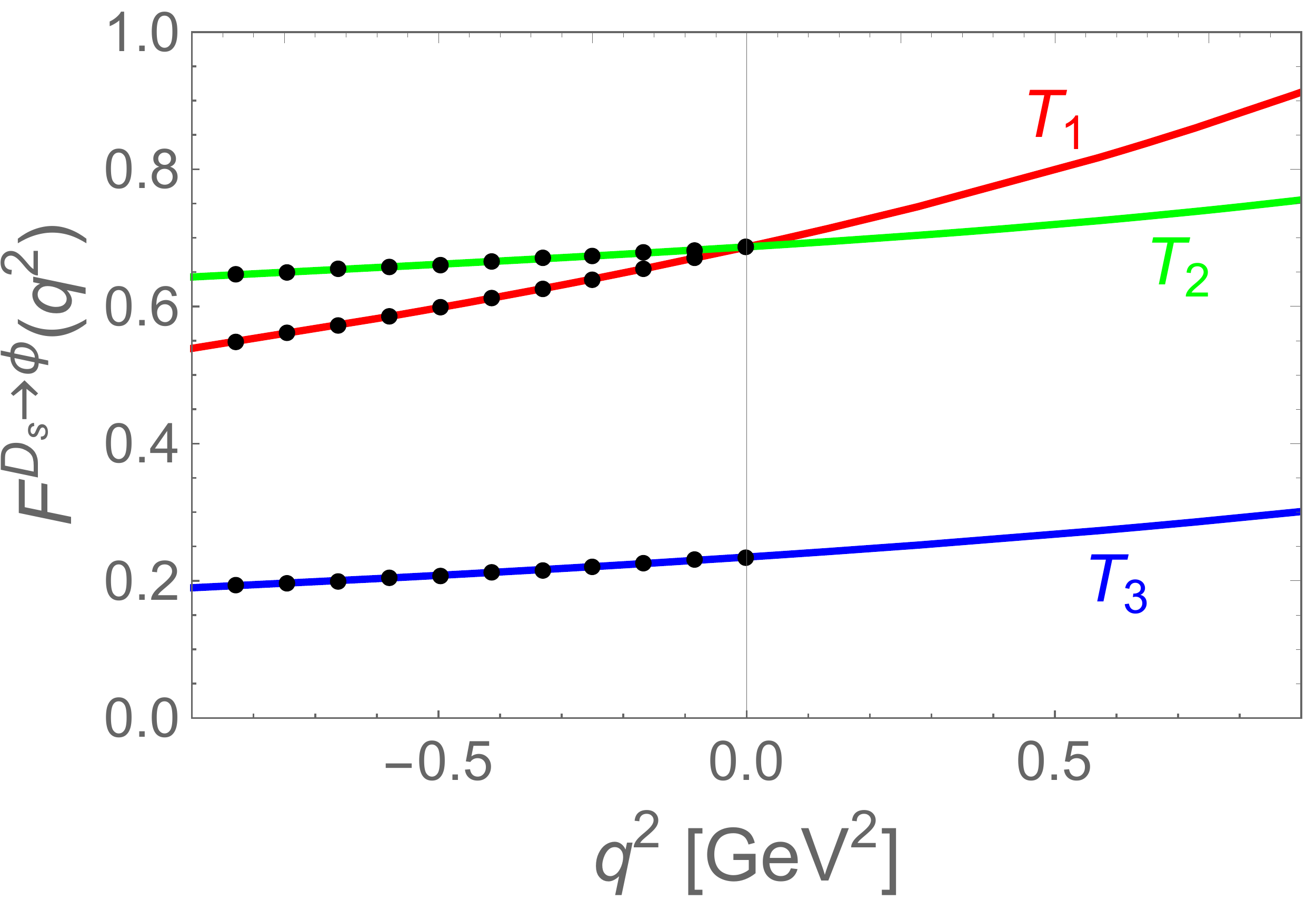}}\\
\vspace{0.3cm}
{\includegraphics[width=0.25\textwidth]{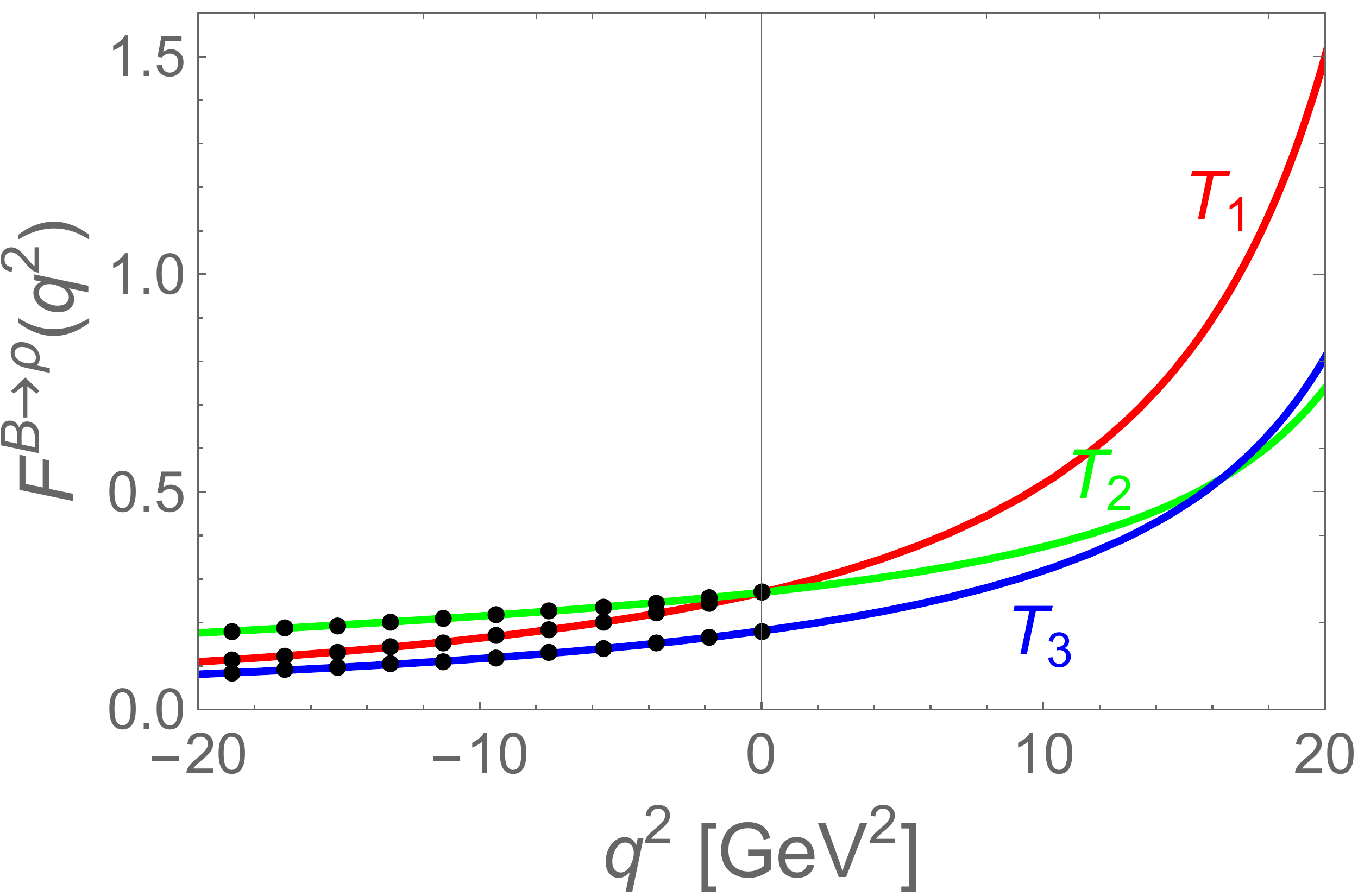}\quad}
{\includegraphics[width=0.25\textwidth]{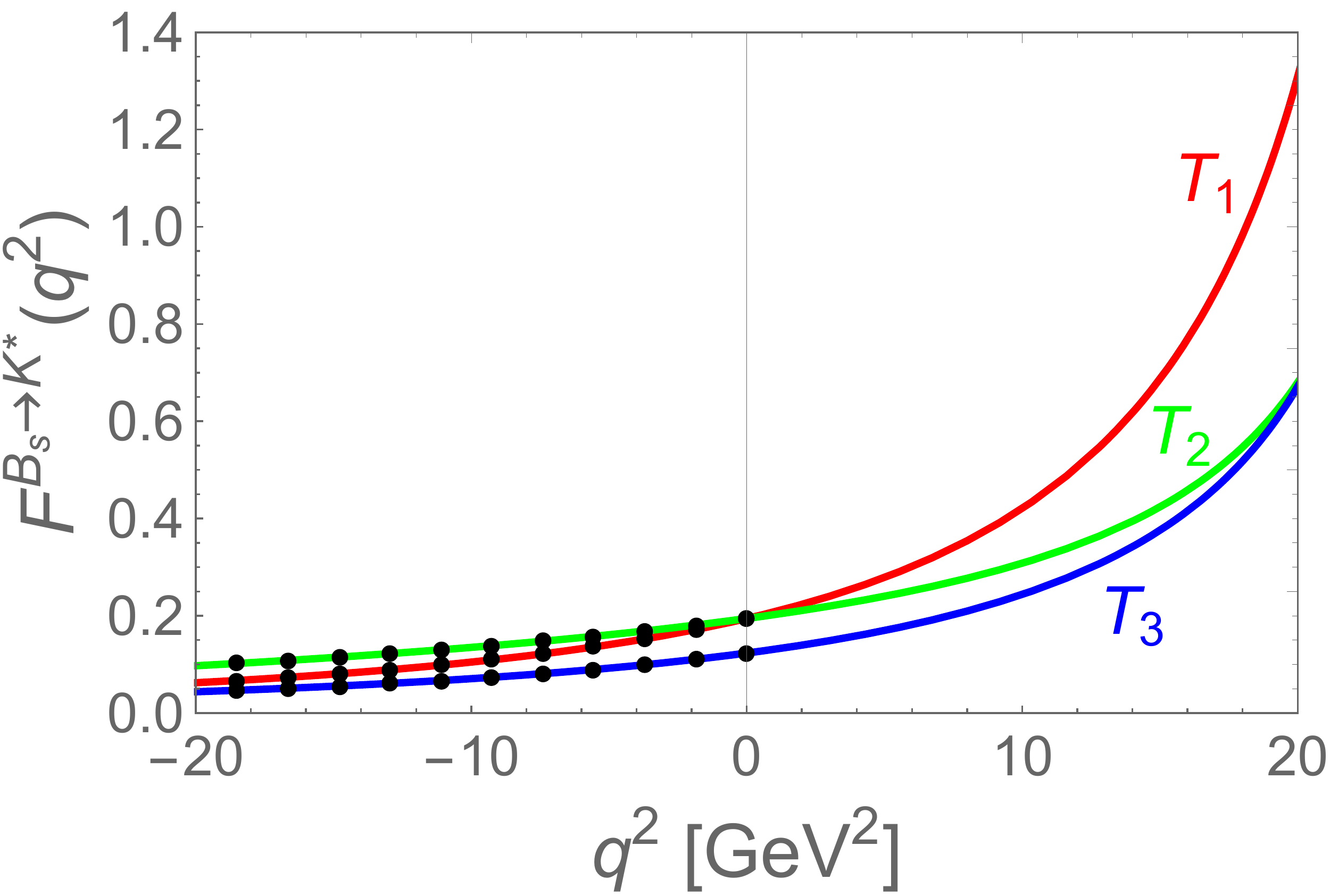}\quad}
{\includegraphics[width=0.25\textwidth]{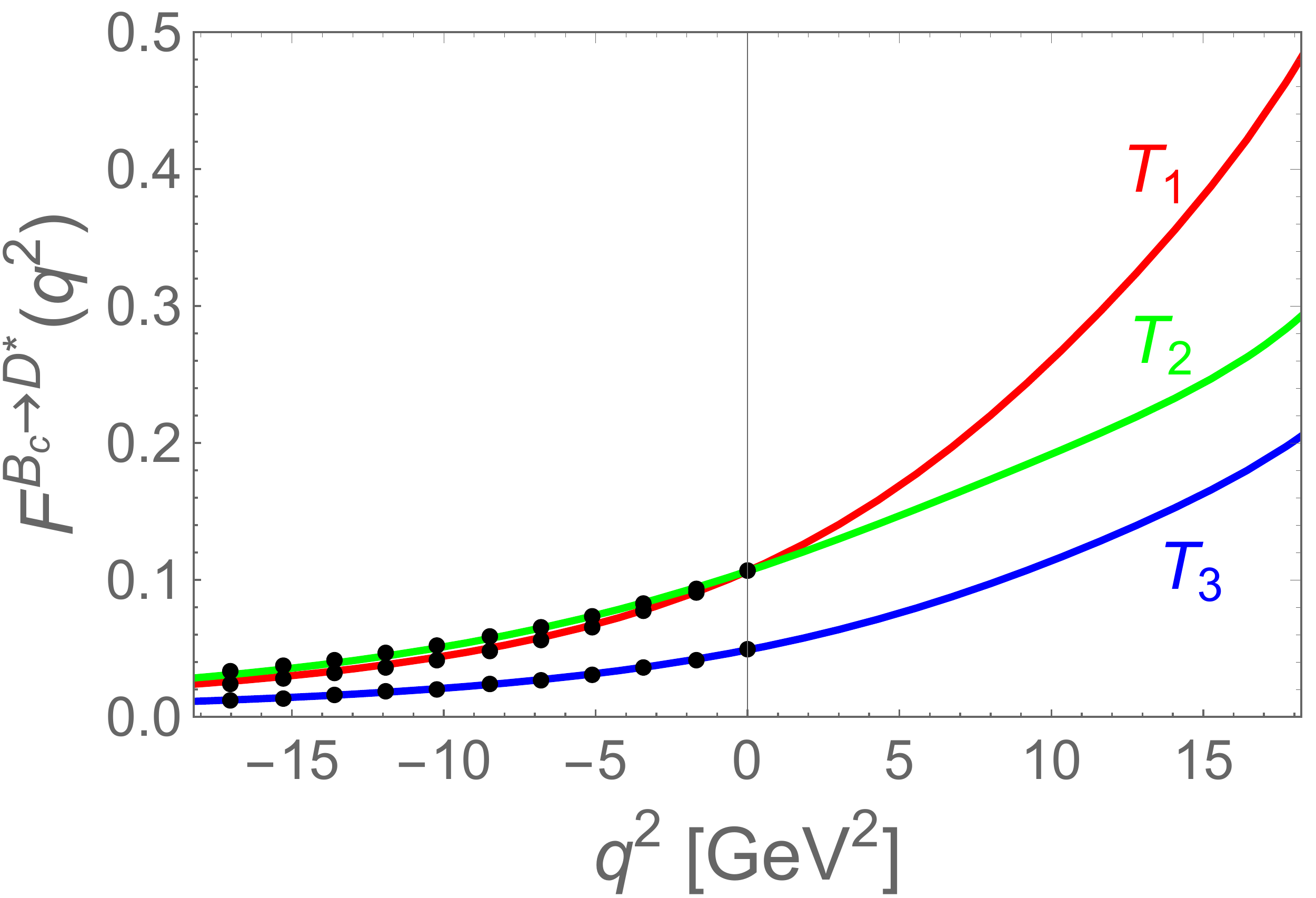}}\\
\vspace{0.3cm}
{\includegraphics[width=0.25\textwidth]{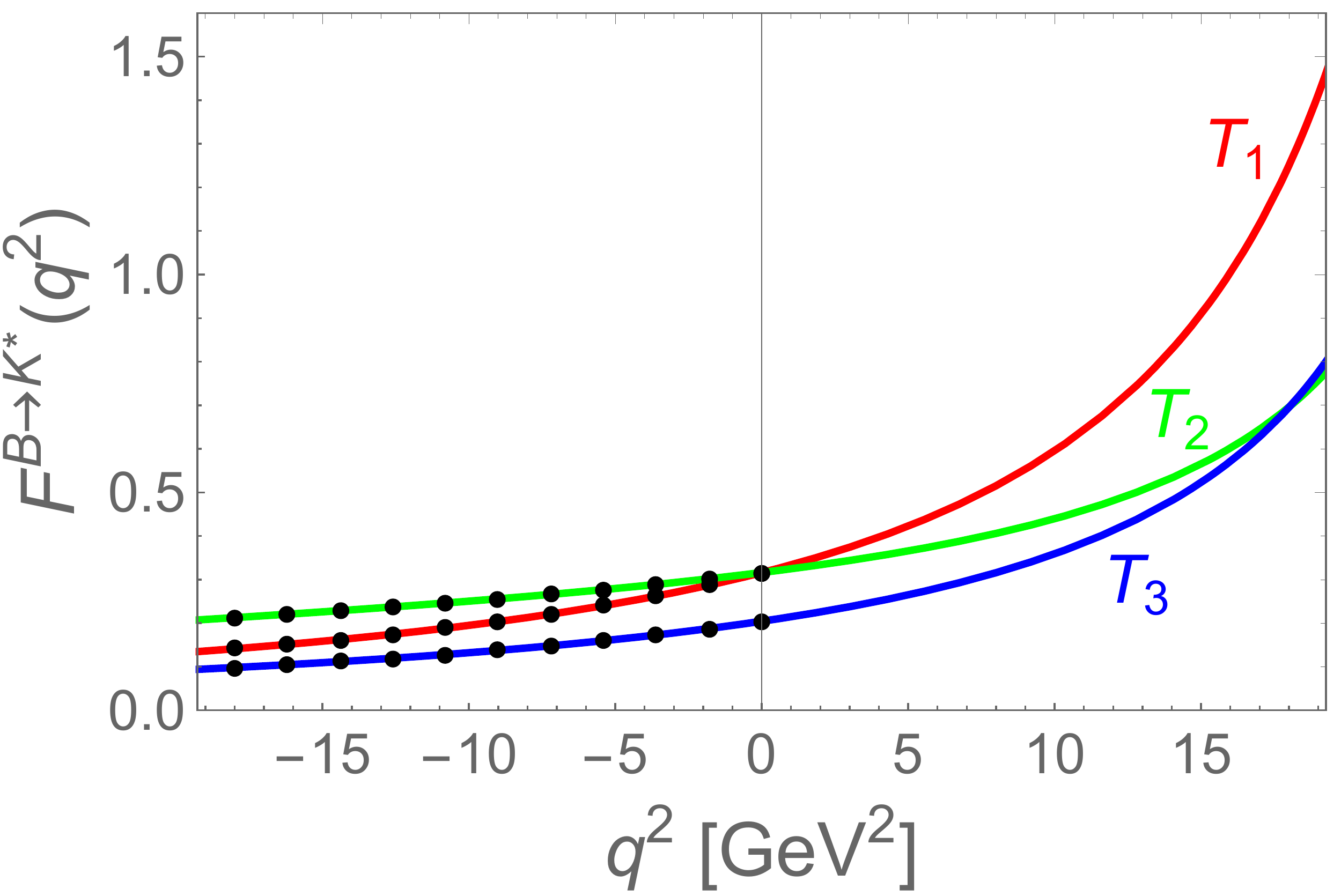}\quad}
{\includegraphics[width=0.25\textwidth]{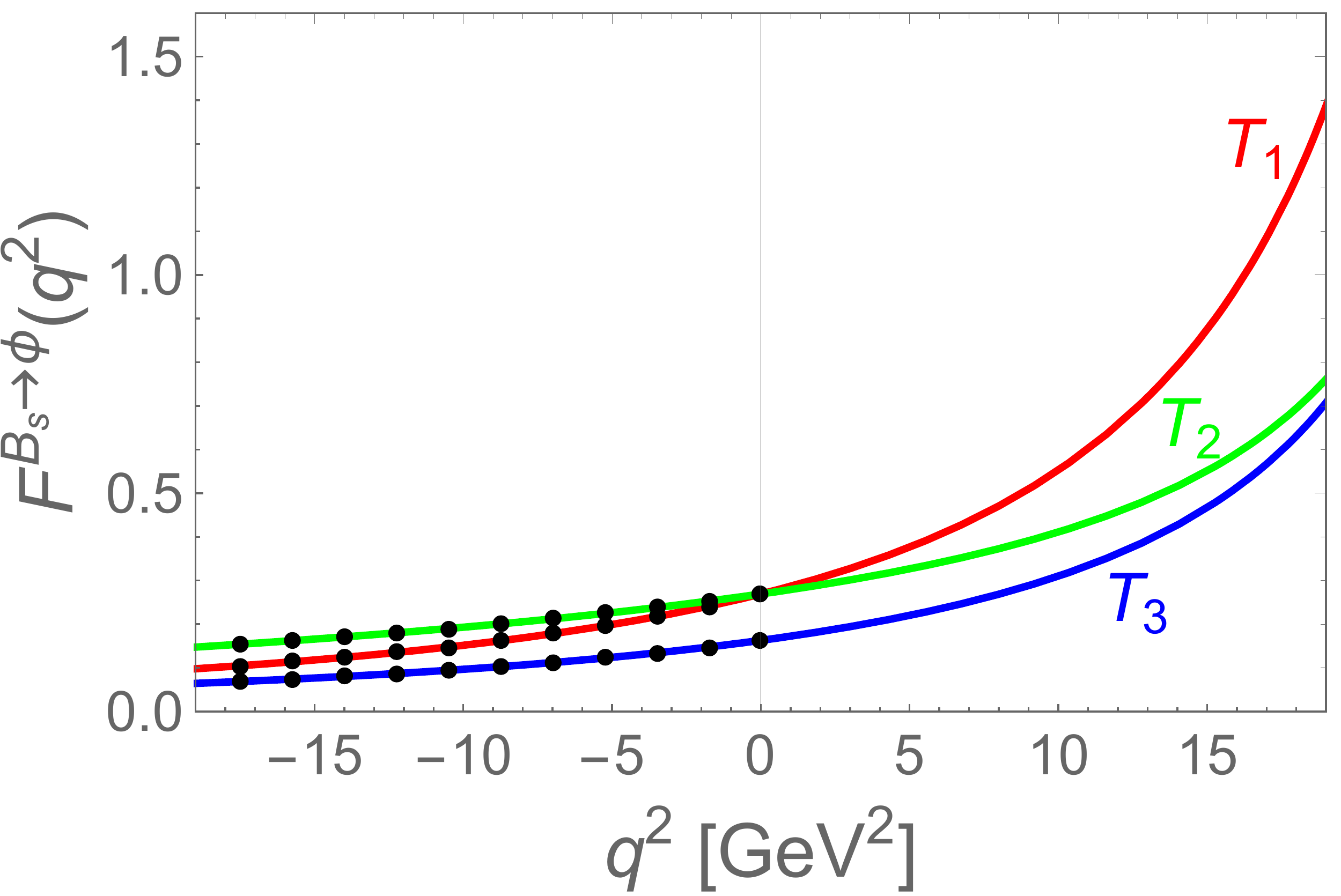}\quad}
{\includegraphics[width=0.25\textwidth]{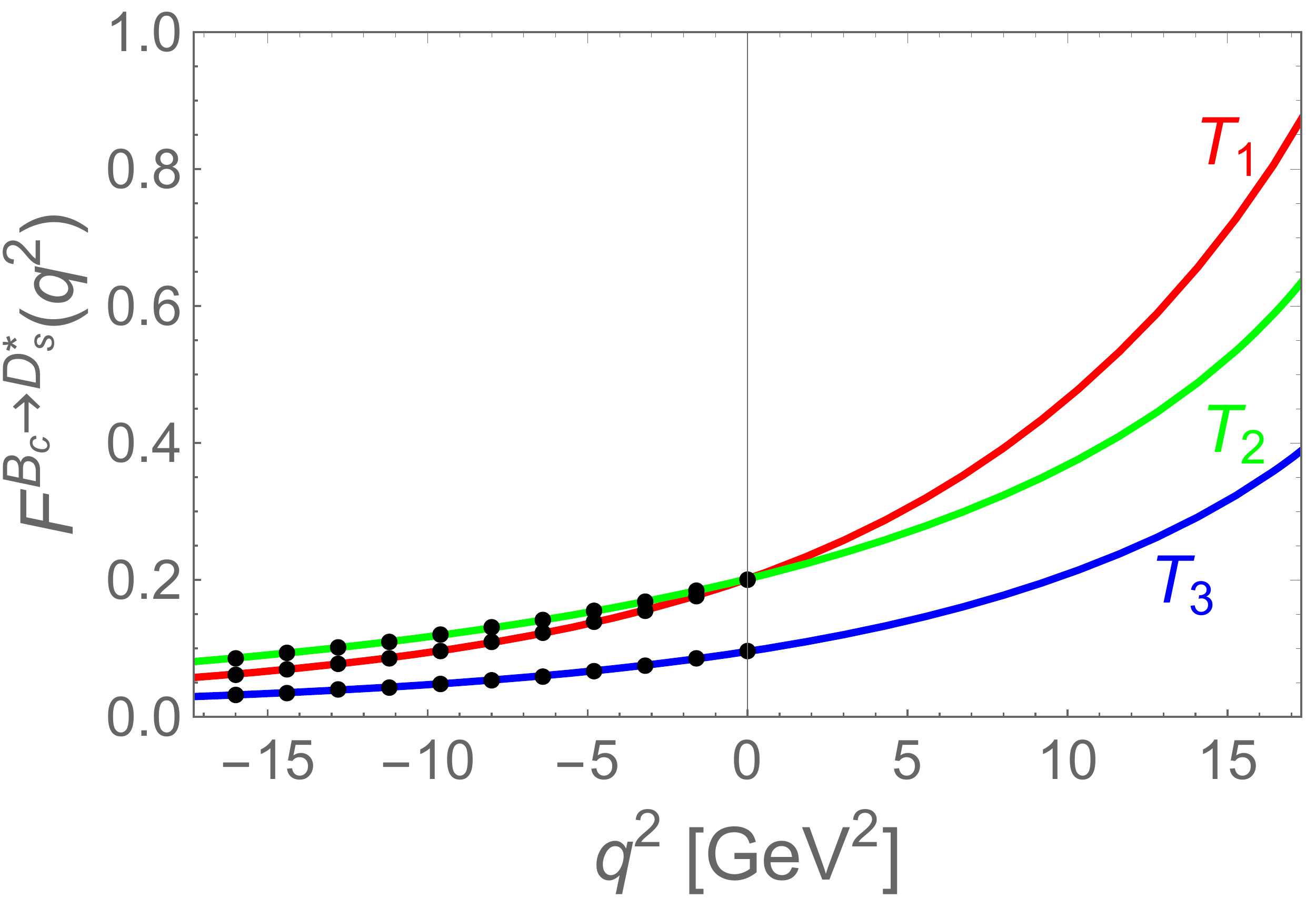}}\\
\vspace{0.3cm}
{\includegraphics[width=0.25\textwidth]{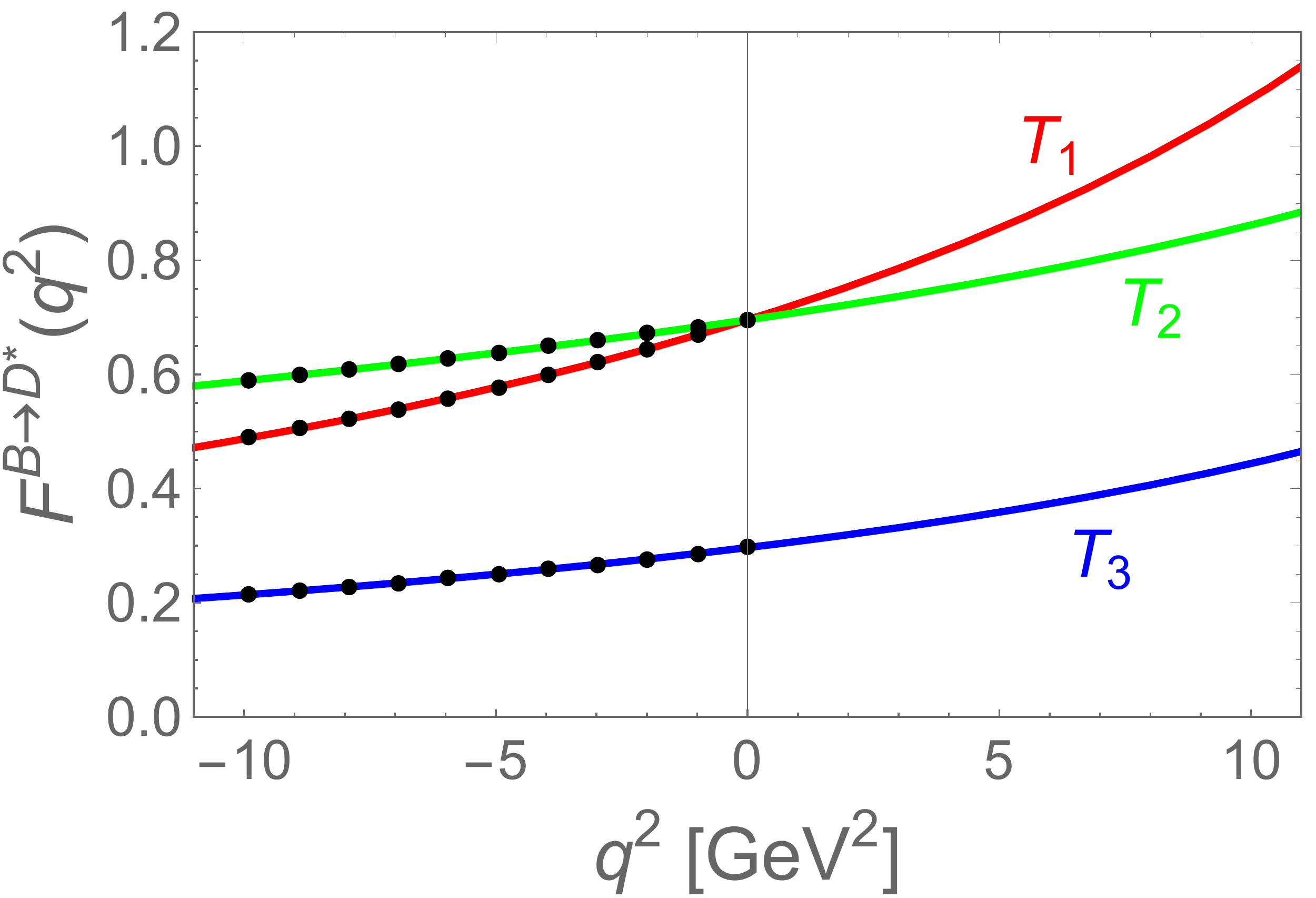}\quad}
{\includegraphics[width=0.25\textwidth]{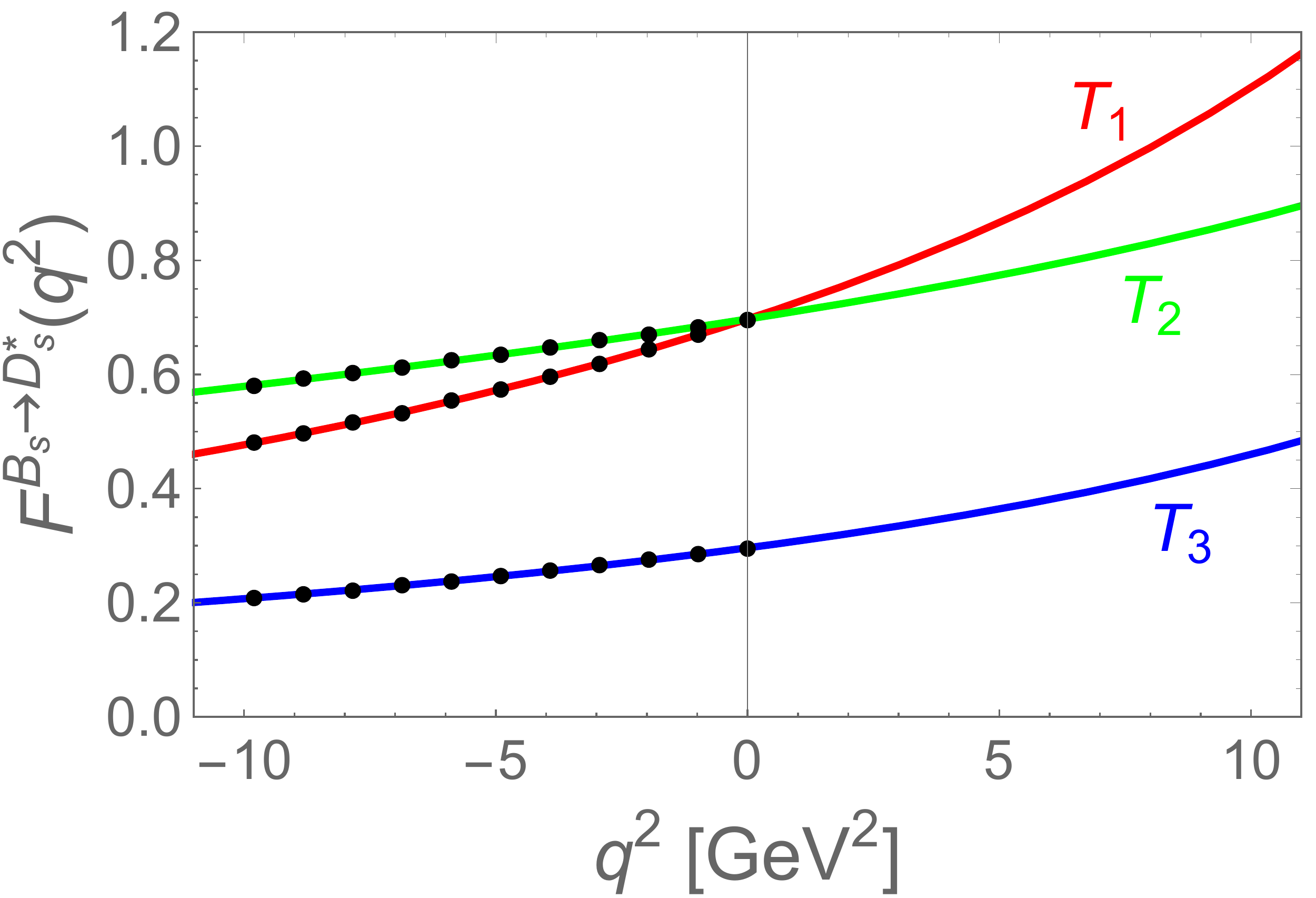}\quad}
{\includegraphics[width=0.25\textwidth]{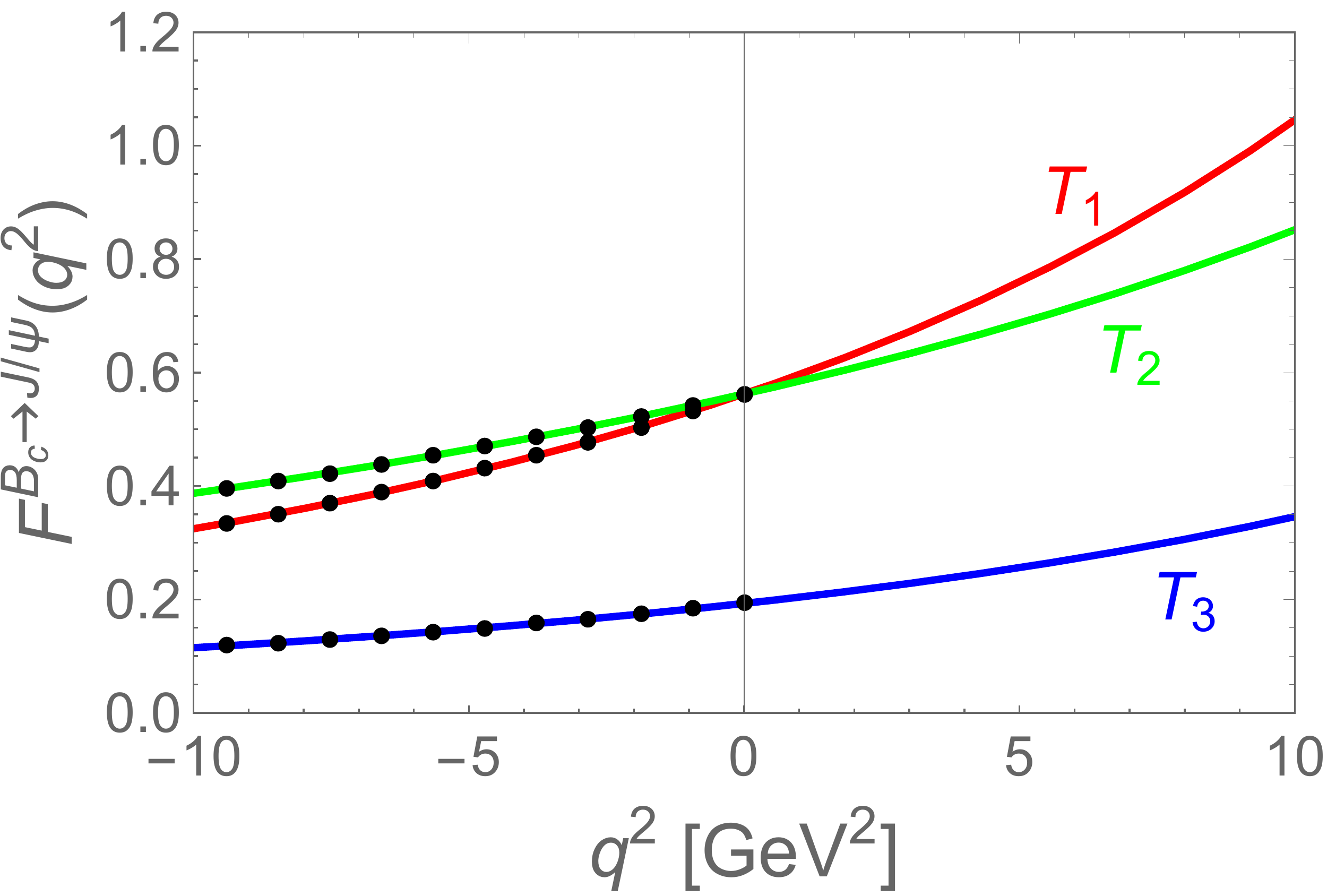}}
  \caption{Same as Fig. \ref{fig:PP} except for $D_{q,s}\to V$ and $B_{q,s,c}\to V$ transitions.}\label{fig:PV}
\end{figure}

\begin{figure}[t]
  \centering
{\includegraphics[width=0.24\textwidth]{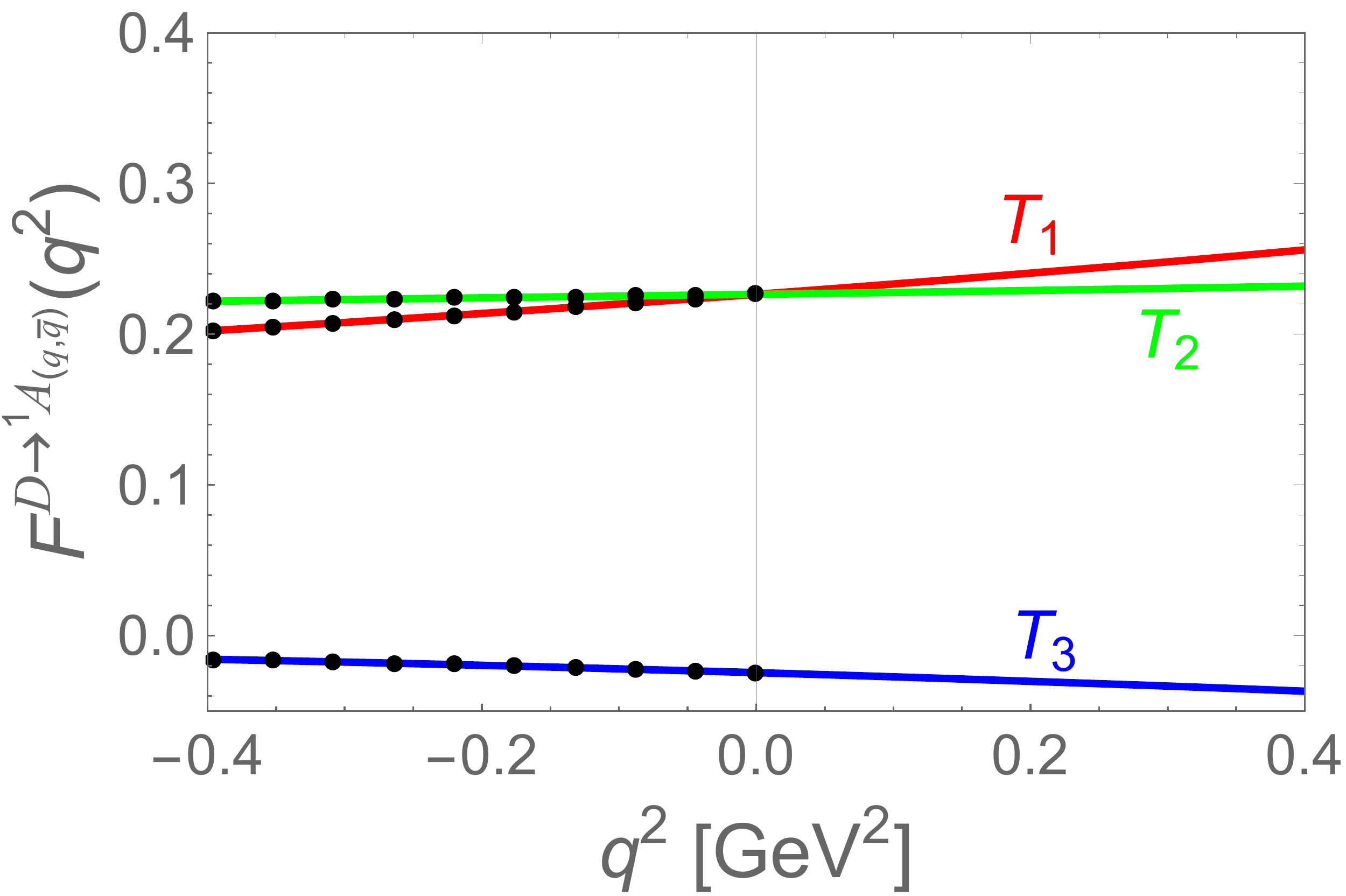}}
{\includegraphics[width=0.24\textwidth]{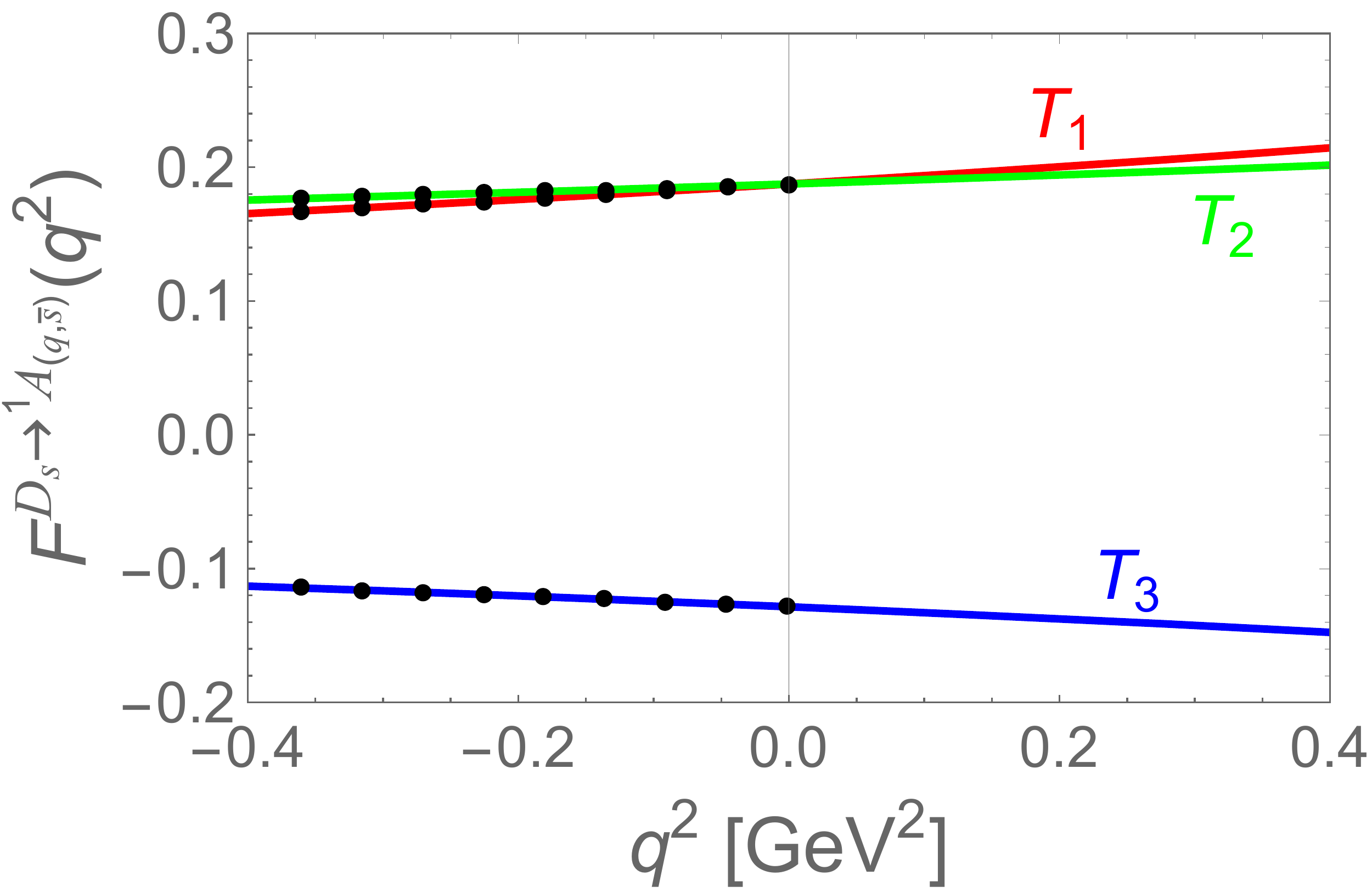}}
{\includegraphics[width=0.24\textwidth]{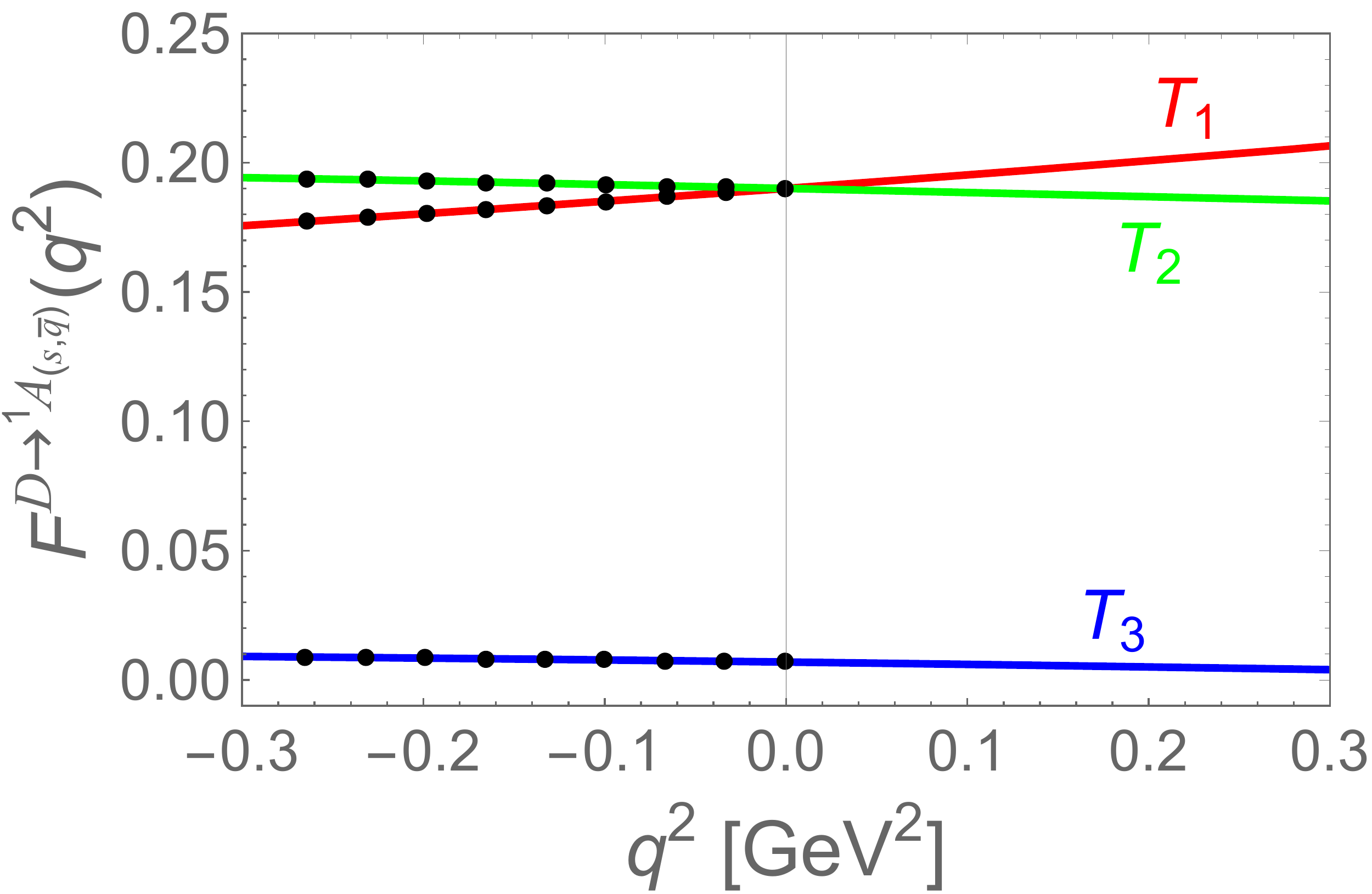}}
{\includegraphics[width=0.24\textwidth]{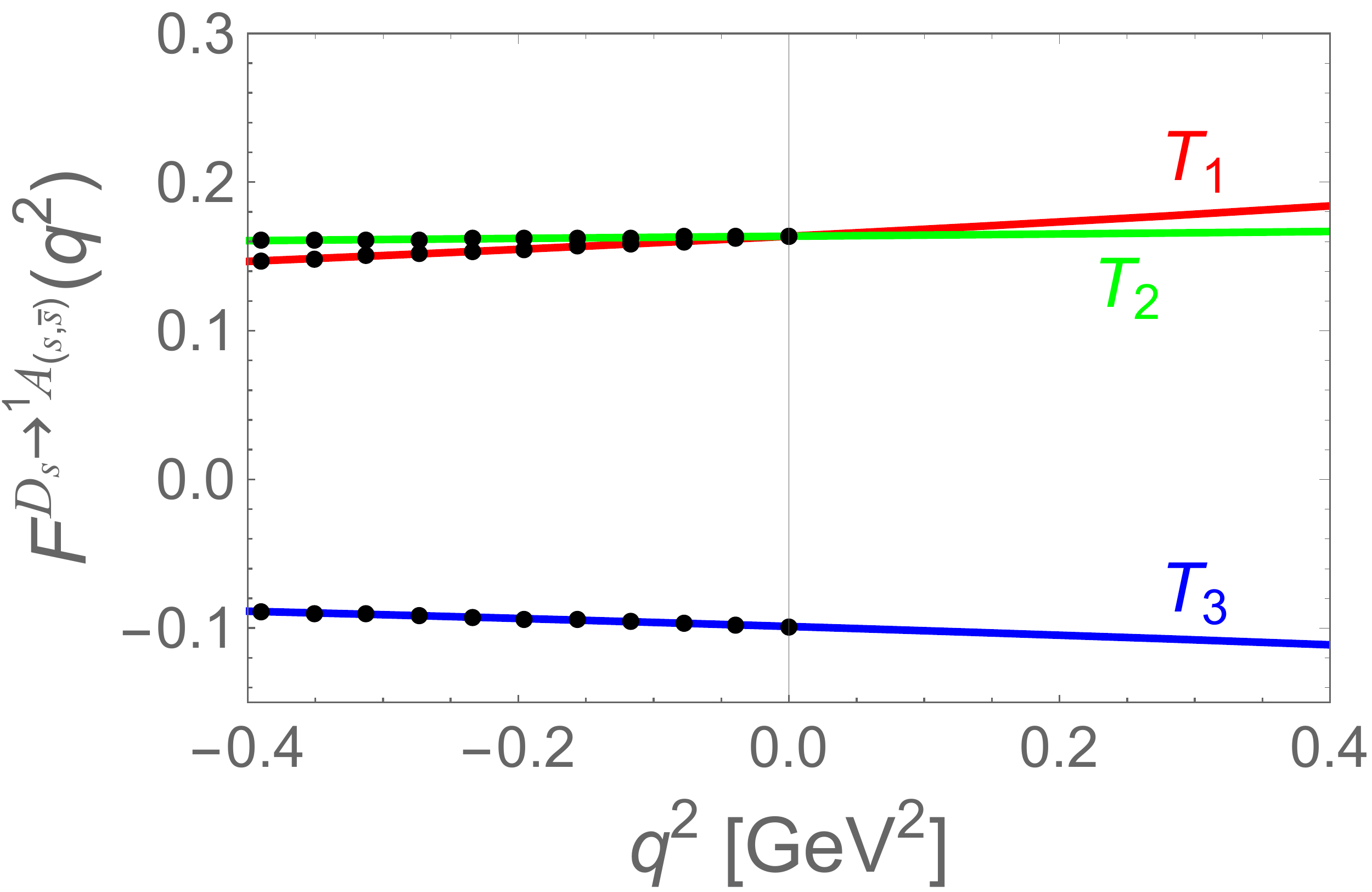}}\\
  \vspace{0.3cm}
{\includegraphics[width=0.25\textwidth]{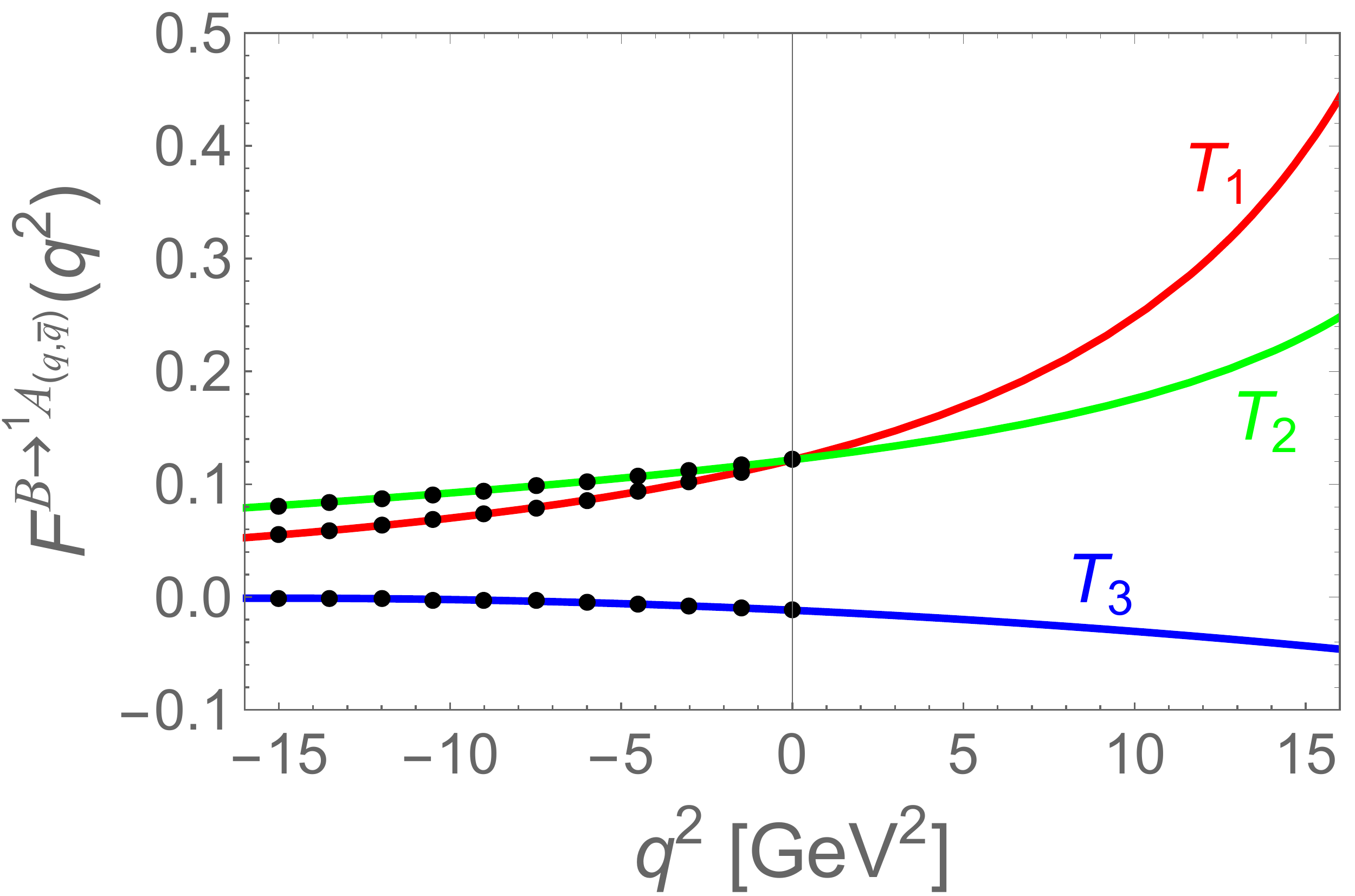}\quad}
{\includegraphics[width=0.25\textwidth]{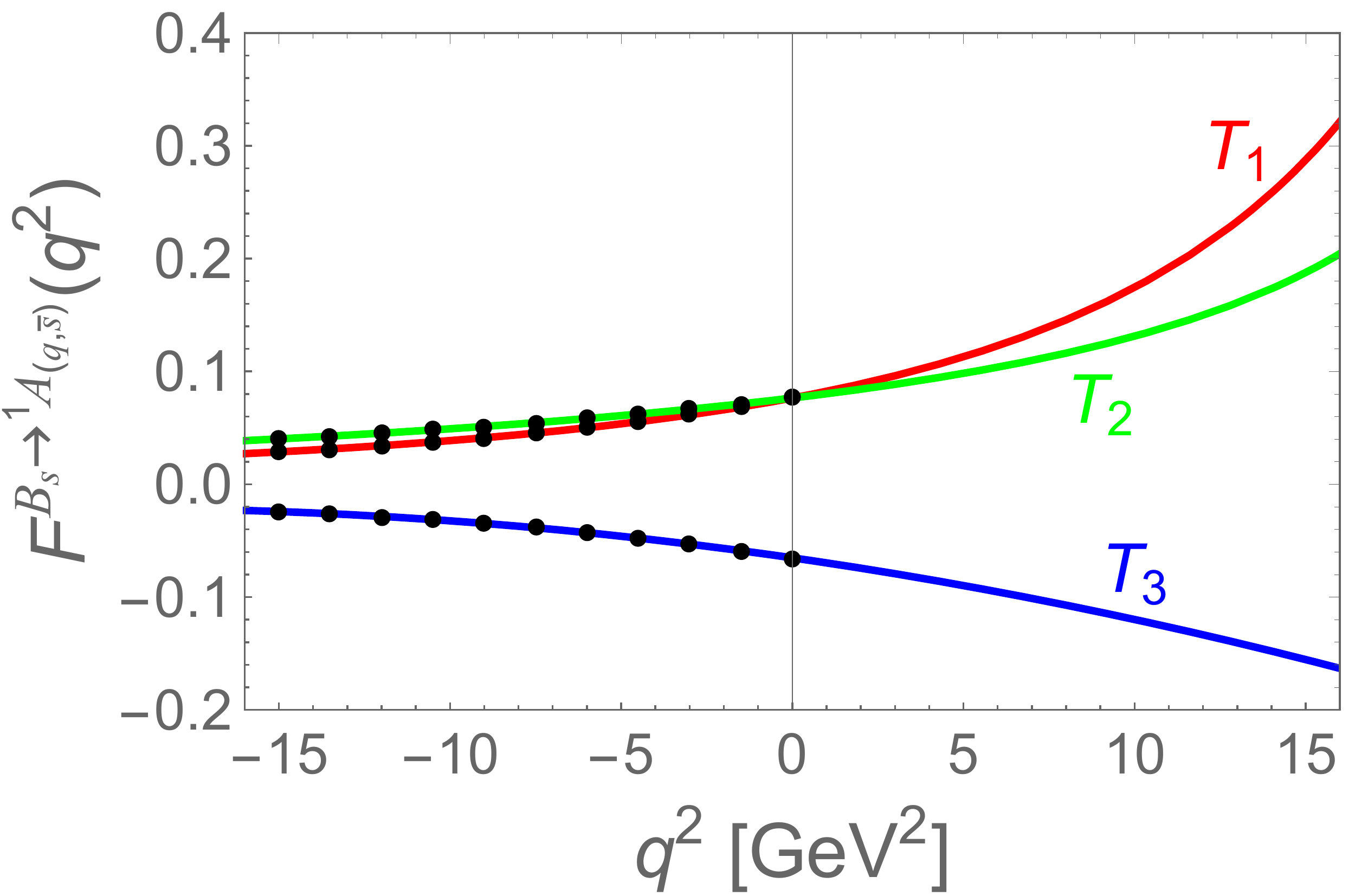}\quad}
{\includegraphics[width=0.25\textwidth]{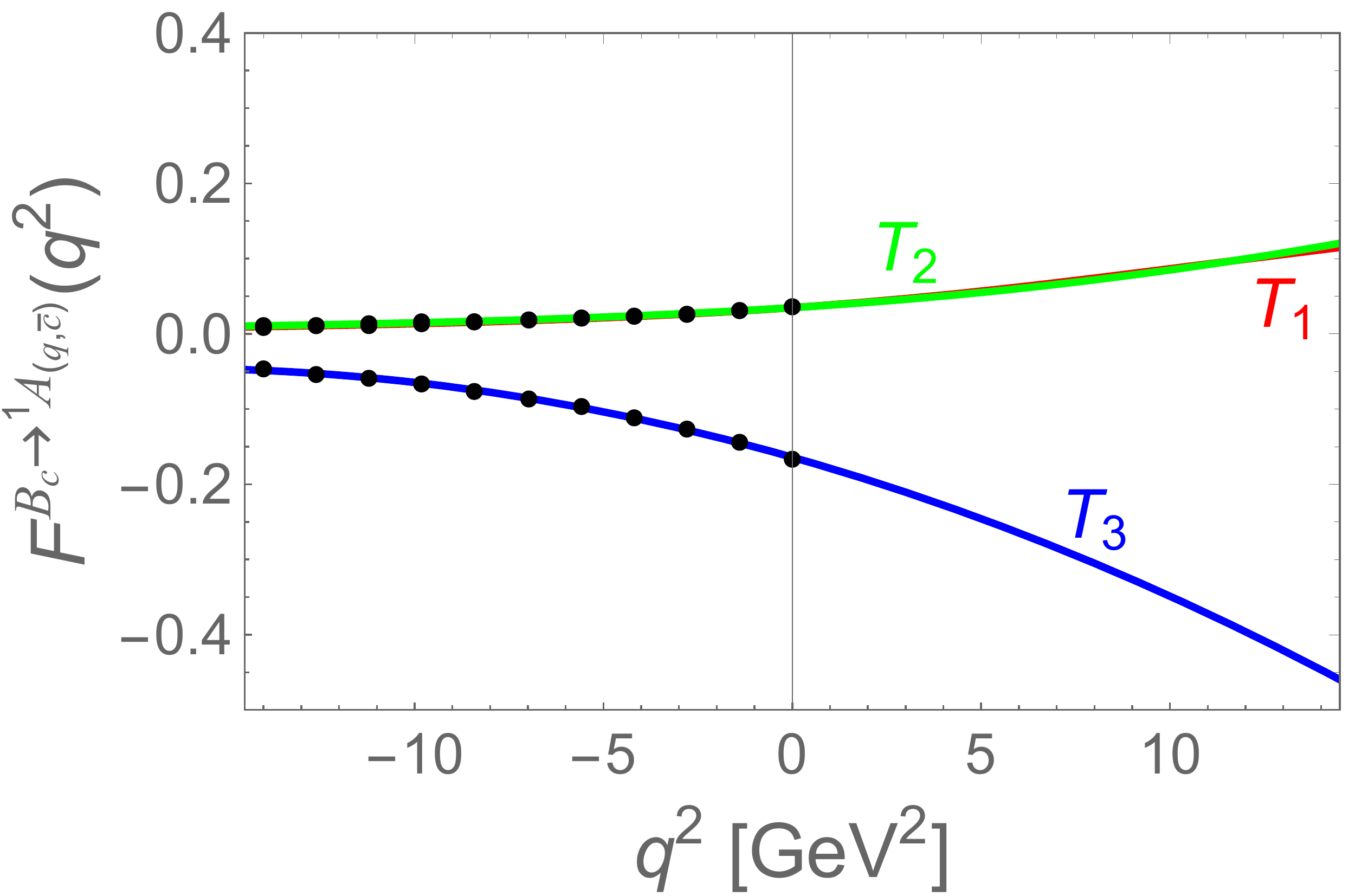}}\\
  \vspace{0.3cm}
{\includegraphics[width=0.25\textwidth]{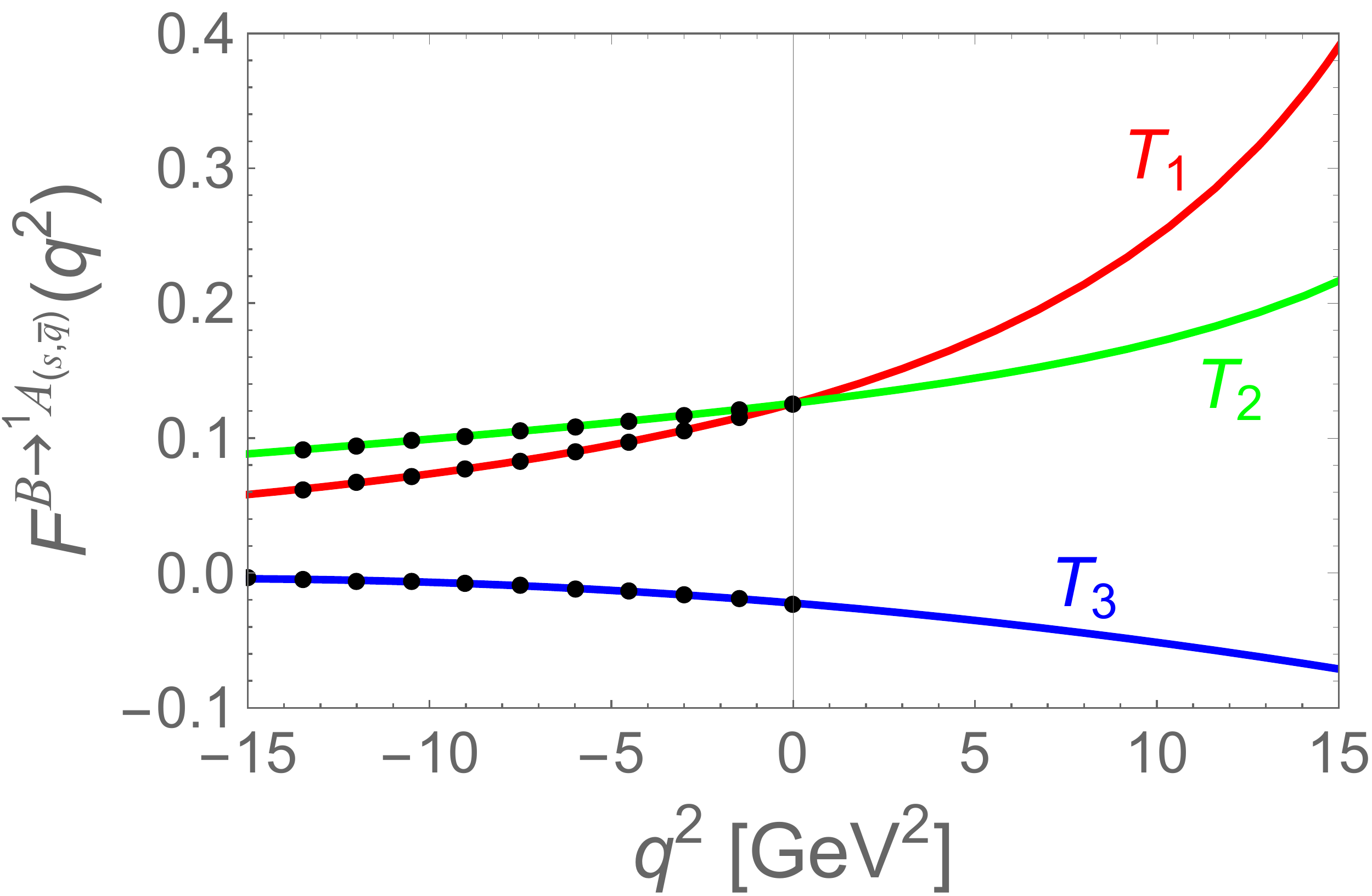}\quad}
{\includegraphics[width=0.25\textwidth]{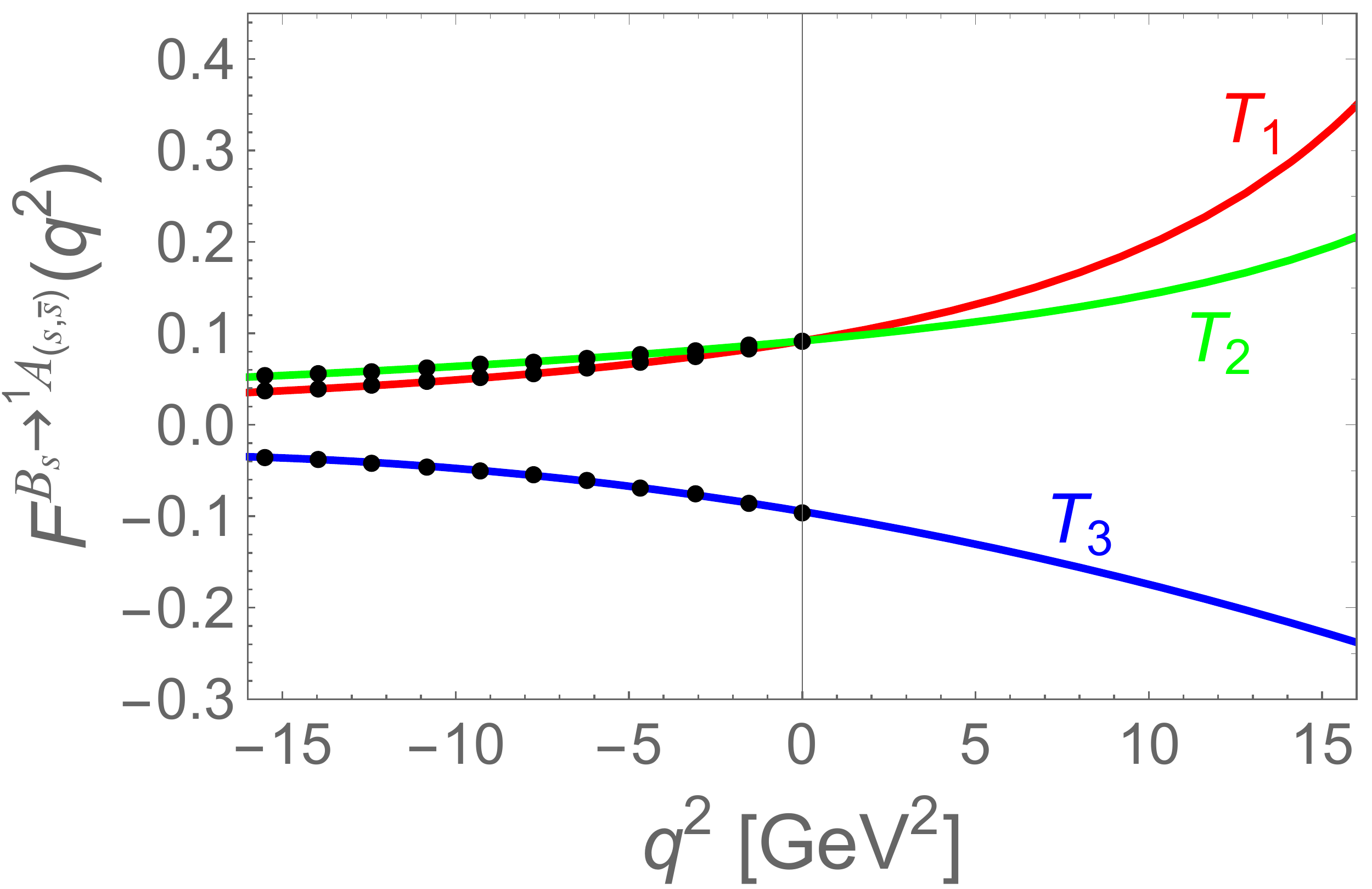}\quad}
{\includegraphics[width=0.25\textwidth]{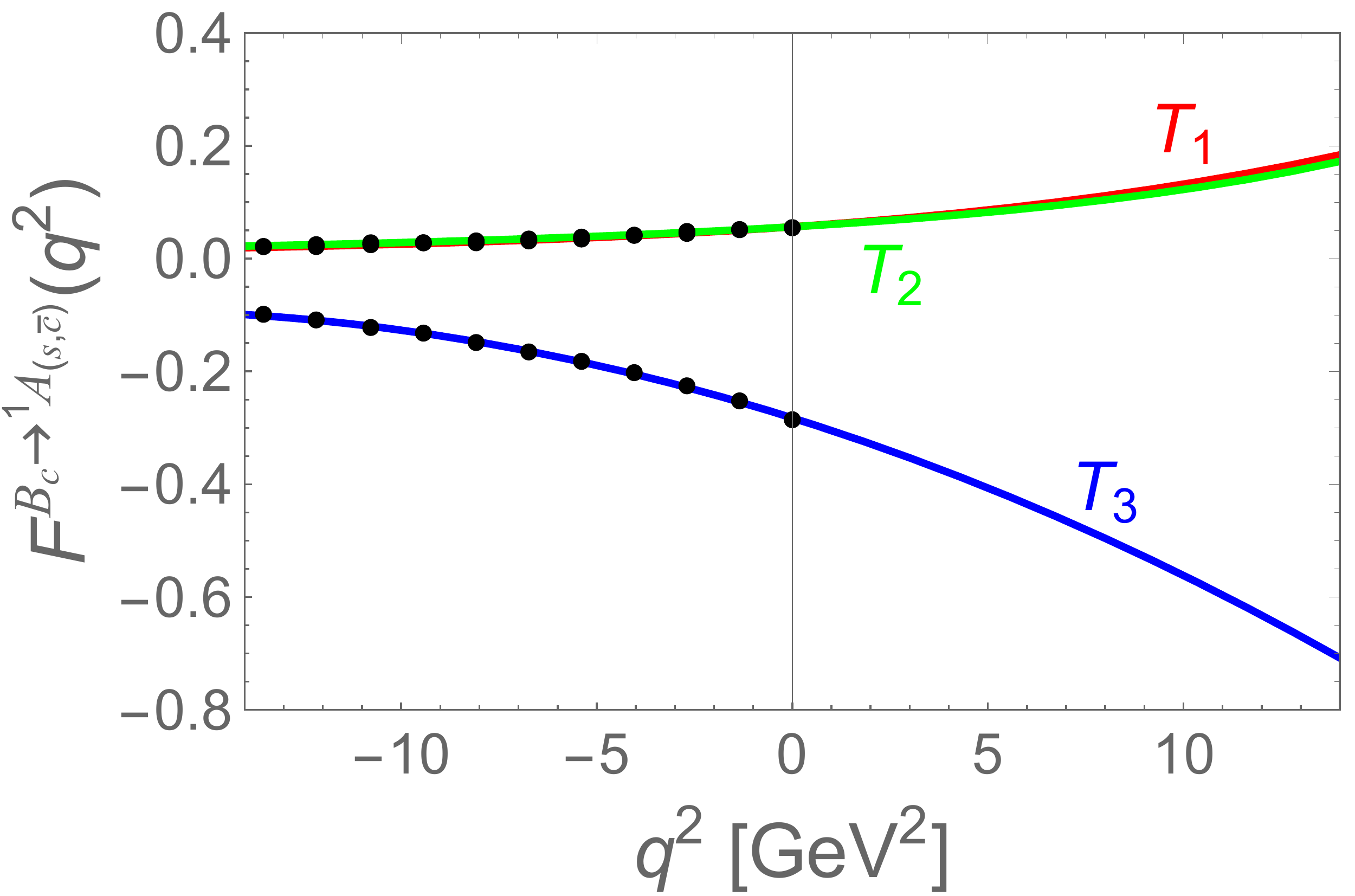}}\\
  \vspace{0.3cm}
{\includegraphics[width=0.25\textwidth]{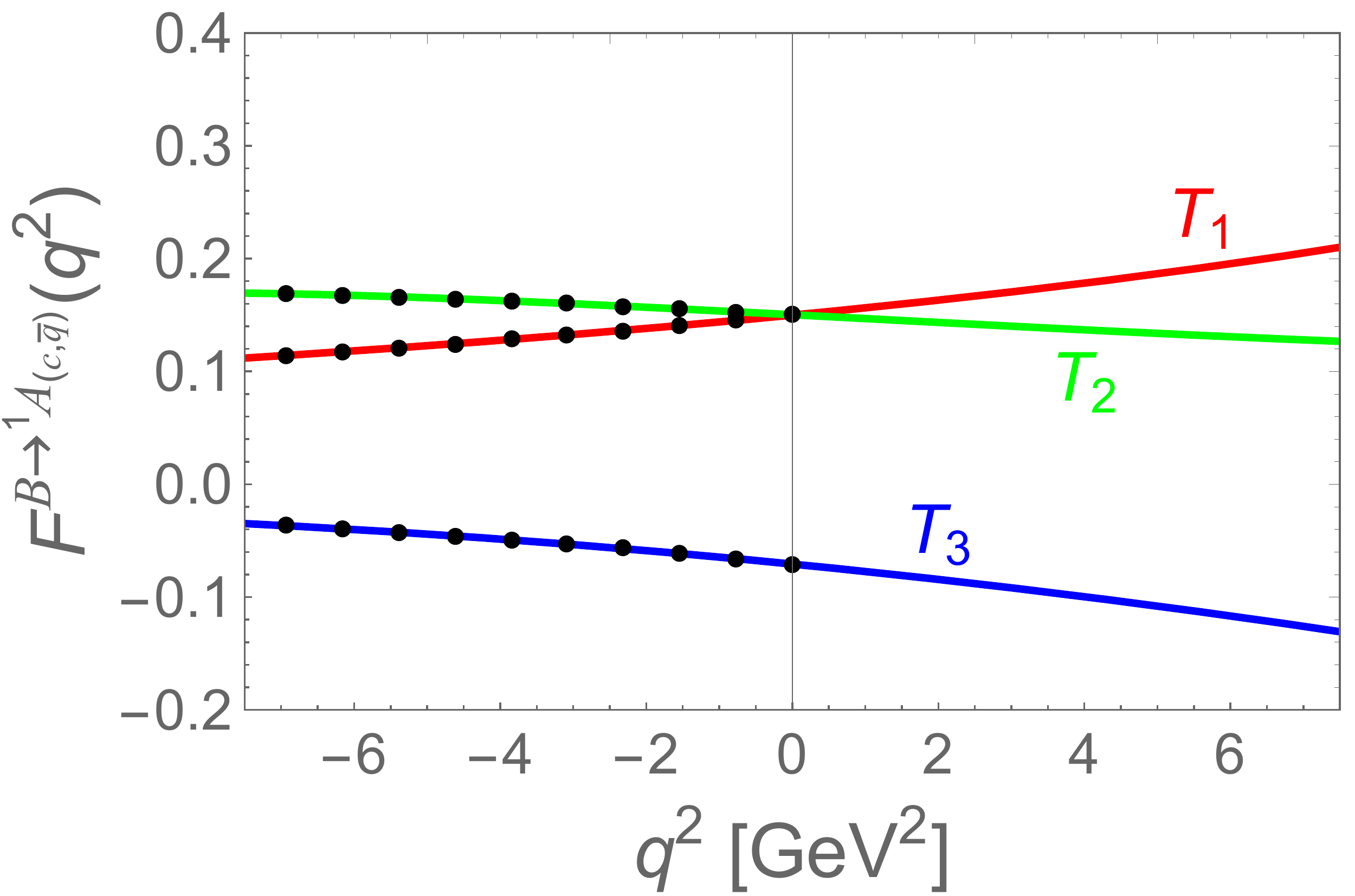}\quad}
{\includegraphics[width=0.25\textwidth]{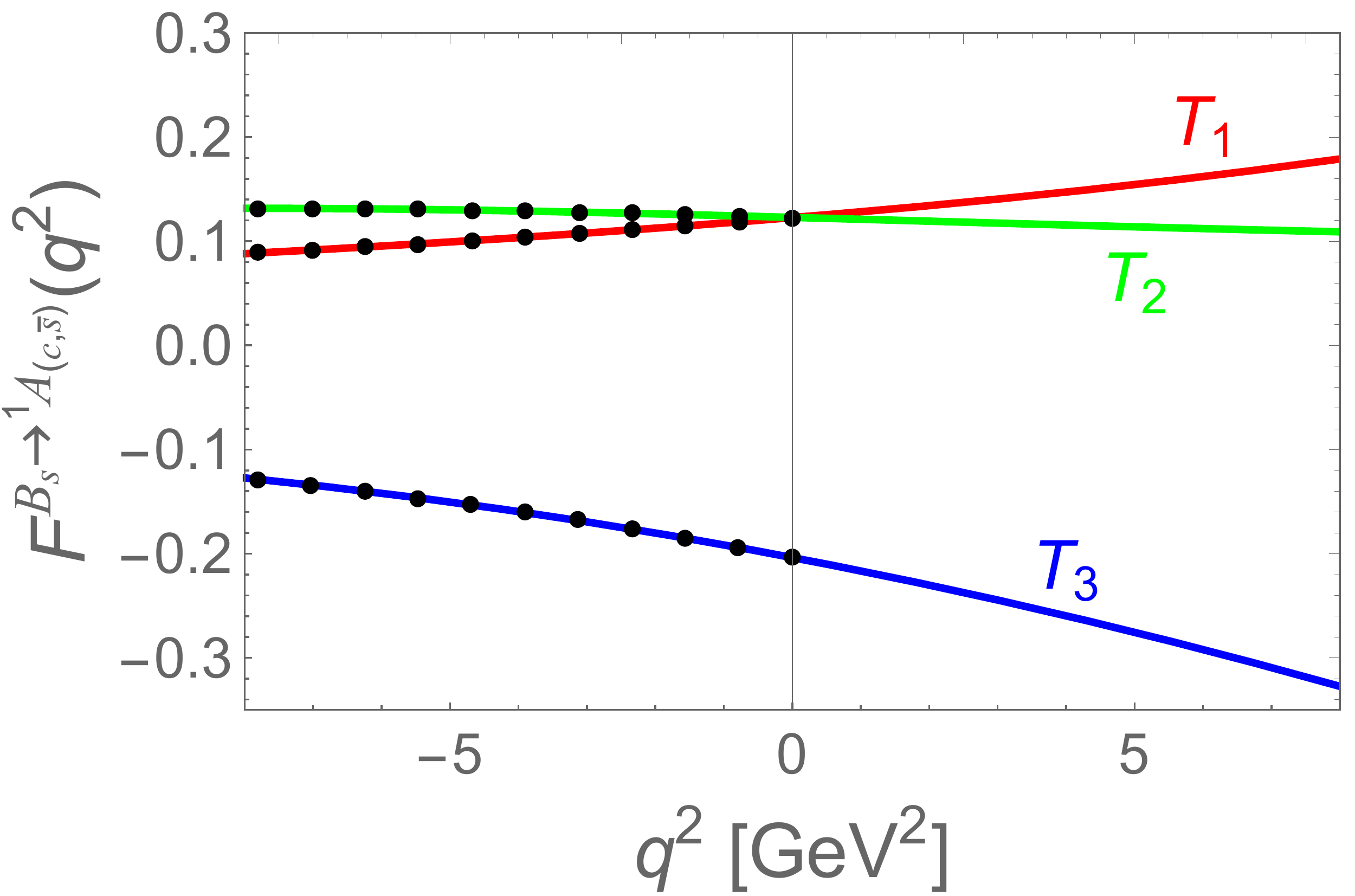}\quad}
{\includegraphics[width=0.25\textwidth]{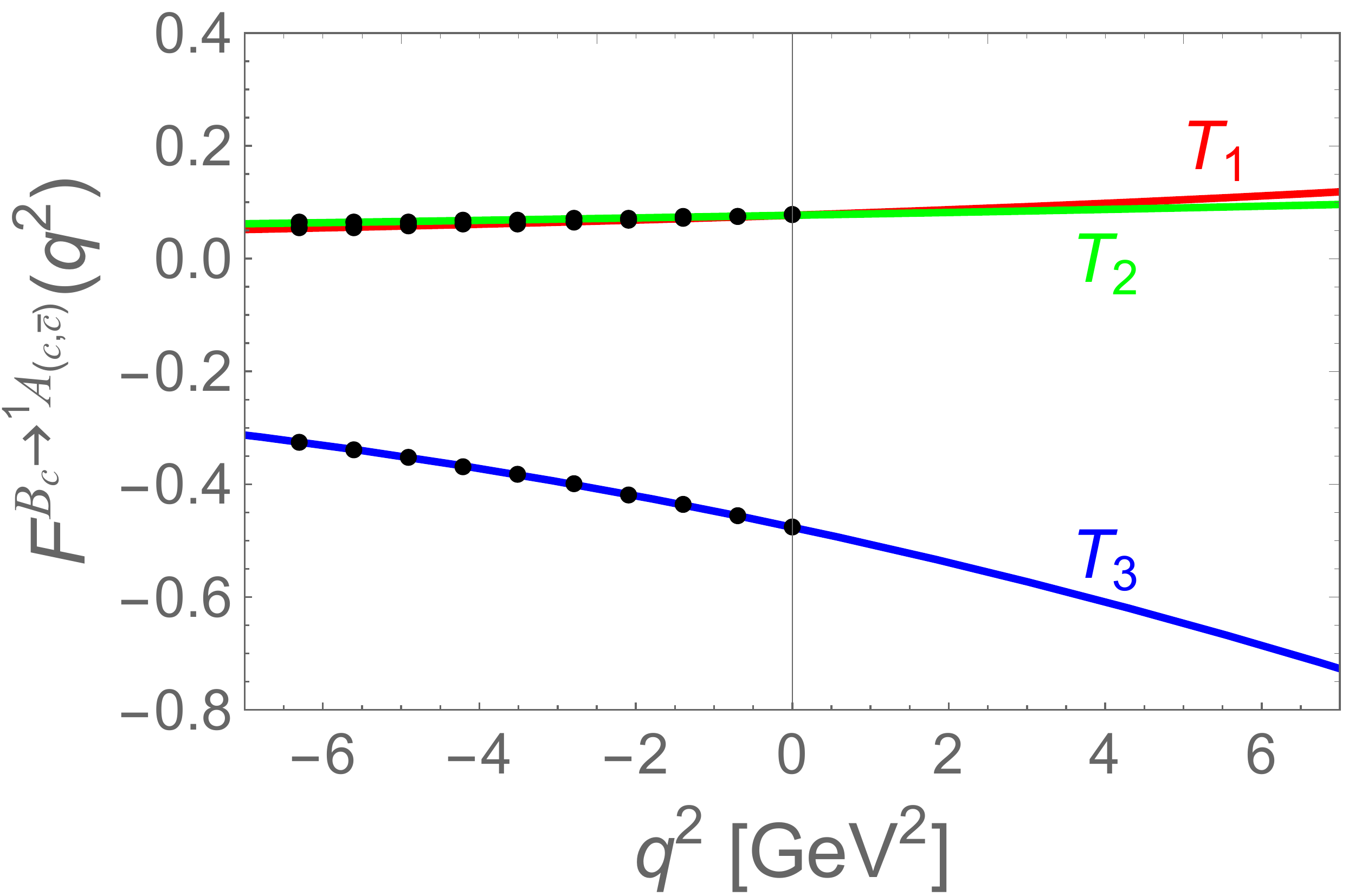}}
 \caption{\label{fig:P1A}Same as Fig. \ref{fig:PP} except for $D_{q,s}\to {^1\!A} $ and $B_{q,s,c}\to {^1\!A} $ transitions.}
\end{figure}

\begin{figure}[t]
  \centering
{\includegraphics[width=0.24\textwidth]{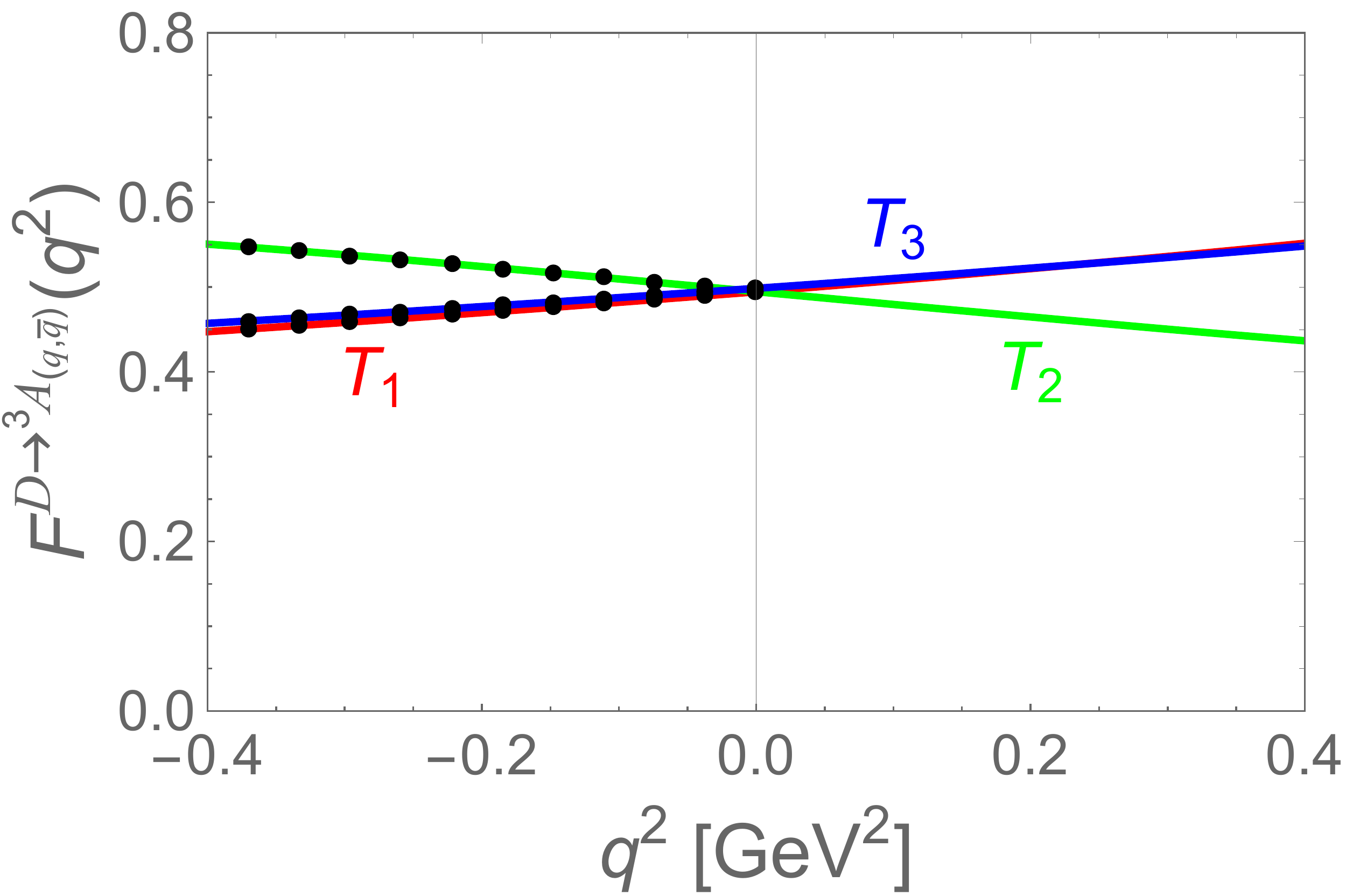}}
{\includegraphics[width=0.24\textwidth]{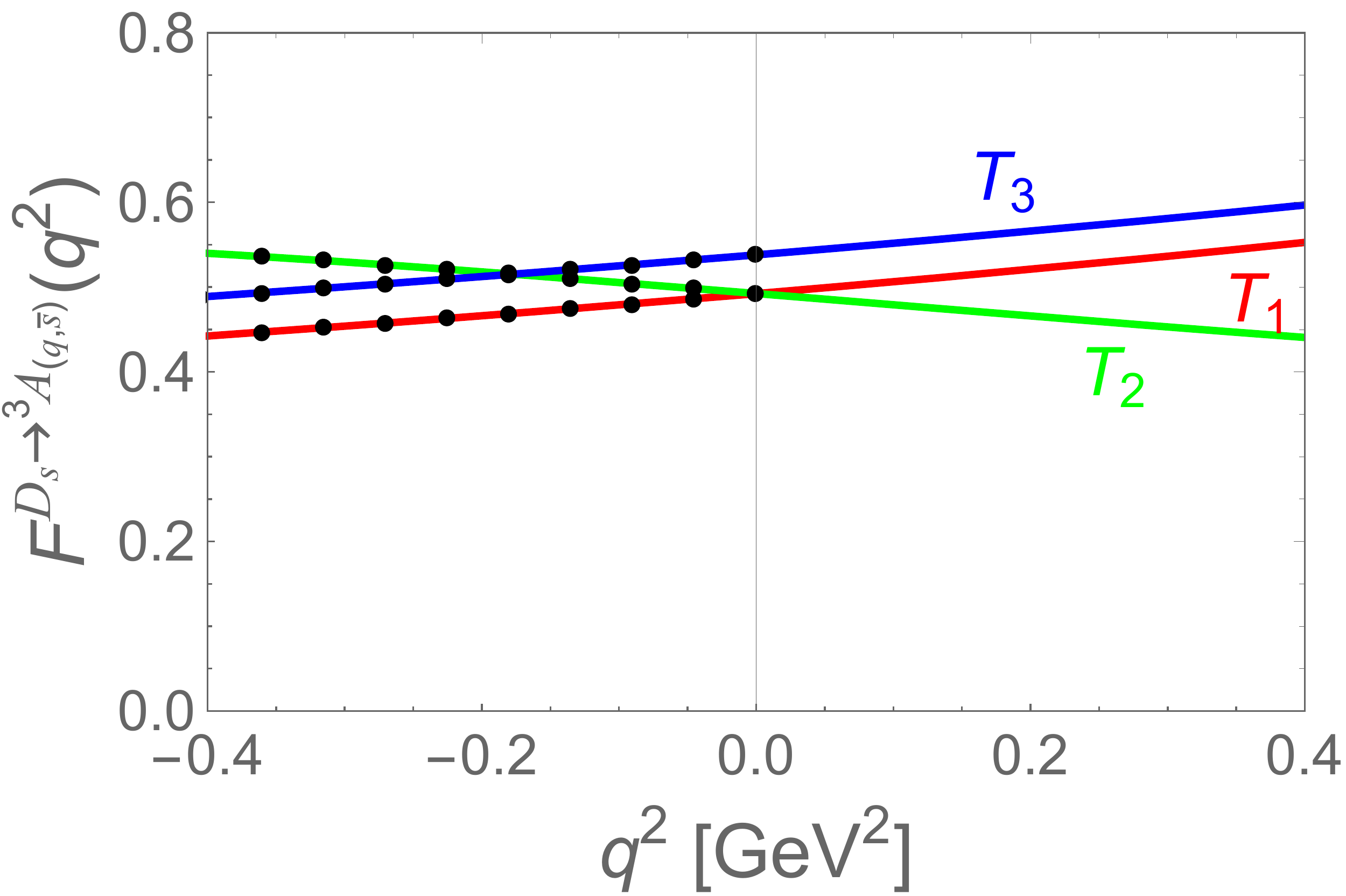}}
{\includegraphics[width=0.24\textwidth]{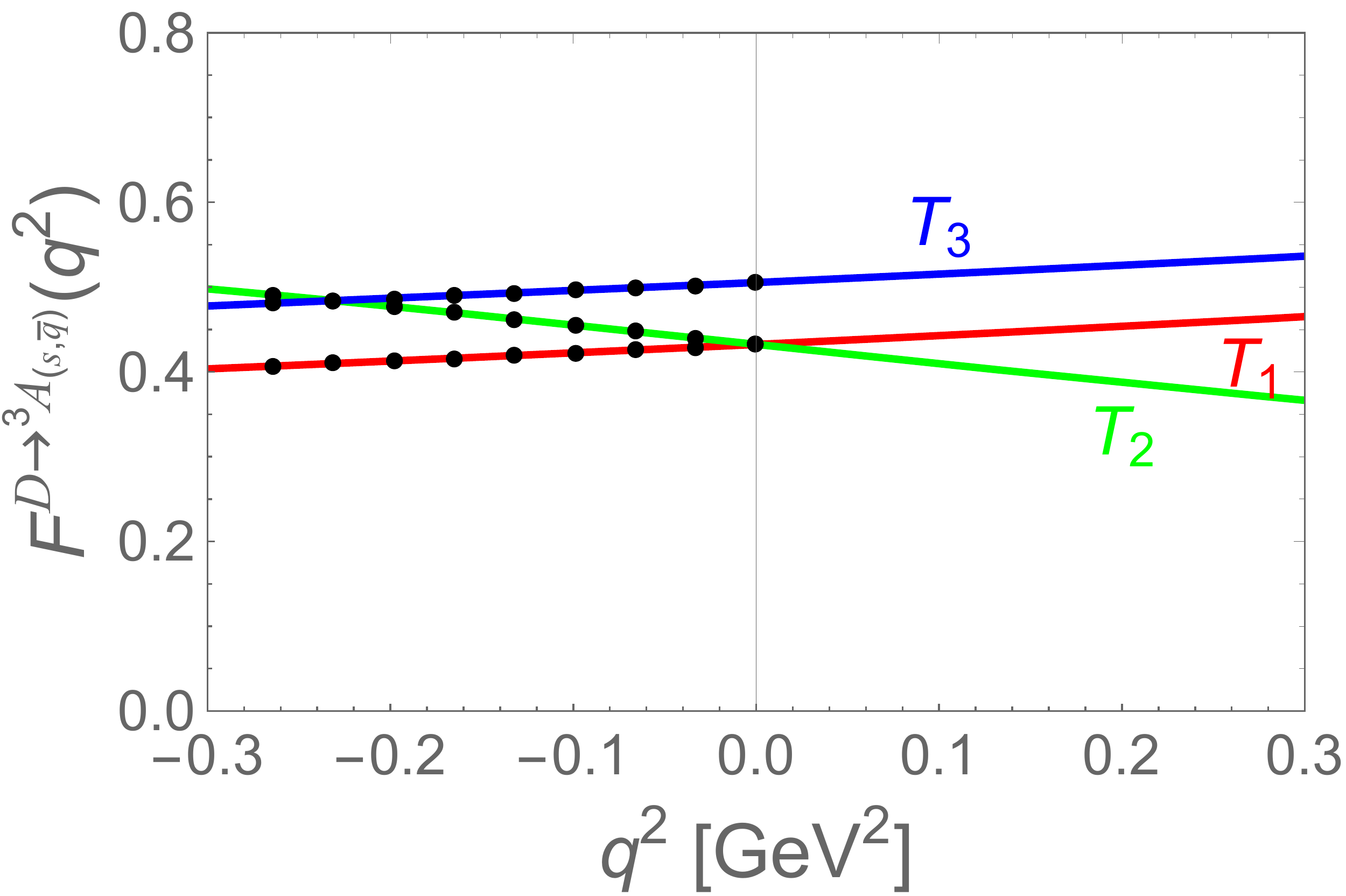}}
{\includegraphics[width=0.24\textwidth]{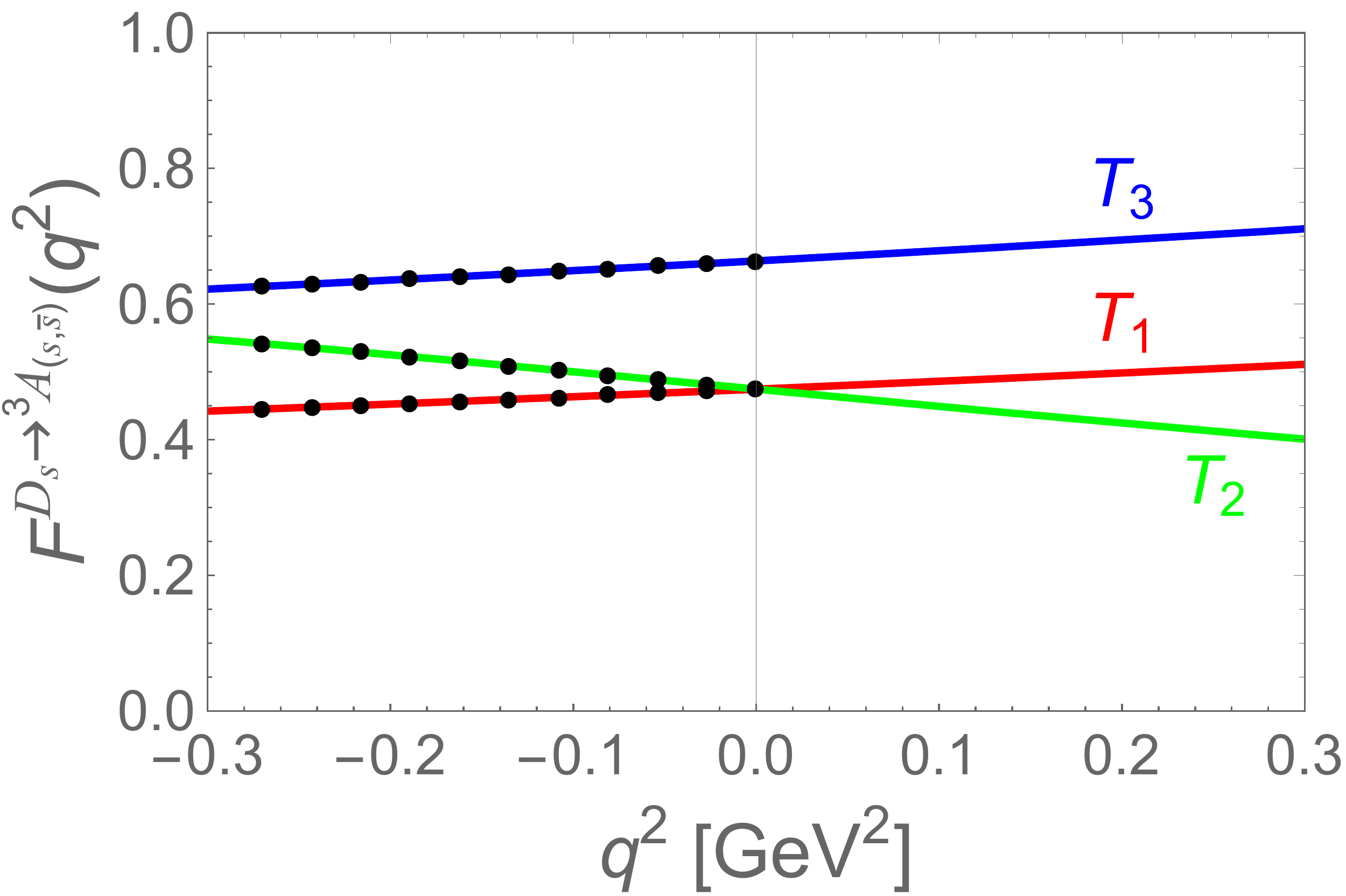}}\\
  \vspace{0.3cm}
{\includegraphics[width=0.25\textwidth]{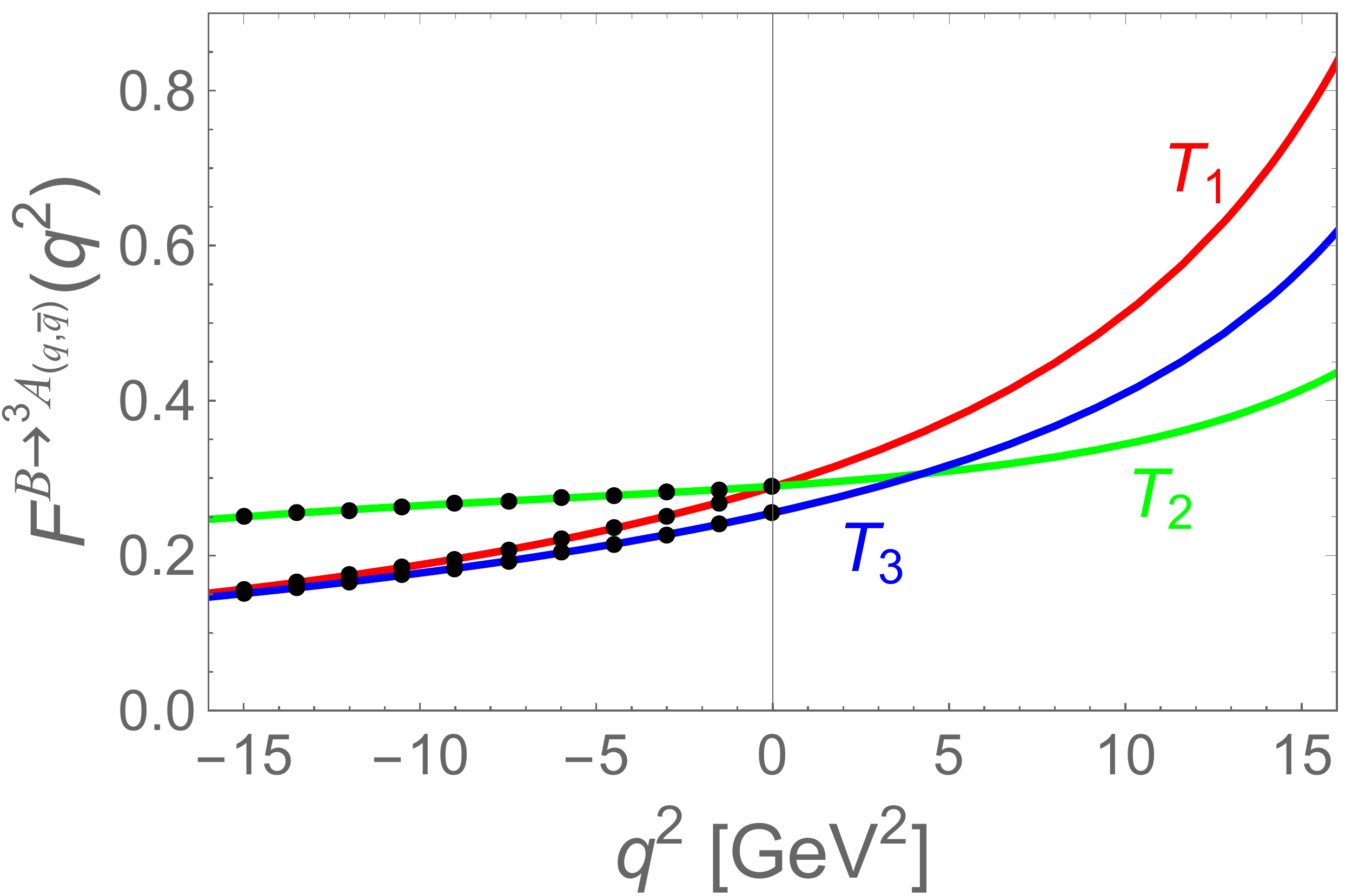}\quad}
{\includegraphics[width=0.25\textwidth]{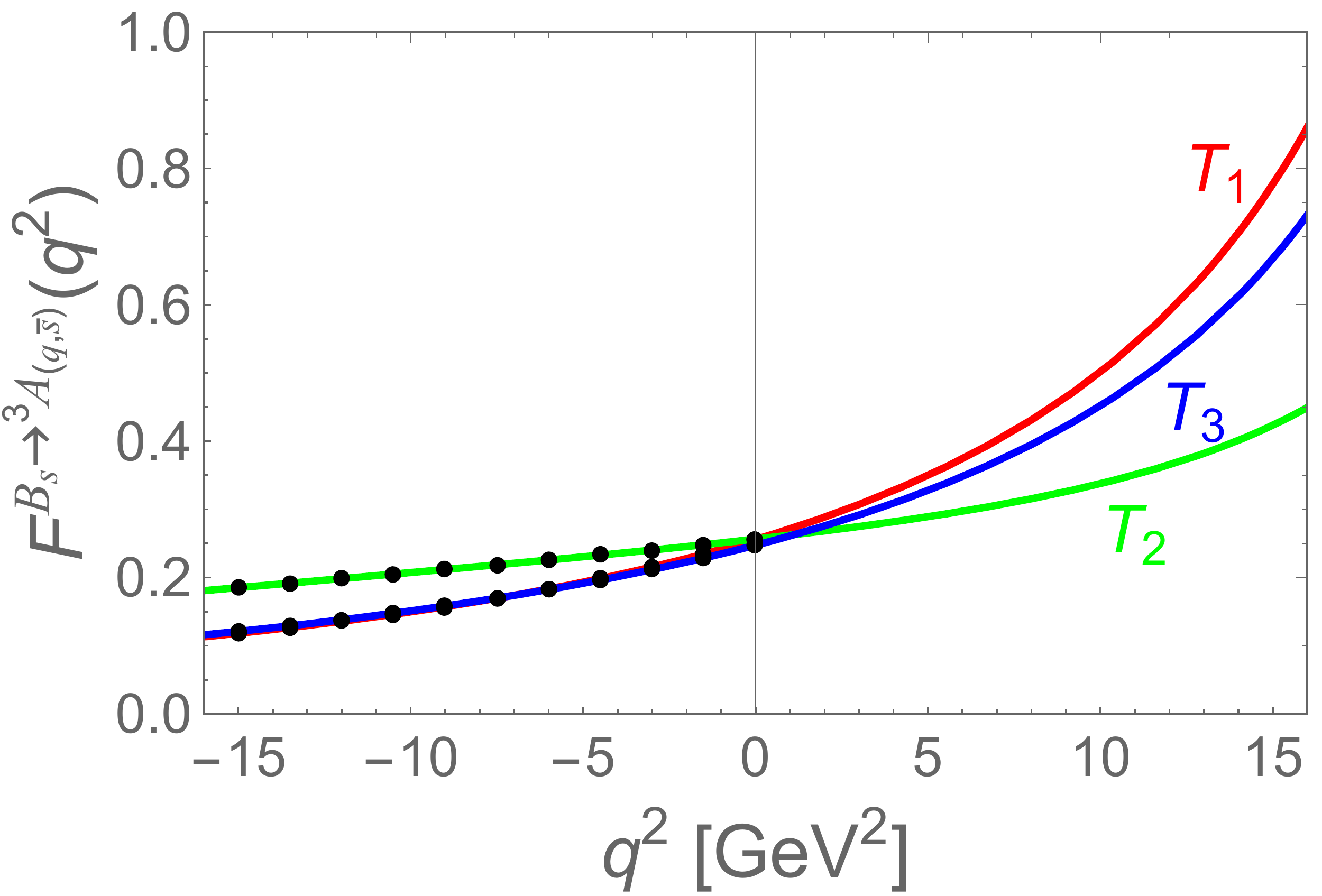}\quad}
{\includegraphics[width=0.25\textwidth]{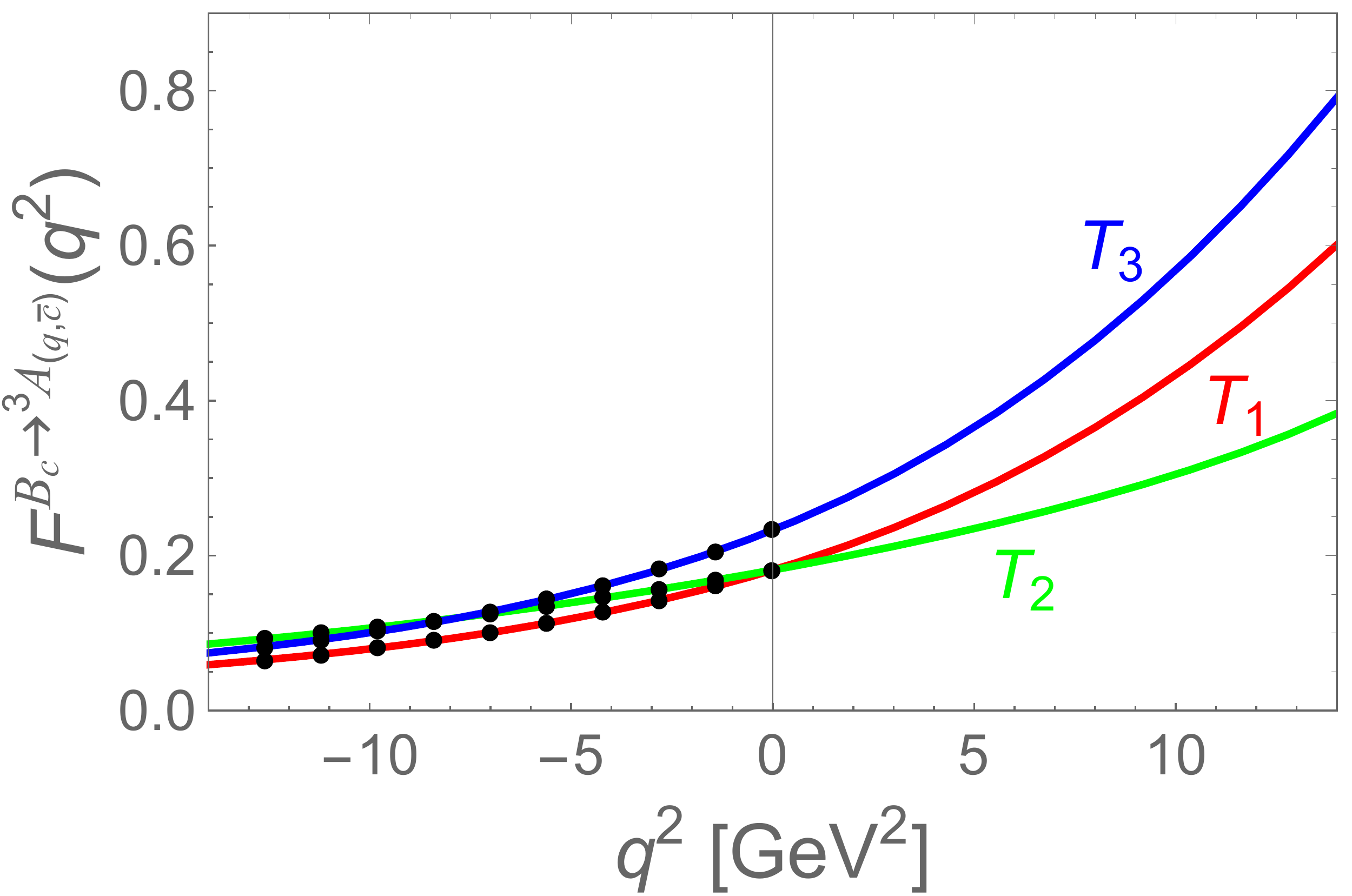}}\\
  \vspace{0.3cm}
{\includegraphics[width=0.25\textwidth]{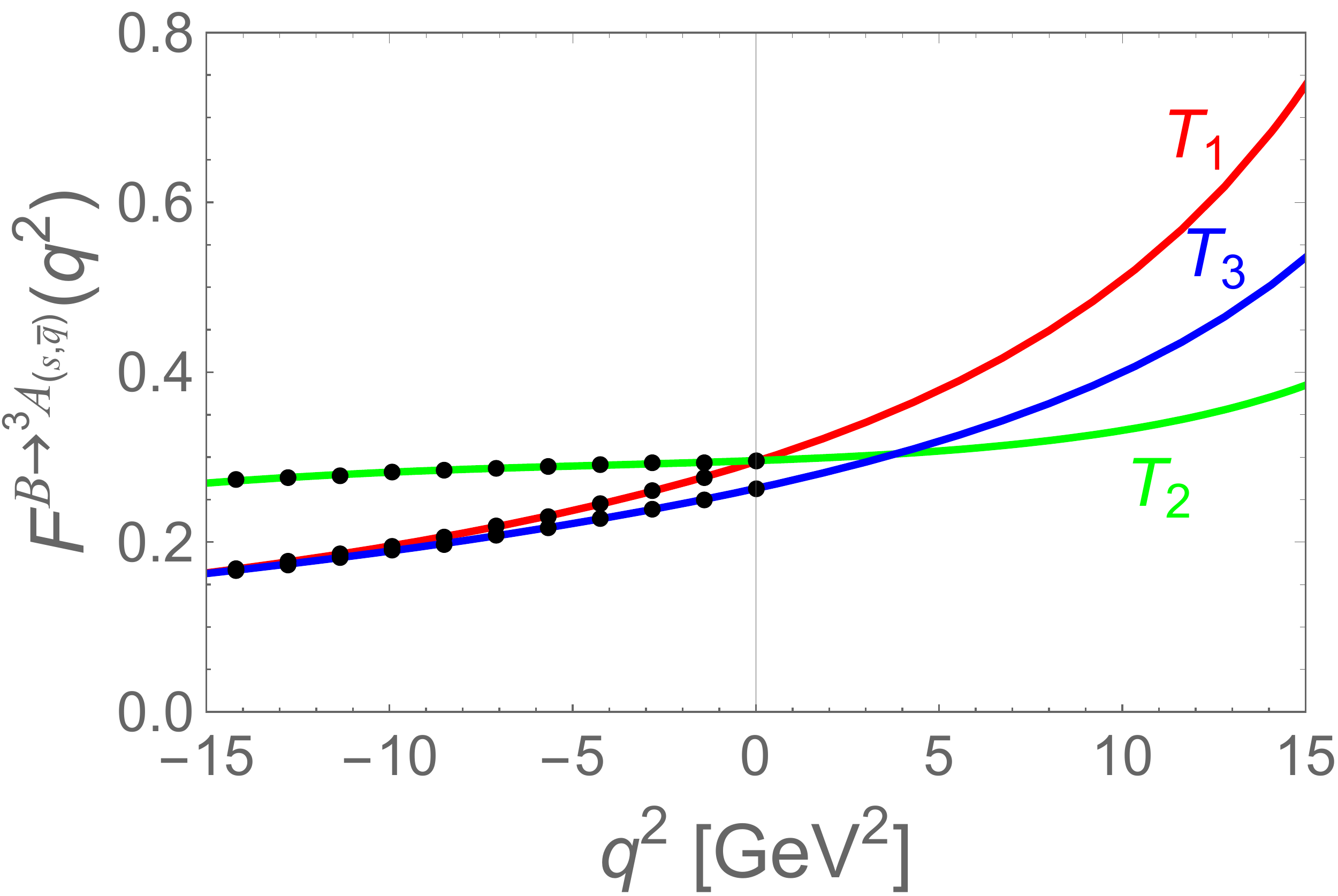}\quad}
{\includegraphics[width=0.25\textwidth]{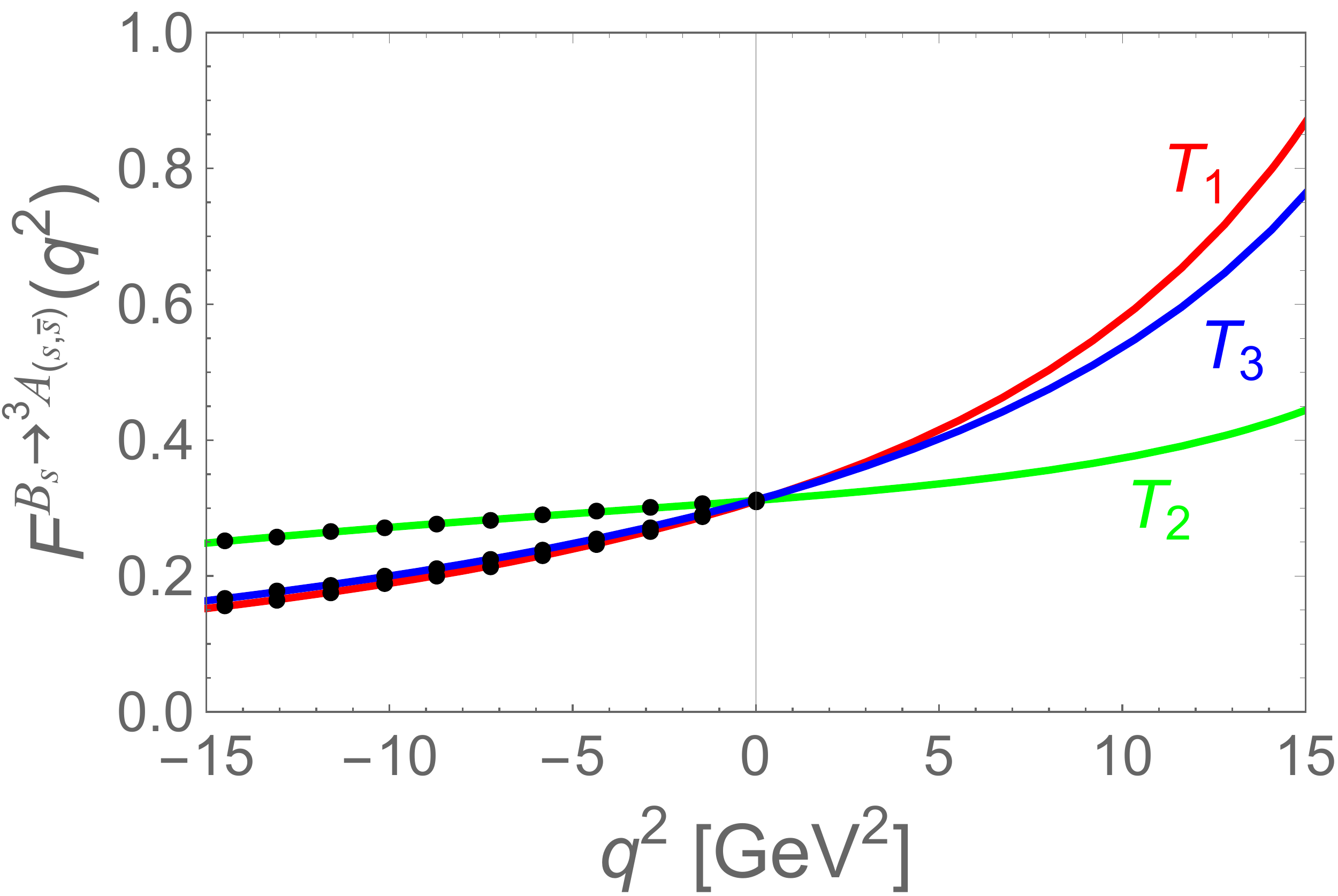}\quad}
{\includegraphics[width=0.25\textwidth]{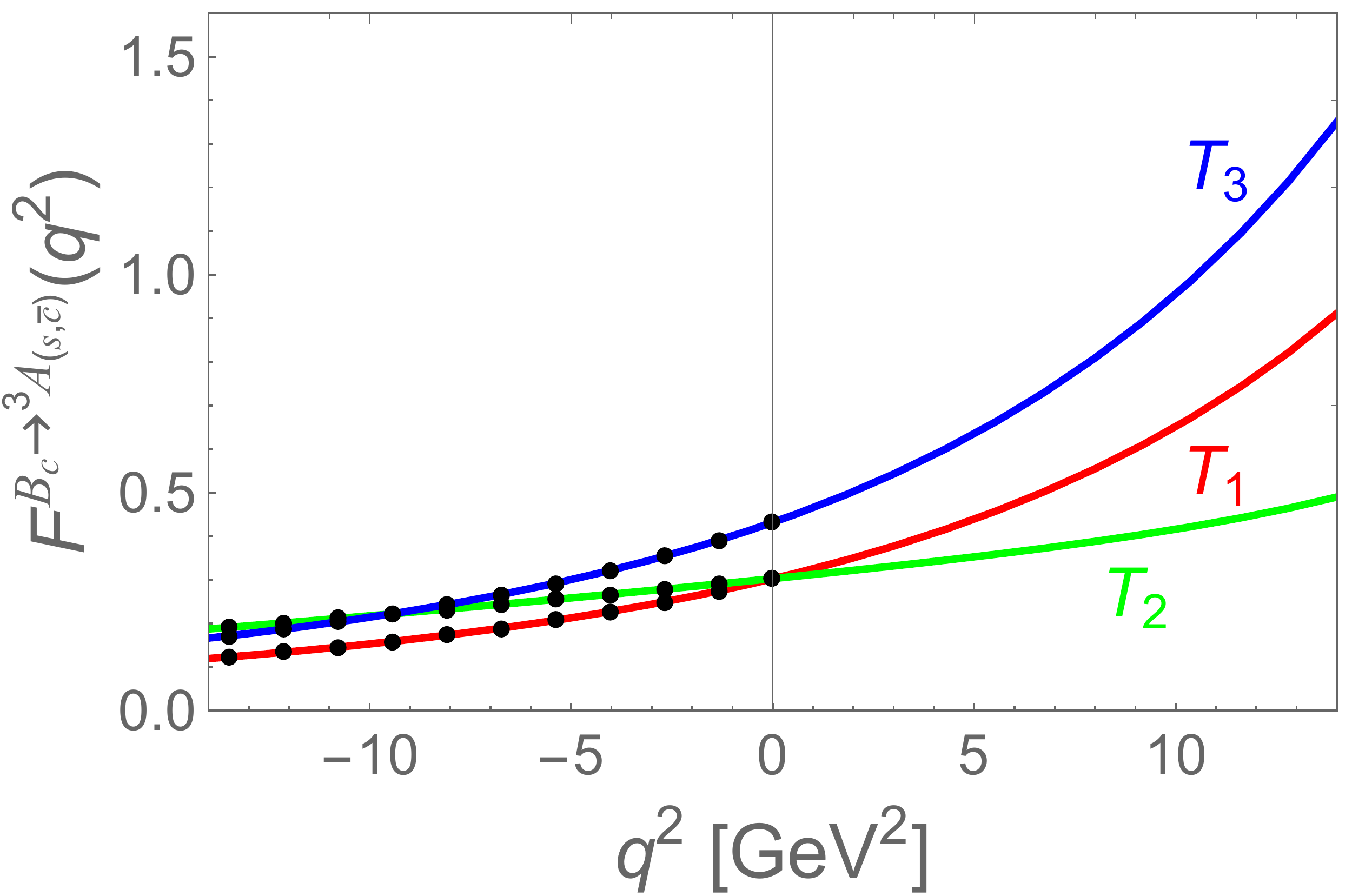}}\\
  \vspace{0.3cm}
{\includegraphics[width=0.25\textwidth]{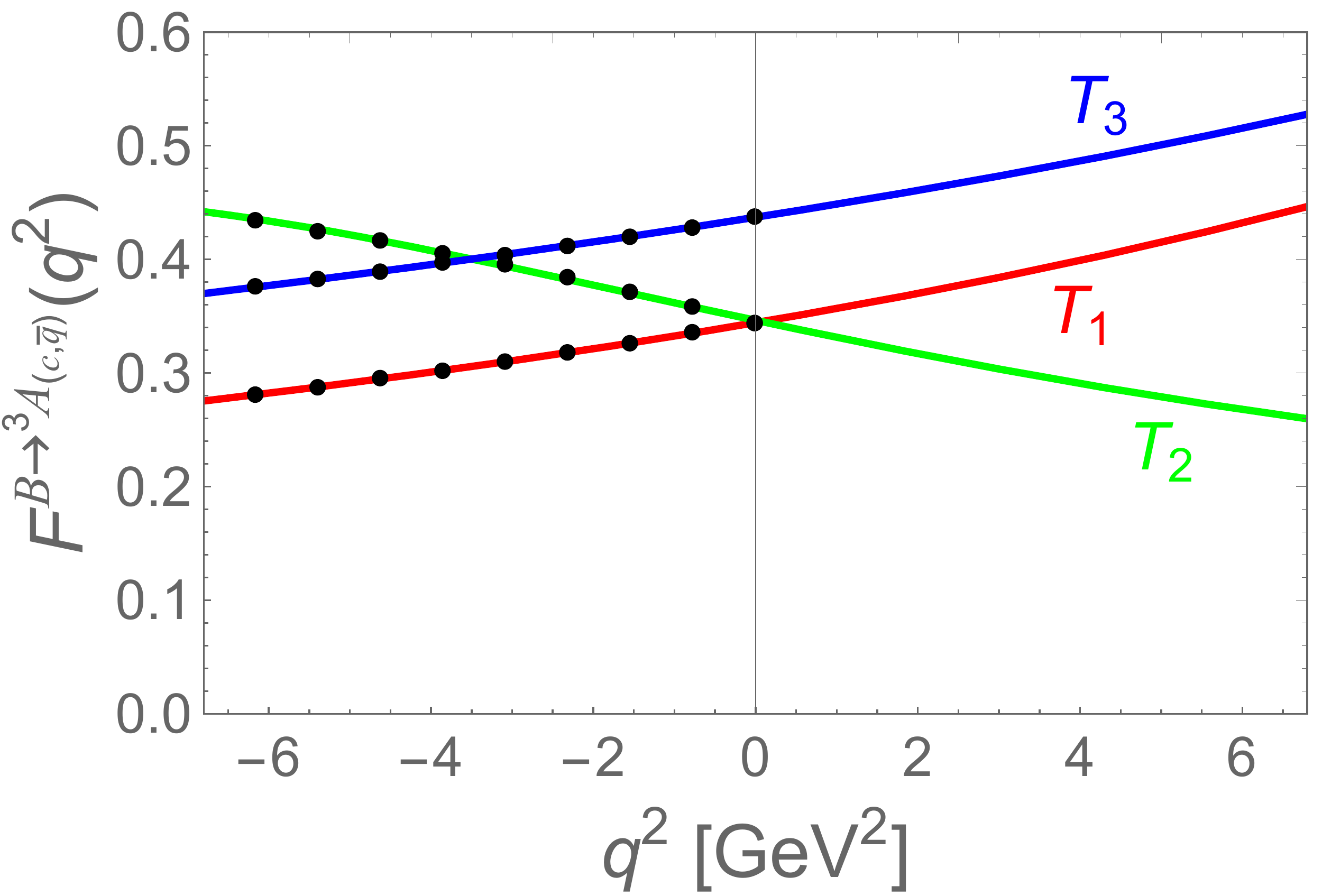}\quad}
{\includegraphics[width=0.25\textwidth]{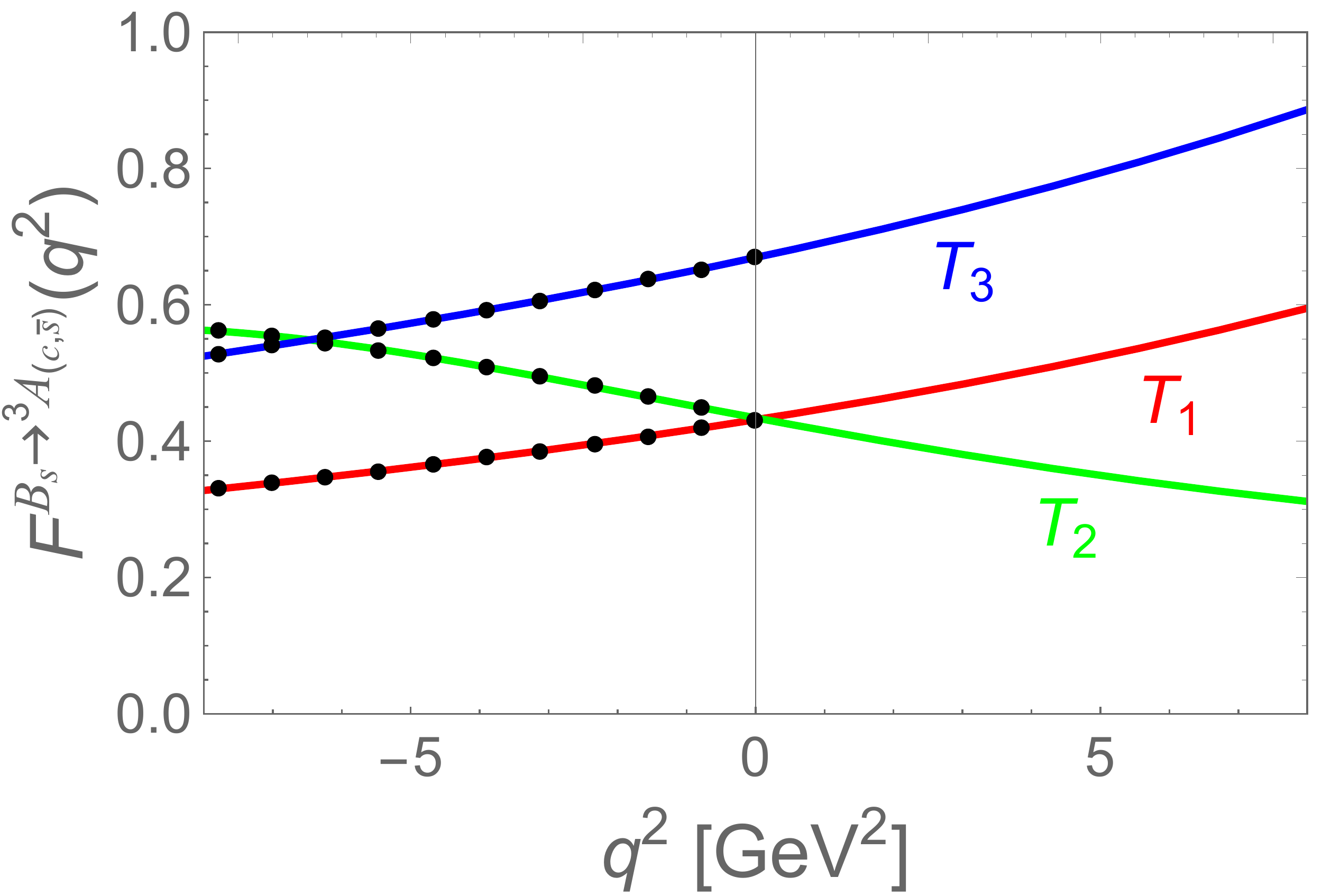}\quad}
{\includegraphics[width=0.25\textwidth]{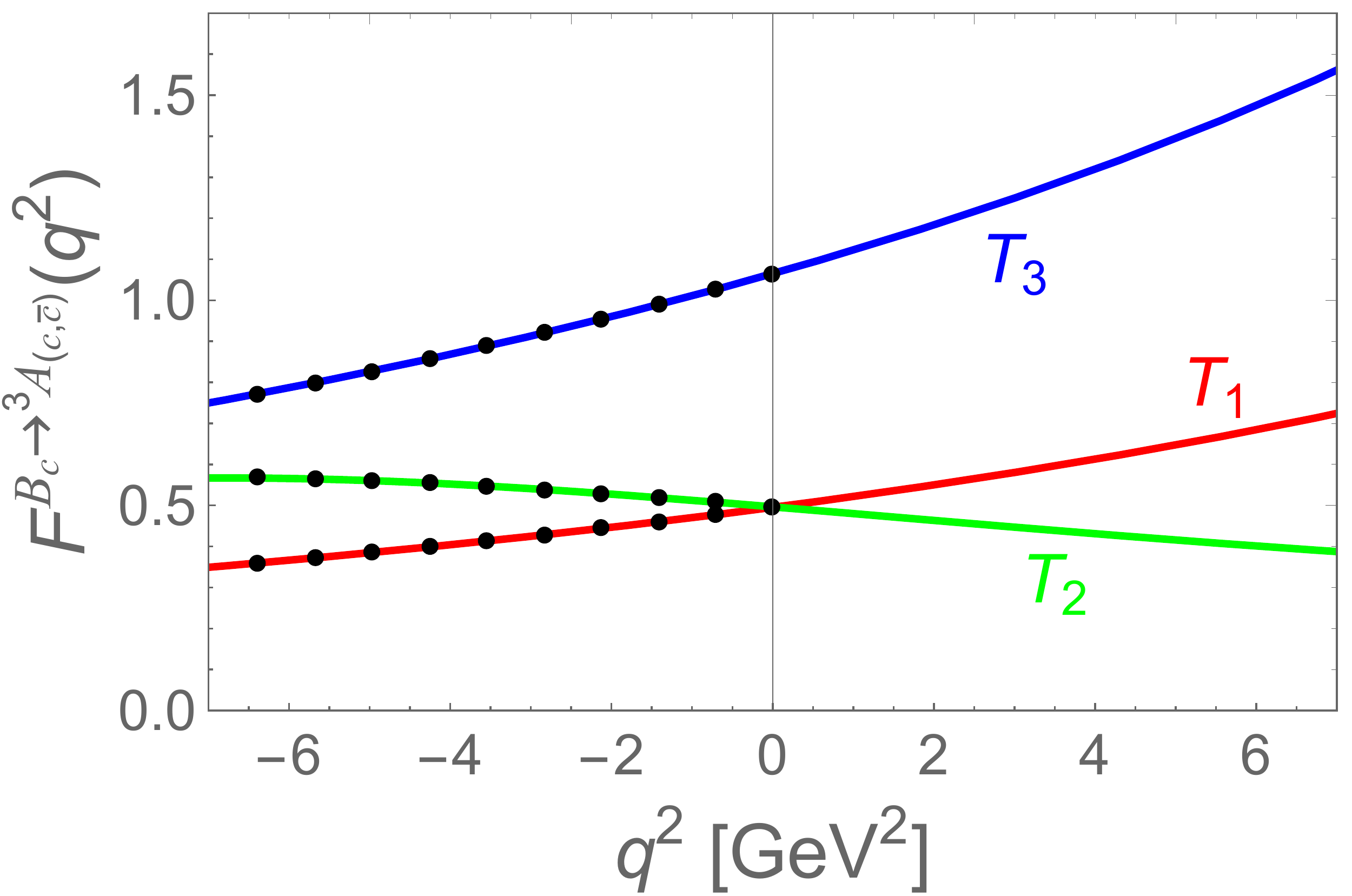}}
\caption{Same as Fig. \ref{fig:PP} except for $D_{q,s}\to {^3\!A} $ and $B_{q,s,c}\to {^3\!A} $ transitions.}\label{fig:P3A}
\end{figure}

From above analyses and discussions, it can be concluded that the type-II scheme provides a feasible solution to the covariance and self-consistency problems of the CLF QM. Therefore, we would like to update the CLF predictions for  the tensor form factors of  some $b\to c,\,s\,,q$  and $c\to s,\,q$~($q=u,d$) induced  $P\to P,\,S,\,V$ and $A$  transitions by employing self-consistent  type-II scheme. The CLF results for the tensor form factors are obtained in the $q^+=0$ frame, which implies that the form factors are known only for space-like momentum transfer, $q^2=-\mathbf{q}_\bot^2\leqslant 0$, and the results in the time-like region need  an additional $q^2$ extrapolation. To achieve this purpose, the three parameters form~\cite{Ball:1998kk}
\begin{align}\label{eq:para1}
{\cal F}(q^2)=\frac{{\cal F}(0)}{1-a(q^2/M_{B,D}^2)+b(q^2/M_{B,D}^2)^2}\,,
\end{align}
is usually employed by the LFQMs.   In Eq.~\eqref{eq:para1}, $M_{B,D}$ is the mass of the relevant $B$ and $D$ mesons, and $M_{B_{q,s,c}}$ and $M_{D_{q,s}}$~($q=u\,,d$) is used  for $b\to (q,s,c)$ and $c\to (q,s)$ transitions respectively; $a$ and $b$ are parameters obtained by fitting to the results computed  directly by LFMQs. 
However, for the case of $b\to$ light-quark transition with a heavy spectator quark, we find that the fitted results for $b$  are very large and some CLF results cannot be well reproduced by using Eq.~\eqref{eq:para1}.  Therefore, instead of  Eq.~\eqref{eq:para1}, we employ an improved form~\cite{Cheng:2004yj}
\begin{align}\label{eq:para2}
{\cal F}(q^2)=\frac{{\cal F}(0)}{\left(  1-q^2/M_{B,D}^2 \right)\left[1-a(q^2/M_{B,D}^2)+b(q^2/M_{B,D}^2)^2\right]}\,,
\end{align}
which is suitable for most of form factors considered in this paper. However, for $T_3^{(1)}$ of some transitions, the coefficient $b$ is rather sensitive to the range of $q^2$. To overcome this difficulty, we fit  $T_3^{(1)}$ to the form~\cite{Cheng:2009ms}
\begin{align}\label{eq:para3}
{\cal F}(q^2)={\cal F}(0)\left[1+a(q^2/M_{B,D}^2)+b(q^2/M_{B,D}^2)^2\right]\,.
\end{align}

Using the values of input parameters collected in appendix B, we then present  our numerical predictions for the tensor form factors in Tables~\ref{tab:PP}-\ref{tab:PA2}; and the $q^2$-dependences are shown in Figs.~\ref{fig:PP}-\ref{fig:P3A}.  From these results, it can be found that the CLF results obtained in the space-like region can be well reproduced by Eqs.~\eqref{eq:para2} and \eqref{eq:para3}, and are further  extrapolated to the time-like space.  In addition, our results for $P\to V$ and $A$  transitions respect the relation that $T_1(0)=T_2(0)$. These numerical results can be applied further in the relevant phenomenological studies of meson decays.

\section{Summary}
In this paper, motivated by the problems of LFQMs, we have investigated the tensor matrix elements and  relevant form factors of  $P\to P,\,S,\,V$ and $A$ transitions within  the SLF and the CLF approaches.  The self-consistency and Lorentz covariance of the CLF predictions for the tensor matrix elements and form factors  are analyzed in detail, and moreover, the zero-mode effects and the relation between valence contribution and SLF result are studied. As has been pointed out in our previous works, the covariance is in fact violated in the CLF QM with the traditional correspondence scheme~(type-I) between  the  manifest covariant BS  and the LF approach; moreover, for $P\to V$ and $A$ transitions, the tensor form factors extracted via $\lbd=0$ and $\pm$ polarization states of $V$ and $A$ mesons are inconsistent with each other,  $[{\cal F}]^{\rm full}_{\lambda=0}\neq [{\cal F}]^{\rm full}_{\lambda=+}\neq [{\cal F}]^{\rm full}_{\lambda=-}$~(type-I) , which implies that CLF QM has a problem of self-consistency.  It is found that such two problems have the same origin~(the non-vanishing $\w$-dependent spurious contributions associated with $B$ functions), and can be resolved simultaneously  by employing the improved type-II correspondence scheme which requires an additional replacement $M\to M_0$  relative to the traditional type-I scheme. Within the type-II  scheme,  the zero-mode corrections are only responsible for neutralizing spurious $\w$-dependent contributions associated with $C$ functions, but do not contribute numerically to the form factors; and the valence contributions in the CLF QM are exactly the same as the SLF results. The findings mentioned above confirm again the main conclusions obtained in Ref.~\cite{Choi:2013mda}  and our previous works~\cite{Chang:2018zjq,Chang:2019mmh,Chang:2019obq} . 

Besides, we find a ``new'' self-consistence problem of CLF approach with traditional type-I scheme. It is found that different strategies for dealing deal with the trace term, $S$,  in the CLF matrix element would result in different formulas for the tensor form factors $T_{2(3)}$ of $P\to V$ and $A$ transitions, and the numerical results are also inconsistent with each other  within type-I scheme; but interestingly, this new inconsistence problem can also be overcome numerically by employing  type-II scheme. Finally, using the CLF approach with the covariant and self-consistent  type-II scheme, the theoretical predictions for the tensor form factors of   $c\to q\,,s$~($q=u\,,d$) induced $D_{q,s}\to P,\,S,\,V,\,A$ and $b\to q\,,s\,,c$ induced $B_{q,s,c}\to P,\,S,\,V,\,A$ transitions are updated.

\section*{Appendix A: the CLF results for the tensor form factors of  $P\to V$ and $P\to A$ transitions given in Refs.~\cite{Cheng:2010yj,Cheng:2009ms}}
 The tensor form factors of  $P\to V$  transition in the CLF QM have  been  obtained in the previous work~\cite{Cheng:2010yj,Cheng:2009ms}, and can also been written as Eq.~\eqref{eq:FCLF} with the integrands,
\begin{align}
\widetilde{T}_1^{\rm{ CLF}}=&2A_1^{(1)}\left[M'^2-M''^2-2m'^2_1-2\hat N'_1+q^2+2\left(m'_1m_2+m''_1m_2-m'_1m''_1\right)\right]\nonumber\\
&-8A_1^{(2)}+(m'_1+m''_1)^2+\hat N'_1+\hat N''_1-q^2+4\left(M'^2-M''^2\right)\left(A_2^{(2)}-A_3^{(2)}\right)\nonumber\\
&+4q^2\left(-A_1^{(1)}+A_2^{(1)}+A_3^{(2)}-A_4^{(2)}\right)-\frac{4}{D''_{V,{\rm con}}}\left(m'_1+m''_1\right)A_1^{(2)}\,,
\label{eq:chengT1}\\
\widetilde{T}_2^{\rm{ CLF}}=&\widetilde{T}_1^{\rm{ CLF}}+\frac{q^2}{M'^2-M''^2}\bigg\{2A_2^{(1)}\left[M'^2-M''^2-2m'^2_1-2\hat N'_1+q^2+2\left(m'_1m_2+m''_1m_2-m'_1m''_1\right)\right]\nonumber\\
&-8A_1^{(2)}-2M'^2+2m'^2_1+\left(m'_1+m''_1\right)^2+2(m_2-2m'_1)m_2+3\hat N'_1+\hat N''_1-q^2+2Z_2\nonumber\\
&+4\left(q^2-2M'^2-2M''^2\right)\left(A_2^{(2)}-A_3^{(2)}\right)-4\left(M'^2-M''^2\right)\left(-A_1^{(1)}+A_2^{(1)}+A_3^{(2)}-A_4^{(2)}\right)\nonumber\\
&-\frac{4}{D''_{V,{\rm con}}}\left(m''_1-m'_1+2m_2\right)A_1^{(2)}
\bigg\}\,,
\label{eq:chengT2}\\
\widetilde{T}_3^{\rm{ CLF}}=&-2A_2^{(1)}\left[M'^2-M''^2-2m'^2_1-2\hat N'_1+q^2+2\left(m'_1m_2+m''_1m_2-m'_1m''_1\right)\right]+8A_1^{(2)}\nonumber\\
&+2M'^2-2m'^2_1-(m'_1+m''_1)^2-2(m_2-2m'_1)m_2-3\hat N'_1-\hat N''_1+q^2-2Z_2\nonumber\\
&-4\left(q^2-M'^2-3M''^2\right)\left(A_2^{(2)}-A_3^{(2)}\right)\nonumber\\
&+\frac{4}{D''_{V,{\rm con}}}\bigg\{\left(m''_1-m'_1+2m_2\right)\left[A_1^{(2)}+\left(M'^2-M''^2\right)\left(A_2^{(2)}+A_3^{(2)}-A_1^{(1)}\right)\right]\nonumber\\
&+\left(m'_1+m''_1\right)\left(M'^2-M''^2\right)\left(A_2^{(1)}-A_3^{(2)}-A_4^{(2)}\right)+m'_1\left(M'^2-M''^2\right)\left(A_1^{(1)}+A_2^{(1)}-1\right)
\bigg\}\,.
\label{eq:chengT3}
\end{align}
The results for  $P\to A$  transition  can be obtained via. the relations given by Eqs.~\eqref{eq:CLFT1A} and \eqref{eq:CLFT3A}.

\section*{Appendix B: Input parameters}
 The masses of valence quark and  Gaussian parameters $\beta$ are essential inputs for computing the form factors. For the former, we take~\cite{Chang:2019obq}
\begin{align}
m_q&=230\pm40\,{\rm MeV}\,,\quad  m_s=430\pm60\,{\rm MeV}\,,\nonumber\\
m_c&=1600\pm300\,{\rm MeV}\,,\quad  m_b=4900\pm400\,{\rm MeV}\,,
\label{eq:mq}
\end{align}
which can cover properly the fitting results and suggested values given in the previous works, for instance,  the result obtained via variational analyses of meson mass spectra  for the Hamiltonian with a smeared-out hyperfine interaction~\cite{Choi:2015ywa},  the values obtained by the variational principle for the linear and harmonic oscillator~(HO) confining potentials, respectively~\cite{Choi:2009ai}, the fitting results obtained via   decay constants and mean square radii of mesons~\cite{Hwang:2010hw}, some commonly used values in the LFQMs~\cite{ Cheng:2003sm,Verma:2011yw} and so on. For the later, its value for a given meson can be obtained by  fitting  to the  data of decay constant.  Using  the  data of decay constant, $f_{P,V}$, collected in Ref.~\cite{Chang:2018zjq} and the default values of quark masses given by Eq.~\eqref{eq:mq}, we obtained the values of $\b$ collected in Table~\ref{tab:input}, in which it have been assumed that $\beta_{q_1\bar{q}_2}$ is universal for $P(V)$ and $S(A)$ mesons due to the lack of  data for $f_{S,A}$. In addition, the self-consistent type-II scheme is employed in computing decay constants.

\begin{table}[t]
\begin{center}
\caption{\label{tab:input} \small The values of Gaussian parameters $\beta$ (in units of MeV).}
\vspace{0.2cm}
\let\oldarraystretch=\arraystretch
\renewcommand*{\arraystretch}{1}
\setlength{\tabcolsep}{8.8pt}
\begin{tabular}{lcccccccccc}
\hline\hline
  &$\beta_{q\bar{q}}$    &$\beta_{s\bar{q}}$   &$\beta_{s\bar{s}}$
  &$\beta_{c\bar{q}}$    &$\beta_{c\bar{s}}$
  \\ \hline
$P$ ($S$)  &$348\pm1$ &$365\pm2$ &$384\pm3$ &$473\pm12$ &$543\pm10$
  \\ \hline
$V$ ($A$)    &$312\pm6$ &$313\pm10$ &$348\pm6$ &$429\pm13$ &$530\pm19$
   \\\hline\hline
  &$\beta_{c\bar{c}}$    &$\beta_{b\bar{q}}$   &$\beta_{b\bar{s}}$   &$\beta_{b\bar{c}}$   &$\beta_{b\bar{b}}$
    \\ \hline
$P$ ($S$)  &$753\pm14$ &$552\pm10$ &$606\pm12$ &$939\pm11$ &$1394\pm12$
\\  \hline
$V$ ($A$)   &  $703\pm7$ &$516\pm15$ &$568\pm10$ &$876\pm20$ &$1390\pm12$\\
\hline\hline
\end{tabular}
\end{center}
\end{table}

\newpage
\section*{Acknowledgements}
This work is supported by the National Natural Science Foundation of China (Grant No. 11875122) and the Program for Innovative Research Team in University of Henan Province (Grant No.19IRTSTHN018).


\end{document}